\documentclass[12pt]{article}
\usepackage{graphicx}
\usepackage{tabularx, booktabs}
\newcolumntype{Y}{>{\centering\arraybackslash}X}
\newcolumntype{B}{>{\centering\arraybackslash\hsize=1.5\hsize}X}
\newcolumntype{Q}{>{\centering\arraybackslash\hsize=2\hsize}X}
\usepackage{titlesec}
\usepackage{wrapfig}
\usepackage{epstopdf}
\usepackage{setspace}
\usepackage[top=1.5in, bottom=1.5in, left=1in, right=1in]{geometry}
\usepackage{amsmath}
\usepackage[title,titletoc,toc]{appendix}
\usepackage[table]{xcolor}
\usepackage{caption, subcaption}
\usepackage[round]{natbib}
\usepackage{lipsum}
\usepackage{tikz}
\usepackage{changepage}

\usepackage{indentfirst}
\usepackage{hyperref}           
\hypersetup{
	colorlinks=true,
	citecolor=blue
}
\graphicspath{ {Figures/} }

\usepackage{dsfont}
\usepackage{algorithm}
\usepackage{algorithmic}
\usepackage{leading} 
\usepackage{setspace}

\usepackage[normalem]{ulem}

\title{\bf Cross-Temporal Forecast Reconciliation at Digital Platforms with Machine Learning\footnote{Correspondence to Ines Wilms, Maastricht University, School of Business and Economics, P.O. Box 616, 6200 MD Maastricht, The Netherlands, Email: i.wilms@maastrichtuniversity.nl. 
We are very grateful to Benjamin Wolter, Pablo Perez Piskunow, Roger Caminal and Afonso Rodrigues for expert advice, to Rob Hyndman, Artem Prokhorov and Roberto Ren\`o for comments  provided on earlier versions of the paper, and to the participants of  the IMS International Conference on Statistics and Data Science (ICSDS 2023) for helpful discussions. 
The last author was financially supported by the Dutch Research Council (NWO) under grant number VI.Vidi.211.032. 
}}

\author{Jeroen Rombouts$^{a}$, Marie Ternes$^b$, Ines Wilms$^b$ \\ \normalsize $^{a}$Essec Business School, France \\ \normalsize $^b$Maastricht University, School of Business and Economics, The Netherlands}

\date{\today }

\begin{document}

\begin{titlepage}
\clearpage\thispagestyle{empty}
\maketitle

\begin{singlespace} 
\noindent 
{\bf Abstract.} Platform businesses operate on a digital core and their decision making requires high-dimensional accurate forecast streams at different levels of cross-sectional (e.g., geographical regions) and temporal aggregation (e.g., minutes to days). It also necessitates coherent forecasts across all levels of the hierarchy to ensure aligned decision making across different planning units such as pricing, product, controlling and strategy. Given that platform data streams feature complex characteristics and interdependencies, we introduce a non-linear hierarchical forecast reconciliation method that produces cross-temporal reconciled forecasts in a direct and automated way through the use of popular machine learning methods. The method is sufficiently fast to allow forecast-based high-frequency decision making that platforms require. We empirically test our framework on unique, large-scale streaming datasets from a leading on-demand delivery platform in Europe and a bicycle sharing system in New York City.
\end{singlespace}

\bigskip


\thispagestyle{empty}
\end{titlepage}

\doublespacing

\section{Introduction \label{sec:introduction}}
Time series to be forecasted are oftentimes naturally part of a hierarchical structure, where higher frequency and granular series are added together to form lower frequency aggregated series. Separate forecasts of each series rarely conserve this hierarchy, and \textit{forecast reconciliation} methods are therefore required. 
In this paper, we consider novel forecast reconciliation methods for on-demand delivery platforms that require forecasts at different levels of cross-sectional and temporal aggregation. We introduce non-linear forecast reconciliation based on popular machine learning methods to capture the complex interdependencies of platform data.

Platforms such as Uber, Lyft,  GrubHub, UberEats or DoorDash  are nowadays omnipresent in the global economy. They operate in high frequency and on a large number of verticals (regions, product categories, etc.). 
Indeed, their market place is typically split up in different geographical regions, so a natural \textit{cross-sectional} aggregation scheme arises from many individual delivery areas over zones towards few market places.
Moreover, a \textit{temporal} aggregation scheme naturally arises since, on the granular end, fast operational decisions (think in terms of minutes) are needed to ensure the platform’s service couriers are at the right time and location to serve consumer demand promptly and to determine compensation schemes for couriers through dynamic pricing. 
On the coarser end, strategic business decisions also require long-term planning since the budget available for each delivery area is set typically using daily demand forecasts. 
Accurate and coherent, i.e.\ \textit{reconciled}, demand forecasts across all levels of the cross-sectional and temporal hierarchy are therefore key to the business' success and to support aligned decision making across different planning units.

The literature to reconcile forecasts has been pioneered by \cite{Hyndman_2011_CSDA} and \cite{Hyndman_2009_INTFOR}, see  \cite{hyndman2021forecasting} for a textbook introduction and \cite{hyndman2014optimally} for an introduction to practitioners given the strong interest from industry.
Typical initial applications concern tourism where demand series are subject to a cross-sectional hierarchy across different geographical regions, though many other cases appear more recently, see for instance \cite{caporin2023exploiting} for a financial application to realized volatility forecasting.  
While forecast reconciliation methods initially focused on producing coherent forecasts across \textit{cross-sectional} hierarchies, the literature on forecast reconciliation in the \textit{temporal} direction (\citealp{athanasopoulos2017temporal}) as well as combinations of both in \textit{cross-temporal} frameworks is much sparser  though in recent years it has proliferated \citep{athanasopoulos2023review}. 
Early work on cross-temporal forecast reconciliation can be found in \cite{kourentzes2019cross} which has then been  formalized further in \cite{difonzo2023cross,difonzo2023spatio}, and extended to a probabilistic forecast setting in \cite{GIROLIMETTO2023}. 
We refer the interested reader  to \cite{athanasopoulos2023review}
for an extensive review on forecast reconciliation in cross-sectional, temporal and cross-temporal hierarchies, to \cite{HOLLYMAN_EJOR_2021149} and \cite{DIFONZO2024490} for the link with the forecast combination literature, and to \cite{ATHANASOPOULOS2024} for recent innovations in hierarchical forecasting.

Methodological considerations to forecast reconciliations are discussed in \cite{hyndman2016fast}, whereas theoretical results are first established by 
\cite{wickramasuriya2019optimal} by putting forward an optimal linear  forecast reconciliation method, minimizing the sum of forecast error variances across the collection of time series.
\cite{van2015game} 
offer a game-theoretic approach to forecast reconciliation, \cite{panagiotelis2021forecast} provide  
a unification of the former two works, and \cite{girolimetto2023point}  present a generalization of point and probabilistic forecast
reconciliation for general, linearly constrained multiple time series.
Traditionally, forecast reconciliation is thus typically achieved by \textit{linearly} combining base forecasts of all the time series in the hierarchy. Recently, however, applications of machine learning  algorithms surface to reconcile forecasts in a \textit{non-linear} way.

Machine learning (ML) algorithms nowadays have become an integral component of every forecaster's toolkit, boasting  a remarkable track record as demonstrated by  various forecasting competitions, such as the M4 and M5 competitions \citep{makridakis2020m4, makridakis2022m5}, and successful economic applications (e.g., \citealp{Medeiros_2021}). ML methods have become standard choices for obtaining forecasts as their algorithms demand minimal user intervention-- typically limited to tuning parameters --and their sophisticated off-the-shelf implementations are readily available, facilitating seamless operationalization in production environments. 
Moreover, ML methods offer flexibility by imposing no constraints on the number of features for a given target, and they come with built-in mechanisms for feature selection and for capturing complex dependencies such as non-linearities. 

In the forecast reconciliation literature, a cross-sectional ML-based reconciliation scheme is proposed by \cite{spiliotis2021hierarchical}, on which our work is mainly built.  
They first collect base forecasts for all series in the cross-sectional hierarchy and then train
a tree-based ML algorithm (random forest or XGBoost) with the bottom level series as response and one-step-ahead forecasts of all the time series in the hierarchy as input features. 
The ML fitted algorithm is hereby used to revise the bottom level forecasts in a  non-linear way, given all base forecasts, and coherent forecasts across the complete hierarchy are obtained via bottom-up aggregation. 
In contrast to standard linear reconciliation approaches that reconcile forecasts by minimizing the reconciliation error between the base and reconciled forecasts, the ML-based reconciliation approach is directly geared towards obtaining the minimum forecast combination error to improve out-of-sample forecast accuracy while still adhering to coherence. This results in substantial forecast improvements as demonstrated on
two applications to tourism and retail industry.
Additionally,  ML-based methods are more
flexible than the standard ones as they allow for non-linear combinations
of the base forecasts and selectively combine them in a direct and automated way without
requiring that the information in the entire hierarchy must be used for producing reconciled forecasts.

\cite{anderer2022hierarchical} use deep learning and LightGBM tree-based methods to propose a variant of the bottom-up method proposed by \cite{spiliotis2021hierarchical}, accounting for biases that are difficult to observe at the bottom level. The latter forecasts are adjusted to obtain a higher forecasting accuracy on the upper levels of the hierarchy. Their approach works well on the M5 competition dataset  consisting of over ten thousand hierarchical retail sales time series (see \citealp{makridakis2022m5hypothesizing} and \citealp{makridakis2022m5accuracyresults} for details).
\cite{abolghasemi2022machine} use tree-based methods and lasso regression in their  dynamic top-down bottom-up  method applied to fast-moving consumer goods and the M5 competition dataset. Finally, a new stream of research uses neural networks for both forecasting and reconciliation;  examples of such end-to-end modeling approaches are \cite{theodosiou2021forecasting} and  \cite{wang2022end}, which bring promising evidence on various datasets.

Our paper also proposes an ML-based reconciliation approach,  extending the work of \cite{spiliotis2021hierarchical} in the following ways.
From a methodological point of view, 
we generalize their ML-based forecast reconciliation method  for cross-sectional hierarchies to cross-temporal hierarchies.
Additionally considering reconciliation across the temporal direction requires important design choices with respect to the ML model used for reconciliation since the inputs consist of mixed-frequency time series, namely all temporal frequencies in the temporal hierarchy. 
Besides, since the ML algorithm is a key backbone in our forecast reconciliation approach, we carefully investigate the 
effect of (i) the ML algorithm used, thereby considering not only random forest and XGBoost as in \cite{spiliotis2021hierarchical} but also LightGBM due to its successful application in the M5 competition; and
(ii) tuning hyperparameters 
to make careful considerations on the trade-off between improvements in forecast accuracy and computational time.
Finally, we consider an extensive set of base forecasts, ranging from academic standards to industry standards (for our platform applications) as well as a 
combination of several base forecasts methods. 

From an application viewpoint, we showcase the performance of our forecast reconciliation method on three prime platform demand streaming datasets, thereby bringing novel insights into the performance of forecast reconciliation methods on modern and unique datasets. 
In our first two platform application, we
use UK demand data (September 2022-September 2023) from a last mile logistics platform where the bottom level series are 30-minute demand series for 117 delivery areas in London (dataset 1) and 24 delivery areas in Manchester (dataset 2), their two largest markets. We aggregate to hourly and daily frequency in the temporal dimension, and towards zone and/or market level in the cross-sectional dimension. 
In our third platform application, we use New York demand rental data (January 2023-December 2023) from a bicycle sharing system where the bottom level series are 30-min demand series for six H3 cells-- a geographical hexagon unit proposed and used by Uber --in New York City. Here, we consider the set of all possible temporal aggregation orders for a 30-min/1 day temporally grouped time series, and we aggregate towards the New York City level in the cross-sectional dimension.

Platform data  have several features that are distinct from the traditional datasets considered in the forecast reconciliation literature.
First, they are high-frequency with, 
our top, hence lowest frequency, level (i.e.\ daily) consisting of the bottom level considered in earlier work such as recent M5 competition (other works even consider lower frequency settings such as monthly to yearly data). Indeed, computational speed and fast
decision making at digital platforms are key to their success, thereby requiring forecasts at the intra-day level.
A second distinctive feature of platform data is their streaming character which means that they face non-stationarities or shifts for example due to expansion, competition among platforms, sales of certain clients, changing customer's preferences or non-regular events such as sports events.
The data on Manchester highlight the performance of the ML forecast reconciliation method in presence of data shifts.
We compare our approach with state-of-the art reconciliation benchmarks from the recent literature. 

We obtain several insights based on our applications.
First, our main empirical finding is that ML-based forecast reconciliation for platform data can result in substantial forecast accuracy improvements compared to existing linear reconciliation methods.
However, ML-based reconciliation is not uniformly superior to linear methods. In fact, for the most important series, i.e.\ high frequency  
area level, random forest based reconciliation typically performs best, for market level series LightGBM based reconciliation yields the smallest forecast errors. On the other hand, linear reconciliation methods  outperform XGBoost based reconciliation on the 
bottom level series. The latter result holds for XGBooost without tuning its hyperparameters to keep computing time low. However, ignoring the huge computational cost, we also study the impact of hyperparameter tuning of the ML algorithms and find for example that XGBoost performs in this case at par with random forest in the lowest levels of the hierarchy.
A second empirical finding is that ML-based reconciliation 
seems to work equally well irrespective of the quality of the base forecasts. Hence, exploring a large model space for obtaining base forecasts is unnecessary. 
Finally, data streams are  prone to data shifts so we compare the 
forecast reconciliation methods on the Manchester market
that is impacted by a major shift in our evaluation period. The results show that, while forecast performance of  all methods suffer during a data shift, ML-based reconciliation is able to  revert quickly back to pre-shift forecast accuracies.

The rest of the paper is organized as follows. 
Section \ref{sec:data} introduces the platform data we analyze in this paper. 
Section \ref{sec:methodology} presents our ML-based forecast reconciliation approach for cross-temporal hierarchies.
Section \ref{sec:setup} describes the forecast set-up we use to evaluate the out-of-sample forecast performance of our ML-based reconciliation approach versus state-of-the-art benchmarks whereas Section \ref{sec:results} presents the results. 
Finally, Section \ref{sec:conclusion} concludes.

\section{Platform Data \label{sec:data}}
We describe the  three datasets that  we use to demonstrate the performance of the ML-based reconciliation methods.
All are prime examples of platform data, the first two come from an on-demand last-mile logistics platform (Section \ref{subsec:stuart}), the third from a bicycle sharing platform (Section \ref{subsec:citibike}).

\subsection{Last-Mile Logistics Platform Data} \label{subsec:stuart}
We use two unique datasets of demand for deliveries  from a leading on-demand last-mile logistics platform in Europe, Stuart, which connects businesses to a fleet of geographically independent couriers.\footnote{The data are provided to us in the context of ongoing research collaborations. Due to a confidentiality agreement, we are not allowed to distribute or report actual demand data. Demand data across all figures are therefore normalized between 0-100. All analyses were carried out with the original data.} We analyze UK London and Manchester demand 30-minute frequency data from September 5, 2022 to September 24, 2023.
To efficiently organize parcel deliveries, London is split into 18 zones which are constituted by the 117 bottom-level delivery areas, whereas Manchester cross-sectionally consists of 24 delivery areas at the bottom-level and Manchester as the entire market - given its size there is no intermediate level consisting of zones in Manchester. 
Planning at the delivery area level is of particular importance because it determines the remuneration paid to the couriers and   it therefore directly helps to optimize the platform's efficiency.

Deliveries are carried out on a daily basis between 7am and 11:30pm in all areas.\footnote{Occasional overnight demand is integrated in the last 30-min of the business day.} This creates a balanced dataset of $13,090$ 
observations at the 30-minute frequency for each of the 136 (=1+18+117) time series  in London and for each of the 25 (=1+24) time series in Manchester. For business planning purposes, the series also require  temporal aggregation to the hourly and the daily frequency, thereby resulting in temporally aggregated time series that consist of 
$13,090/2=6,545$ and $13,090/34=385$
observations respectively. 
We restrict ourselves to these three temporal frequencies since 
these are used for decision making, and 
doing so permits the platform to keep the computational burden under control in a production environment where speed is vital.
For the London dataset, the forecast reconciliation therefore consists of three cross-sectional and three temporal layers implying 408 distinct time series. 
For the Manchester dataset, there are two cross-sectional and three temporal layers implying 75 distinct time series.
The cross-temporal hierarchies of both datasets are summarized in Table \ref{overview:datasets}.

\begin{table}
	\centering
	\caption{Overview of the cross-temporal hierarchy for the three datasets.} \label{overview:datasets}
	\begin{tabular}{lccccccc} \hline
		\textbf{Hierarchy} &&& \textbf{London} && \textbf{Manchester} && \textbf{Citi Bike} \\ \hline 
		\textbf{Cross-sectional} &&& && && \\
		number of levels &&& 3 && 2 && 2 \\
		$n_b$ bottom-level series &&& 117 && 24 && 6 \\
		$n_a$ upper-level series &&& 19 && 1 &&  1\\
		$n$ total series &&& 136 && 25 && 7 \\
		&&& && && \\
		\textbf{Temporal} &&& && && \\
		number of  levels &&& 3 && 3 && 10 \\ \hline
		
		\hline
	\end{tabular}
\end{table}
\begin{figure}[t]
	\includegraphics[width = \textwidth]{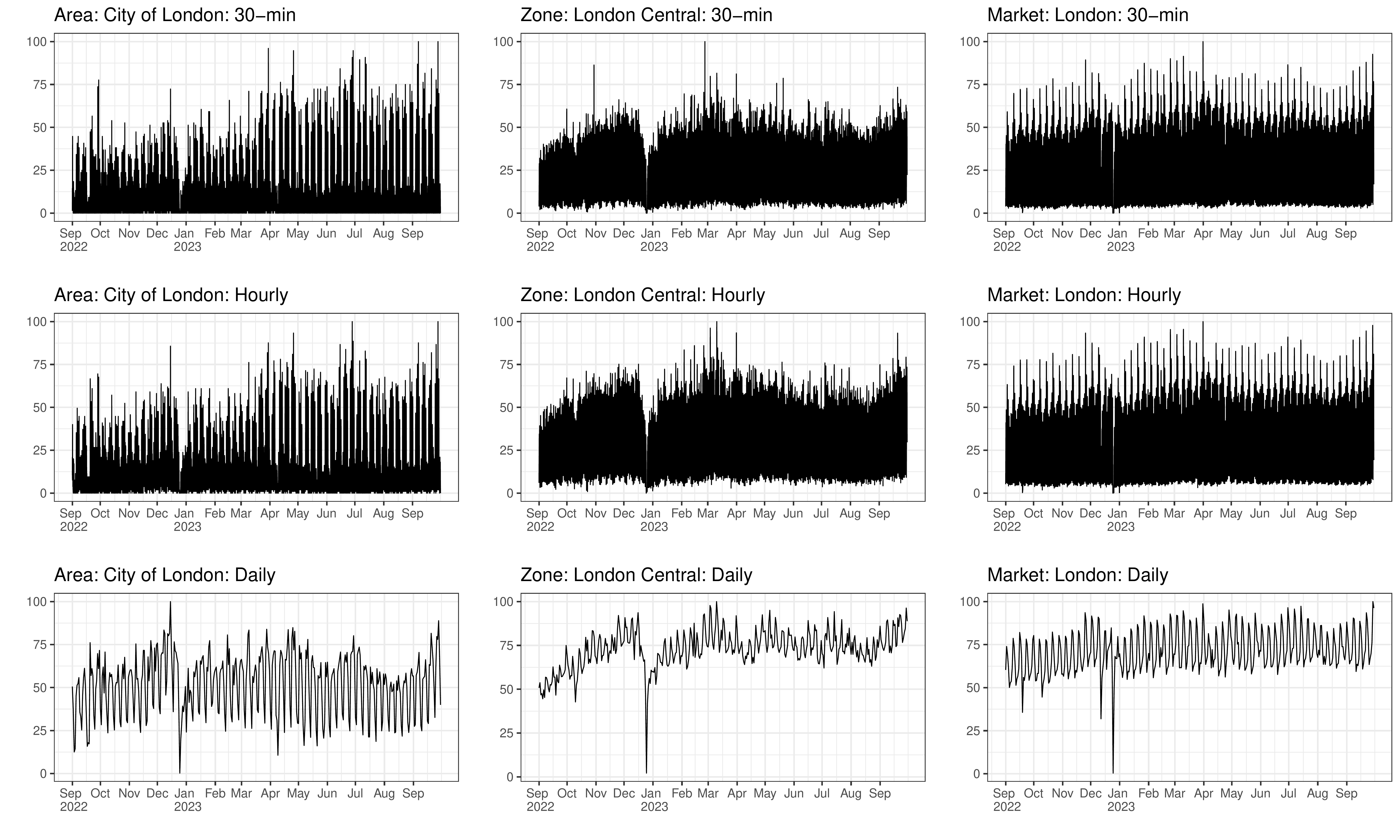}
	\caption{Time series plots for full sample of the area City of London belonging to the zone London Central and the market London.Demand data are normalized between 0-100 in each panel. 
		To relate the scale of the series at the different cross-sectional levels (area, zone, market), note for reference that 
		demand at the market level is on average 15 times larger than at the zone level which is, in turn, on average 18 times higher than at the area level.
	}
	\label{fig:map_London_series_full_smample}
\end{figure} 

\begin{figure}[t]
	\includegraphics[width = \textwidth]{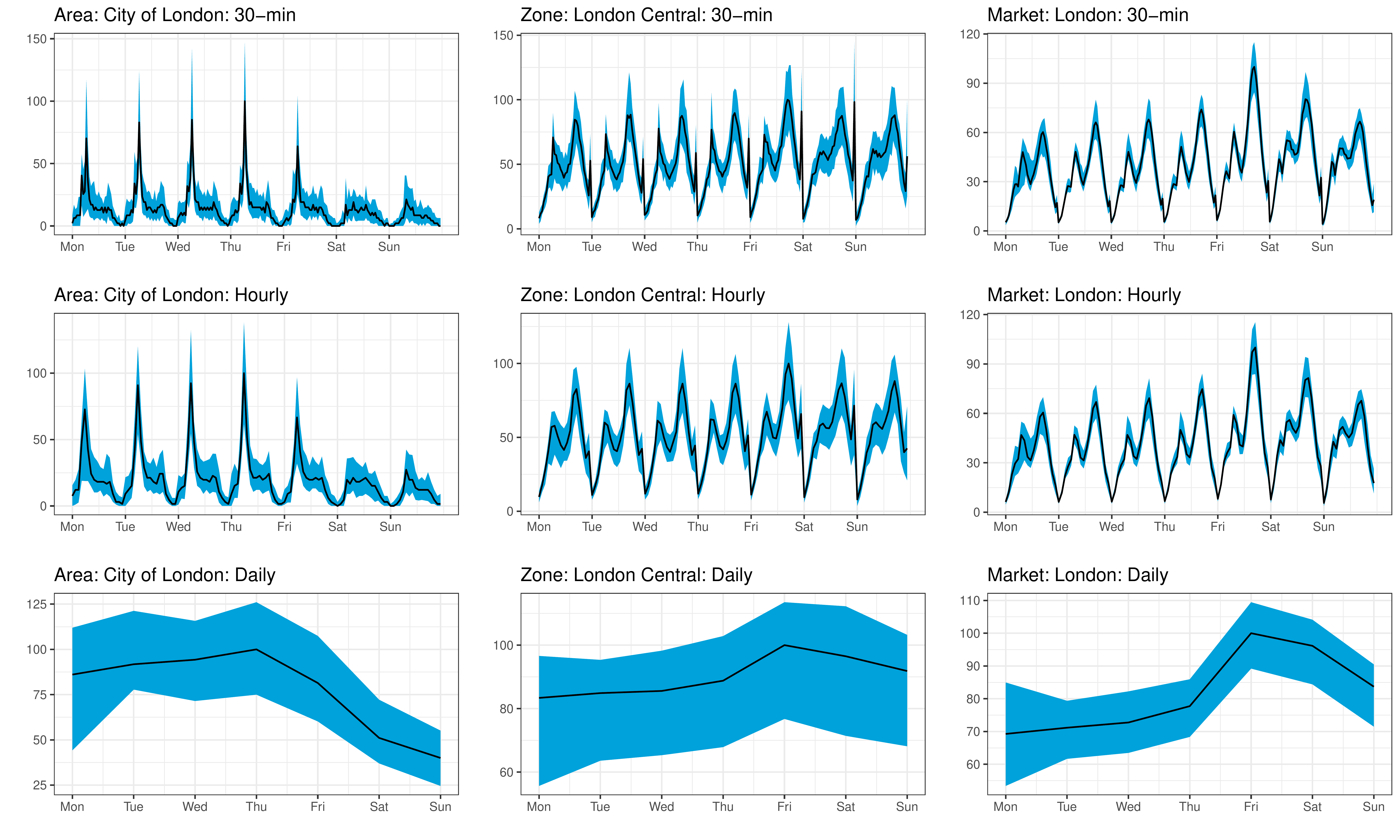}
	\caption{Time series plots for 30-min, hourly and daily demand (in the rows) for the  area City of London belonging to the zone London Central and the market London (in the columns).
		The black solid line is the pointwise median. The shaded blue area displays the 5\% and 95\% pointwise quantiles.}
	\label{fig:map_London_seasonalities_hierarchy}
\end{figure}

\textit{London Dataset}.
Figure \ref{fig:map_London_series_full_smample} plots demand for the City of London delivery area, London Central zone, and London market at 30-minute, hourly and daily frequencies over the sample period. We observe some typical characteristics of platform data. First, intra-day area data is spiky because of peak demand periods triggered by events such as flash sales, food promotions, bad weather. These peaks are largely tempered at daily frequency. Holidays can also cause large drops in delivery demand, this is particularly noticeable in the end-of-year period. 
Second, there are strong seasonality patterns at all levels of the hierarchy. Figure \ref{fig:map_London_seasonalities_hierarchy} 
displays these seasonality patterns. We clearly see that the intra-day seasonality (visible from  both the 30-minute and hourly time series) is mainly driven by food deliveries. The bottom row of the plots traces daily seasonality, and illustrates that there can be substantial differences between weekends and the rest of the week. In general, London faces higher weekend demand while the opposite is true for the delivery area City of London. This implies that there exists heterogeneity in demand patterns at the delivery area level.  In fact, unreported plots for other delivery areas reveal that very different dynamics appear.  

\begin{figure}
	\makebox[\textwidth][c]{
		\begin{subfigure}{0.52\textwidth}
			\includegraphics[height = 6.5cm, width = 8.5cm]{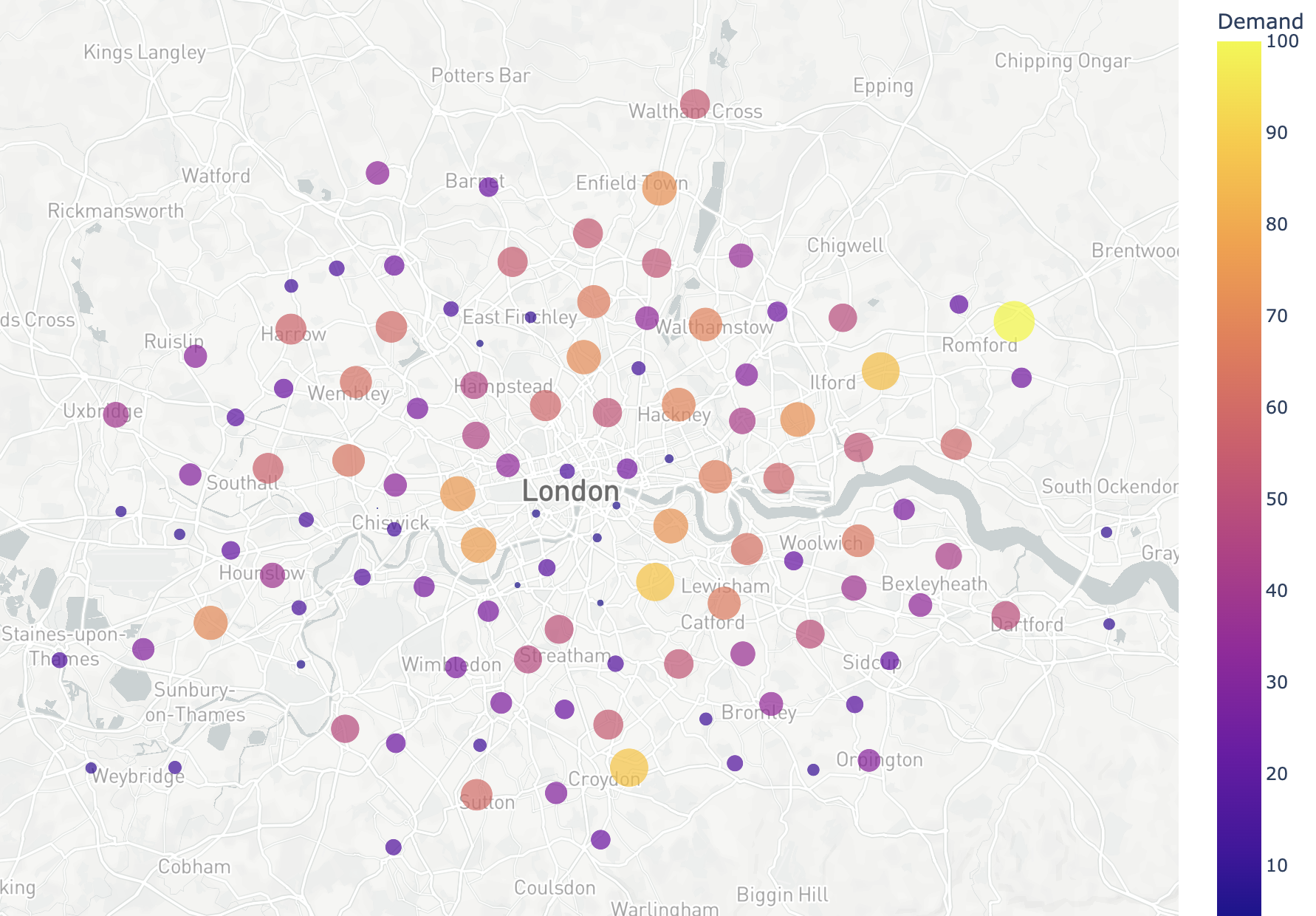}
		\end{subfigure}
		
		\begin{subfigure}{0.47\textwidth}
			\includegraphics[width=0.97\textwidth]{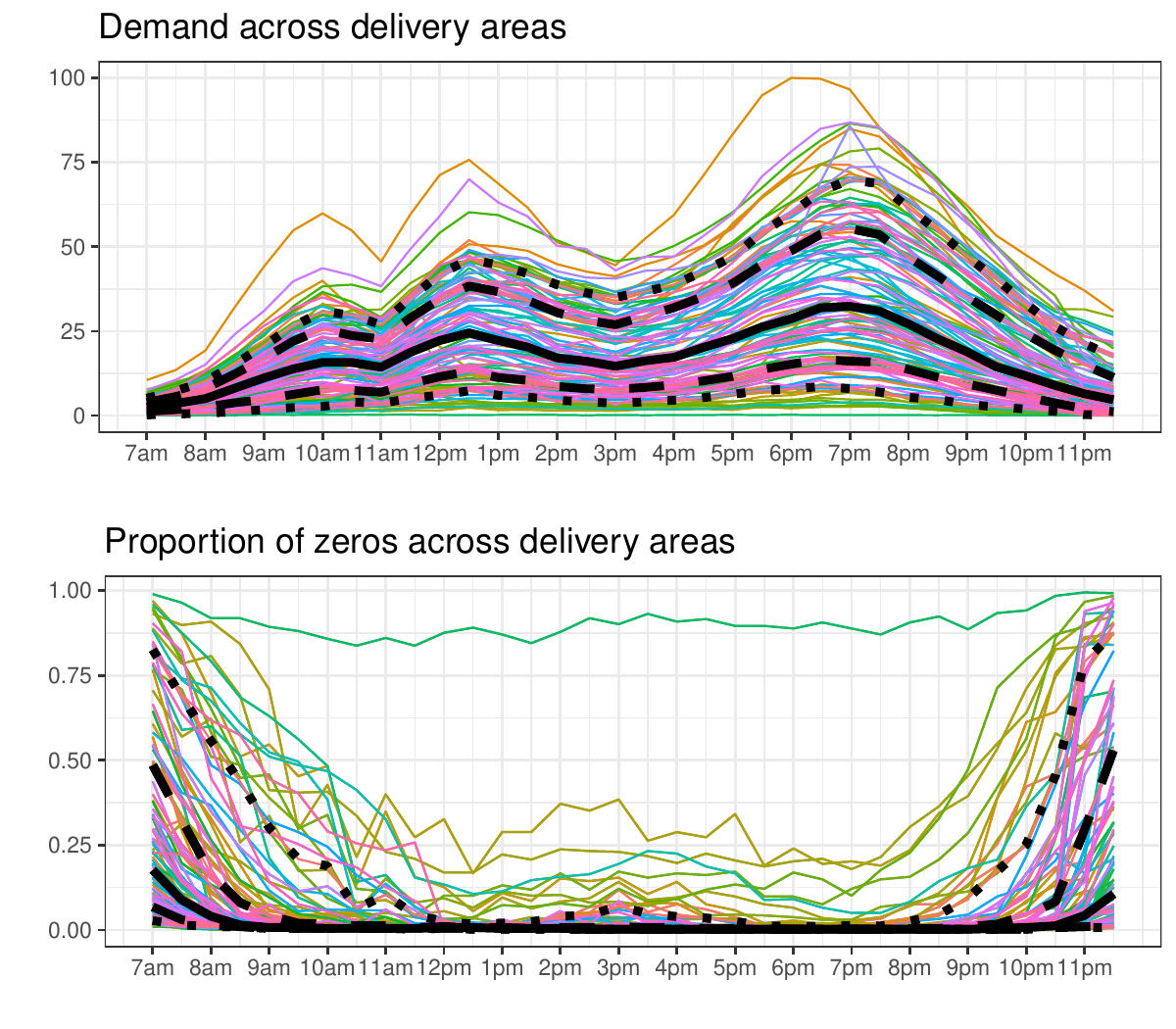}
		\end{subfigure}
	}
	\caption{Left: Aggregated demand in each area (circle) in  the London market.
		Right: Average demand (top) and proportion of zeros (bottom) across areas; 
		the median is displayed in solid black, 10\% and 90\% quantiles in dotted black, 25\% and 75\% in dashed black.}
	\label{fig:map_London_totalareademand}
\end{figure}

Figure \ref{fig:map_London_totalareademand} illustrates this heterogeneity across areas.
In the left panel, we visualize the aggregated demand over the sample period for each delivery area. Apart from larger activity above River Thames, there is no specific spatial pattern in demand volume, low- and high-volume delivery areas co-exist. The largest volume area is Romford, outside the city center. The City of London and Westminster area volumes respectively represent about 15\% and 5\% of Romford. 
In the right top panel of Figure \ref{fig:map_London_totalareademand}, we plot
the 30-minute average demand throughout the business day for each delivery area. The solid black median line visualizes well the lunch and dinner periods, the colored  area curves stress the large variability, except for early morning demand which is overall low. The right bottom panel of  Figure \ref{fig:map_London_totalareademand} zooms into the proportion of zeros across areas. Between 10am and 9pm, the proportion of 30-minute zero demand spells is low for most delivery areas, Stirling Road (upper green curve) however has mostly zero demand throughout the entire business day. Large diversity appears in the morning and evening hours with delivery areas that are idle and others that are active. 

\begin{figure}[htbp] 
	\centering
	{\includegraphics[width=10cm,height=6cm]{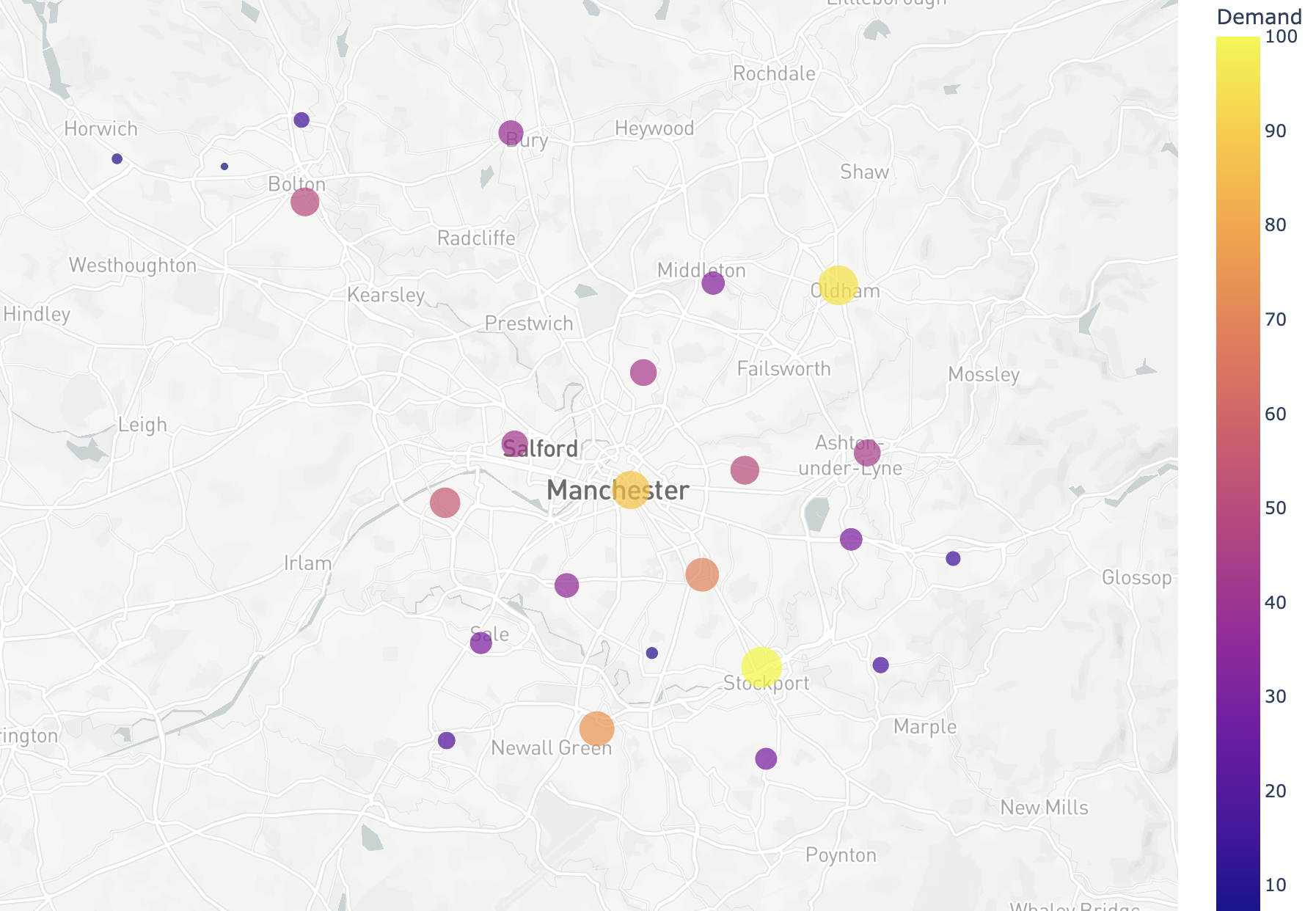} }
	\caption{Aggregated demand in each delivery area in  Manchester. 
	}	
	\label{fig:map_Manchester_totalareademand}
\end{figure}

\begin{figure}[t]
	\centering
	{\includegraphics[width = \textwidth]{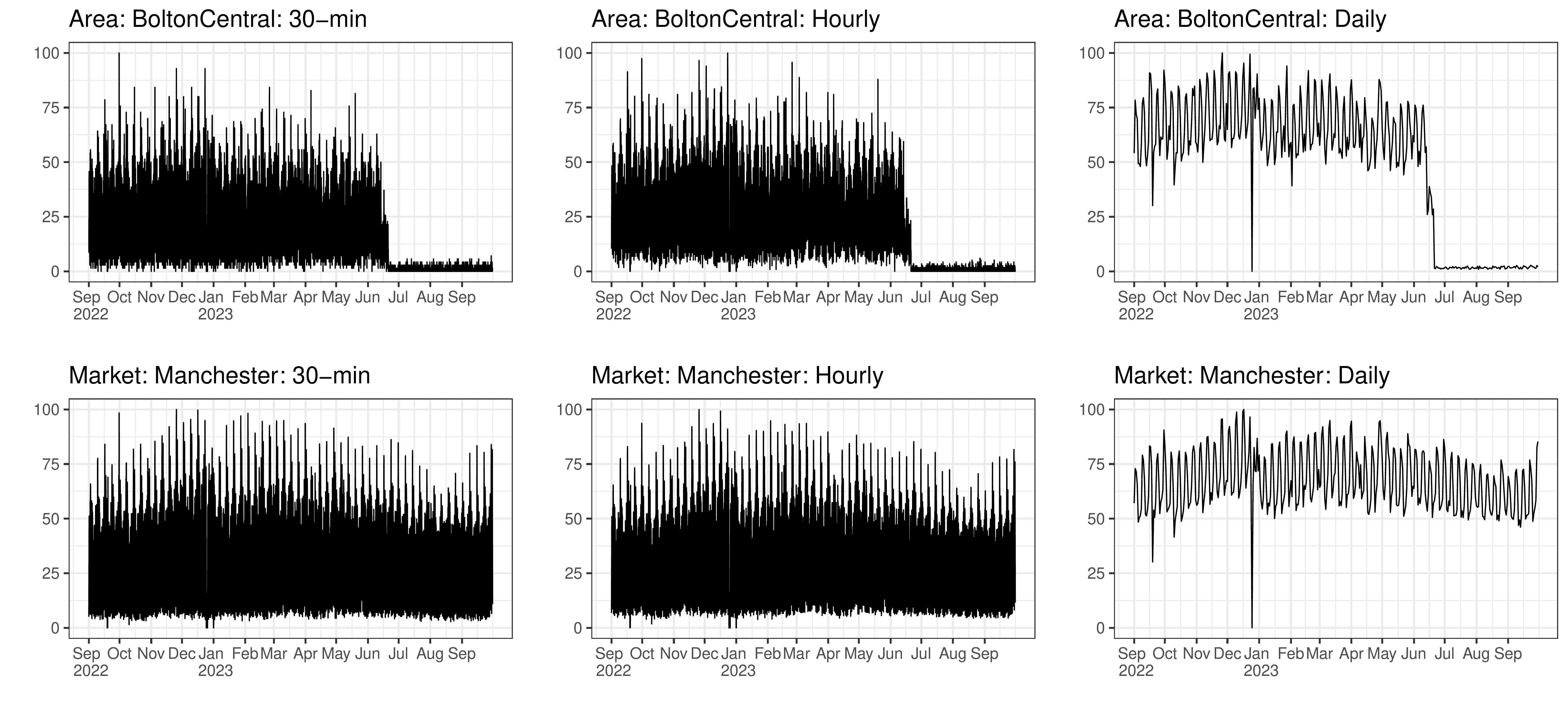} }
	\caption{Time series plots for full sample of area Bolton Central (top) belonging to the market Manchester (bottom), with a data shift for Bolton Central at June 14 2023. Demand data are normalized between 0-100 in each panel. 
		To relate the scale of the series at the different cross-sectional levels (area, market), note for reference that 
		demand at the market level is on average 15 times larger than at the area level before the data shift and on average 250 times higher after the data shift.
	}	
	\label{fig:map_Manchester_shift}
\end{figure}

\textit{Manchester Dataset}.
\label{subsec:result:Manchester}
The streaming platform data for the London market are reasonably stable over time.
The Manchester dataset, in contrast, exhibits shifts in the data streams. Figure \ref{fig:map_Manchester_totalareademand}
displays aggregated demand over the sample period in each delivery area in  Manchester. Figure \ref{fig:map_Manchester_shift}, top panel, illustrates how severe the  data shift is on June 14, 2023 in all considered temporal frequencies for the Bolton Central delivery area situated North-West of Manchester.
However, for the entire market of Manchester, the data shift is not apparent as can be seen from the bottom panel of Figure \ref{fig:map_Manchester_shift}. 

The Manchester dataset thus offers a
compelling scenario to investigate how our ML-based forecast reconciliation procedure performs when facing  data shifts. 
In fact, breaks or data shifts are regularly observed in platform data because of new competitors entering the marketplace and companies moving some of their locations to other on-demand platforms.

Finally, since London and Manchester constitute the two largest UK markets of the delivery platform, one might wonder why 
we keep the two dataset separate as this may hinder the opportunity of further improvement of forecast accuracy.
In fact, we treat both cities separately since the forecasting tasks are considered independently by the platform. 
Indeed,  London and Manchester are  marketplaces  with potential spillover effects between the delivery areas in their own market. Each city operates independently by local decision makers because the business environment, e.g.,\ competitors and demographics, can  drastically change from one city to the other. The demand forecasts are used to determine salary levels of local independent couriers which form the supply-side of the market and are city-specific.

\subsection{Bicycle Sharing Platform Data} \label{subsec:citibike}
The third dataset consists of demand  for bicycle rentals across six H3 cells in New York City. 
The data comes from Citi Bike, owned by the American leading mobility service provider Lyft, and is publicly available at 
\texttt{https://citibikenyc.com/system-data}. 
We download data for the year 2023 (2023-citibike-tripdata.zip) and process them to obtain geo-spatial streaming 30-min time series  measuring demand for bicycle rentals across the six H3 cells from 0am to 11:30pm. H3 cells, Uber’s Hexagonal Hierarchical Spatial Index, are defined using the geospatial indexing system that partitions the world into hexagonal cells.

\begin{figure}
	\includegraphics[height = 6.5cm, width = 8.5cm]{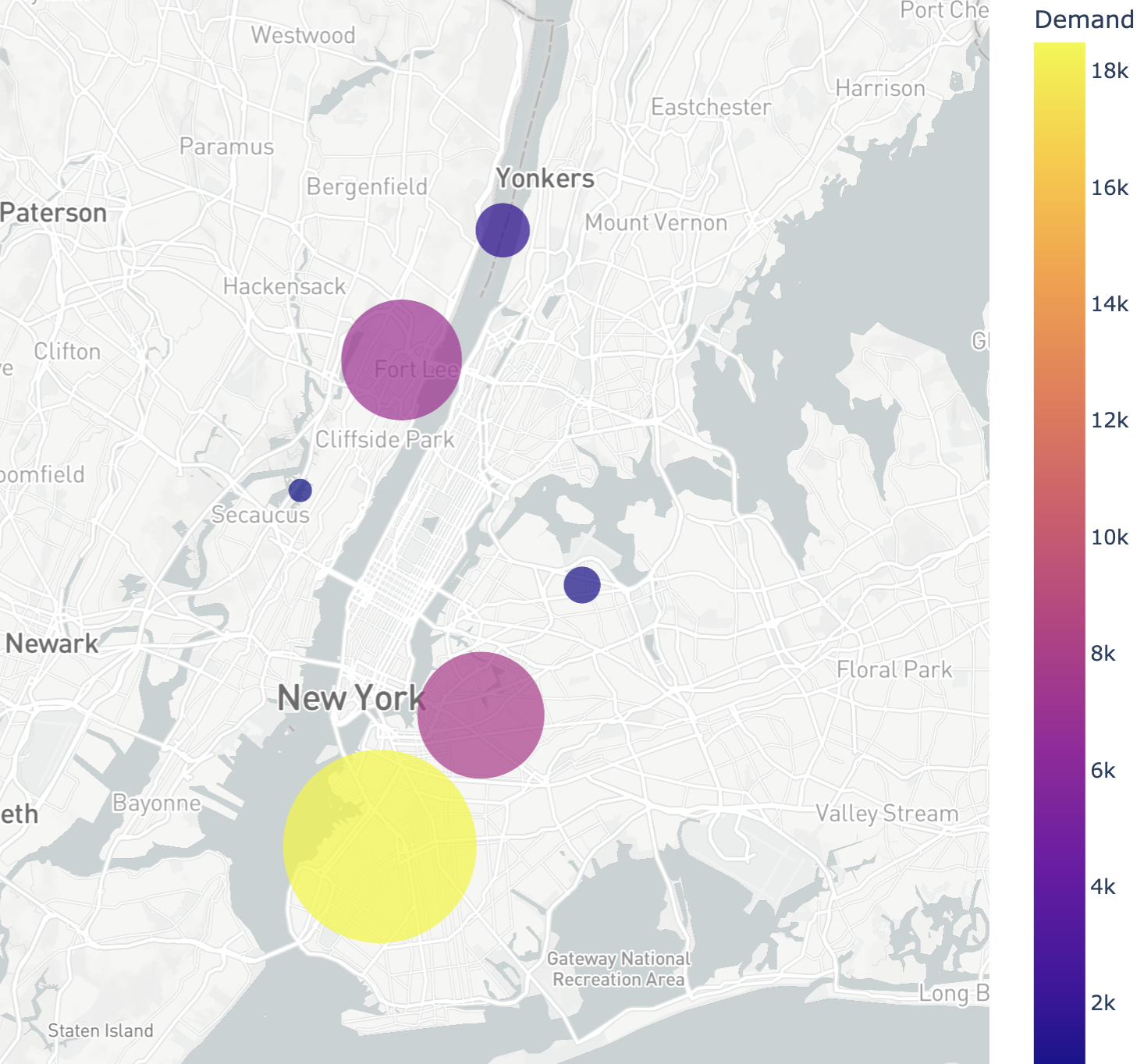}
	\includegraphics[height = 6.5cm, width = 8.5cm]{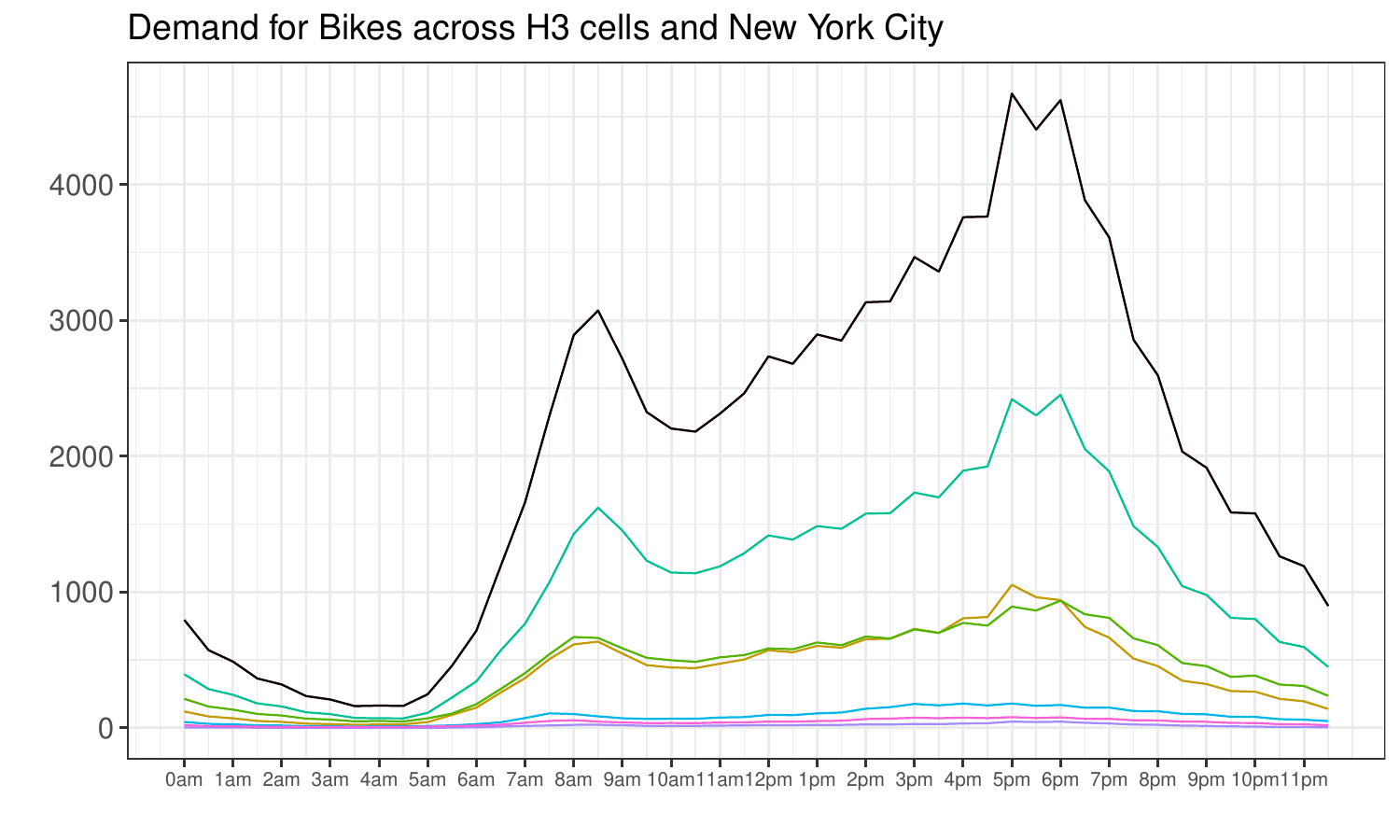}
	\caption{Left: Aggregated demand (in 1000s) in each H3 cell (circle). Right: Average demand across the six H3 cells (colored) and New York City (black). \label{fig:citibike}}
\end{figure}

The cross-temporal hierarchy for our Citi Bike dataset consists of two cross-sectional levels and 10 temporal levels, as summarized in Table \ref{overview:datasets}.
For the cross-section, New York City forms the upper level whereas the bottom level consists of six H3 cells. 
Figure \ref{fig:citibike}, left panel  displays the aggregate demand over the sample period in each H3 cell  whereas  the right panel displays the average 30-min demand in each H3 cell (colored) and in New York City (black) as a whole.  
For the temporal orders, we consider the set of all possible
temporal aggregation orders for a 30-min/1 day temporally grouped time series, namely 
30-min, 1-hour, 1.5-hour, 2-hour, 3-hour, 4-hour, 6-hour, 8-hour, 12-hour and 24-hour. In contrast to the hierarchical (nested) temporal structure of the two last-mile delivery data sets, the temporal structure of the Citi Bike data set is grouped since the bottom-level series are aggregated by attributes that are crossed, see \cite{athanasopoulos2023review} for a discussion.

\section{Machine Learning Based Forecast Reconciliation \label{sec:methodology}}
In Section \ref{subsec:CThierarchy}, we review the cross-temporal framework using the general notation from \cite{athanasopoulos2023review}. 
In Section \ref{subsec:MLreconciliation-for-CT} we introduce our ML-based forecast reconciliation approach for cross-temporal hierarchies.

\subsection{Cross-Temporal Hierarchies } \label{subsec:CThierarchy}
We start with some general notation so that our methodology of the next section can be widely applied. We denote $y_{b,t}$ for  $b=1,\ldots,n_b$  and $t>0$ the high-frequency time series observation for bottom level $b$ at time $t$. The cross-sectional hierarchy is simply obtained by summing over elements in the set of the $n_b$ bottom level time series. This yields $n_a$ additional high-frequency time series resulting in a total of $n=n_a+n_b$ series.

Regarding the temporal hierarchy, note that in a platform setting data is streaming but for simplicity let us consider $t=1,\ldots,m$  with $m$ the number of observations to sum over when going from the highest to the lowest sampling frequency in the hierarchy.
If we define $k = k_1,...,k_p$  the $p$ temporal aggregation levels, with $k_1=1$ and $k_p=m$, then the non-overlapping temporally aggregated series, for each series $i=1,\ldots,n$ and $k = k_1,...,k_p$, are  given by 
$$
x_{i,j}^{[k]}=\sum_{t=(j-1)k + 1}^{jk} y_{i,t} \qquad j=1,...,m/k.
$$
Next, denoting $\tau$ as the observation index of the lowest frequency (top temporal level), we can stack these observations in  ($m_k \times 1$) vectors 
$\boldsymbol{x}_{i,\tau}^{[k]}  = \left[x_{i,m_k(\tau-1)+1}^{[k]}, x_{i,m_k(\tau-1)+2}^{[k]} , \ldots, x_{i,m_k \tau}^{[k]}\right]^\top,$
where $\top$ denotes the transpose of a vector and $m_k = m/k$.

Finally, the cross-temporal hierarchy is obtained by considering for each variable $i=1,\ldots,n$ of the cross-sectional hierarchy, the entire temporal hierarchy. Notation wise, we collect in a ($\sum_{l=1}^p m_{k_l} \times 1$) vector 
all temporally aggregated time series 
$\boldsymbol{x}_{i,\tau}= \left [ x_{i,\tau}^{[m]}, \ldots, {\boldsymbol{x}_{i,\tau}^{[1]}}^\top \right ]^\top,$
and then stack
all $n$ variables into the vector 
$ \boldsymbol{x}_\tau = \left[ \boldsymbol{x}_{1,\tau}^\top, \ldots,  \boldsymbol{x}_{n,\tau}^\top \right ]^\top$.

\begin{figure}[t]
	\resizebox{0.83\textwidth}{!}{\begin{minipage}{\textwidth}
			\begin{tikzpicture}[x=0.75pt,y=0.75pt,yscale=-1,xscale=1]
				
				%Shape: Ellipse [id:dp5016269417722741] 
				\draw  [fill={rgb, 255:red, 74; green, 144; blue, 226 }  ,fill opacity=1 ] (255.29,170.57) .. controls (255.29,160.41) and (263.52,152.17) .. (273.68,152.17) .. controls (283.84,152.17) and (292.07,160.41) .. (292.07,170.57) .. controls (292.07,180.72) and (283.84,188.96) .. (273.68,188.96) .. controls (263.52,188.96) and (255.29,180.72) .. (255.29,170.57) -- cycle ;
				%Shape: Ellipse [id:dp20011679910785007] 
				\draw  [fill={rgb, 255:red, 74; green, 144; blue, 226 }  ,fill opacity=1 ] (297.71,169.95) .. controls (297.71,159.79) and (305.94,151.55) .. (316.1,151.55) .. controls (326.26,151.55) and (334.5,159.79) .. (334.5,169.95) .. controls (334.5,180.1) and (326.26,188.34) .. (316.1,188.34) .. controls (305.94,188.34) and (297.71,180.1) .. (297.71,169.95) -- cycle ;
				%Shape: Ellipse [id:dp38938365177950574] 
				\draw  [fill={rgb, 255:red, 74; green, 144; blue, 226 }  ,fill opacity=0.6 ] (275.72,114.83) .. controls (275.72,104.67) and (283.96,96.43) .. (294.12,96.43) .. controls (304.28,96.43) and (312.51,104.67) .. (312.51,114.83) .. controls (312.51,124.99) and (304.28,133.22) .. (294.12,133.22) .. controls (283.96,133.22) and (275.72,124.99) .. (275.72,114.83) -- cycle ;
				%Shape: Ellipse [id:dp4140322475765519] 
				\draw  [fill={rgb, 255:red, 74; green, 144; blue, 226 }  ,fill opacity=0.2 ] (334.25,59.71) .. controls (334.25,49.55) and (342.48,41.31) .. (352.64,41.31) .. controls (362.8,41.31) and (371.04,49.55) .. (371.04,59.71) .. controls (371.04,69.87) and (362.8,78.1) .. (352.64,78.1) .. controls (342.48,78.1) and (334.25,69.87) .. (334.25,59.71) -- cycle ;
				%Straight Lines [id:da8219027083205008] 
				\draw    (334.25,59.71) -- (294.12,96.43) ;
				%Straight Lines [id:da49650422744714806] 
				\draw    (294.74,133.22) -- (274.3,152.17) ;
				%Straight Lines [id:da523127567790747] 
				\draw    (294.74,133.22) -- (316.1,151.55) ;
				%Shape: Ellipse [id:dp5659357803233509] 
				\draw  [fill={rgb, 255:red, 74; green, 144; blue, 226 }  ,fill opacity=1 ] (371.41,170.88) .. controls (371.41,160.72) and (379.64,152.48) .. (389.8,152.48) .. controls (399.96,152.48) and (408.2,160.72) .. (408.2,170.88) .. controls (408.2,181.03) and (399.96,189.27) .. (389.8,189.27) .. controls (379.64,189.27) and (371.41,181.03) .. (371.41,170.88) -- cycle ;
				%Shape: Ellipse [id:dp477705530473888] 
				\draw  [fill={rgb, 255:red, 74; green, 144; blue, 226 }  ,fill opacity=1 ] (413.21,170.26) .. controls (413.21,160.1) and (421.45,151.86) .. (431.61,151.86) .. controls (441.76,151.86) and (450,160.1) .. (450,170.26) .. controls (450,180.41) and (441.76,188.65) .. (431.61,188.65) .. controls (421.45,188.65) and (413.21,180.41) .. (413.21,170.26) -- cycle ;
				%Shape: Ellipse [id:dp2348406122568243] 
				\draw  [fill={rgb, 255:red, 74; green, 144; blue, 226 }  ,fill opacity=0.6 ] (391.85,115.14) .. controls (391.85,104.98) and (400.08,96.74) .. (410.24,96.74) .. controls (420.4,96.74) and (428.63,104.98) .. (428.63,115.14) .. controls (428.63,125.29) and (420.4,133.53) .. (410.24,133.53) .. controls (400.08,133.53) and (391.85,125.29) .. (391.85,115.14) -- cycle ;
				%Straight Lines [id:da5729927319804675] 
				\draw    (371.04,59.71) -- (410.24,96.74) ;
				%Straight Lines [id:da7961332769062186] 
				\draw    (410.24,133.53) -- (389.8,152.48) ;
				%Straight Lines [id:da13779357454243502] 
				\draw    (410.24,133.53) -- (431.61,151.86) ;
				%Shape: Ellipse [id:dp41061994467890917] 
				\draw  [fill={rgb, 255:red, 248; green, 231; blue, 28 }  ,fill opacity=1 ] (23.94,169.2) .. controls (23.94,162.02) and (29.77,156.2) .. (36.95,156.2) .. controls (44.13,156.2) and (49.95,162.02) .. (49.95,169.2) .. controls (49.95,176.39) and (44.13,182.21) .. (36.95,182.21) .. controls (29.77,182.21) and (23.94,176.39) .. (23.94,169.2) -- cycle ;
				%Shape: Ellipse [id:dp32794452103689387] 
				\draw  [fill={rgb, 255:red, 248; green, 231; blue, 28 }  ,fill opacity=0.5 ] (49.03,117.61) .. controls (49.03,107.45) and (57.26,99.22) .. (67.42,99.22) .. controls (77.58,99.22) and (85.81,107.45) .. (85.81,117.61) .. controls (85.81,127.77) and (77.58,136.01) .. (67.42,136.01) .. controls (57.26,136.01) and (49.03,127.77) .. (49.03,117.61) -- cycle ;
				%Shape: Ellipse [id:dp579997653615411] 
				\draw  [fill={rgb, 255:red, 248; green, 231; blue, 28 }  ,fill opacity=0.17 ] (107.92,57.88) .. controls (107.92,44.35) and (118.89,33.39) .. (132.42,33.39) .. controls (145.95,33.39) and (156.91,44.35) .. (156.91,57.88) .. controls (156.91,71.41) and (145.95,82.37) .. (132.42,82.37) .. controls (118.89,82.37) and (107.92,71.41) .. (107.92,57.88) -- cycle ;
				%Straight Lines [id:da4647514976023097] 
				\draw    (107.92,57.88) -- (67.42,99.22) ;
				%Straight Lines [id:da13335544948742473] 
				\draw    (66.49,136.32) -- (36.95,156.2) ;
				%Straight Lines [id:da45710219529960683] 
				\draw    (66.49,136.32) -- (92.32,156.57) ;
				%Shape: Ellipse [id:dp9755221794093243] 
				\draw  [fill={rgb, 255:red, 248; green, 231; blue, 28 }  ,fill opacity=1 ] (51.81,169.57) .. controls (51.81,162.39) and (57.64,156.57) .. (64.82,156.57) .. controls (72,156.57) and (77.82,162.39) .. (77.82,169.57) .. controls (77.82,176.76) and (72,182.58) .. (64.82,182.58) .. controls (57.64,182.58) and (51.81,176.76) .. (51.81,169.57) -- cycle ;
				%Shape: Ellipse [id:dp7114138469469273] 
				\draw  [fill={rgb, 255:red, 248; green, 231; blue, 28 }  ,fill opacity=1 ] (79.31,169.57) .. controls (79.31,162.39) and (85.13,156.57) .. (92.32,156.57) .. controls (99.5,156.57) and (105.32,162.39) .. (105.32,169.57) .. controls (105.32,176.76) and (99.5,182.58) .. (92.32,182.58) .. controls (85.13,182.58) and (79.31,176.76) .. (79.31,169.57) -- cycle ;
				%Straight Lines [id:da9433783390809796] 
				\draw    (66.49,136.32) -- (64.82,156.57) ;
				%Shape: Ellipse [id:dp24101431831337528] 
				\draw  [fill={rgb, 255:red, 248; green, 231; blue, 28 }  ,fill opacity=0.5 ] (179.08,117.92) .. controls (179.08,107.76) and (187.32,99.53) .. (197.48,99.53) .. controls (207.64,99.53) and (215.87,107.76) .. (215.87,117.92) .. controls (215.87,128.08) and (207.64,136.32) .. (197.48,136.32) .. controls (187.32,136.32) and (179.08,128.08) .. (179.08,117.92) -- cycle ;
				%Straight Lines [id:da5495536372052685] 
				\draw    (156.91,57.88) -- (197.48,99.53) ;
				%Straight Lines [id:da6659308930684014] 
				\draw    (197.48,136.01) -- (183.73,156.57) ;
				%Shape: Ellipse [id:dp8893748106436306] 
				\draw  [fill={rgb, 255:red, 248; green, 231; blue, 28 }  ,fill opacity=1 ] (170.72,169.57) .. controls (170.72,162.39) and (176.55,156.57) .. (183.73,156.57) .. controls (190.91,156.57) and (196.73,162.39) .. (196.73,169.57) .. controls (196.73,176.76) and (190.91,182.58) .. (183.73,182.58) .. controls (176.55,182.58) and (170.72,176.76) .. (170.72,169.57) -- cycle ;
				%Shape: Ellipse [id:dp38493527103629654] 
				\draw  [fill={rgb, 255:red, 248; green, 231; blue, 28 }  ,fill opacity=1 ] (198.22,169.57) .. controls (198.22,162.39) and (204.04,156.57) .. (211.23,156.57) .. controls (218.41,156.57) and (224.23,162.39) .. (224.23,169.57) .. controls (224.23,176.76) and (218.41,182.58) .. (211.23,182.58) .. controls (204.04,182.58) and (198.22,176.76) .. (198.22,169.57) -- cycle ;
				%Straight Lines [id:da5683861320619321] 
				\draw    (197.48,136.01) -- (211.23,156.57) ;
				%Shape: Ellipse [id:dp6680824671323857] 
				\draw  [fill={rgb, 255:red, 248; green, 231; blue, 28 }  ,fill opacity=0.17 ] (593.51,58.63) .. controls (593.51,45.61) and (604.06,35.05) .. (617.08,35.05) .. controls (630.1,35.05) and (640.66,45.61) .. (640.66,58.63) .. controls (640.66,71.65) and (630.1,82.21) .. (617.08,82.21) .. controls (604.06,82.21) and (593.51,71.65) .. (593.51,58.63) -- cycle ;
				%Shape: Ellipse [id:dp05908022197541607] 
				\draw  [fill={rgb, 255:red, 74; green, 144; blue, 226 }  ,fill opacity=1 ] (597.07,68.57) .. controls (597.07,66.5) and (598.75,64.83) .. (600.81,64.83) .. controls (602.87,64.83) and (604.55,66.5) .. (604.55,68.57) .. controls (604.55,70.63) and (602.87,72.31) .. (600.81,72.31) .. controls (598.75,72.31) and (597.07,70.63) .. (597.07,68.57) -- cycle ;
				%Shape: Ellipse [id:dp42339405812400255] 
				\draw  [fill={rgb, 255:red, 74; green, 144; blue, 226 }  ,fill opacity=1 ] (606.58,68.57) .. controls (606.58,66.5) and (608.25,64.83) .. (610.31,64.83) .. controls (612.38,64.83) and (614.05,66.5) .. (614.05,68.57) .. controls (614.05,70.63) and (612.38,72.31) .. (610.31,72.31) .. controls (608.25,72.31) and (606.58,70.63) .. (606.58,68.57) -- cycle ;
				%Shape: Ellipse [id:dp21480194735544855] 
				\draw  [fill={rgb, 255:red, 74; green, 144; blue, 226 }  ,fill opacity=0.6 ] (601.51,55.9) .. controls (601.51,53.83) and (603.18,52.16) .. (605.24,52.16) .. controls (607.31,52.16) and (608.98,53.83) .. (608.98,55.9) .. controls (608.98,57.96) and (607.31,59.63) .. (605.24,59.63) .. controls (603.18,59.63) and (601.51,57.96) .. (601.51,55.9) -- cycle ;
				%Shape: Ellipse [id:dp30856872509916045] 
				\draw  [fill={rgb, 255:red, 74; green, 144; blue, 226 }  ,fill opacity=0.4 ] (612.91,45.13) .. controls (612.91,43.06) and (614.58,41.39) .. (616.65,41.39) .. controls (618.71,41.39) and (620.39,43.06) .. (620.39,45.13) .. controls (620.39,47.19) and (618.71,48.86) .. (616.65,48.86) .. controls (614.58,48.86) and (612.91,47.19) .. (612.91,45.13) -- cycle ;
				%Straight Lines [id:da950570031860019] 
				\draw    (605.24,59.63) -- (610.31,64.83) ;
				%Straight Lines [id:da9498860009720638] 
				\draw    (605.24,59.63) -- (600.81,64.83) ;
				%Shape: Ellipse [id:dp7026826979421745] 
				\draw  [fill={rgb, 255:red, 74; green, 144; blue, 226 }  ,fill opacity=1 ] (619.88,68.57) .. controls (619.88,66.5) and (621.55,64.83) .. (623.62,64.83) .. controls (625.68,64.83) and (627.35,66.5) .. (627.35,68.57) .. controls (627.35,70.63) and (625.68,72.31) .. (623.62,72.31) .. controls (621.55,72.31) and (619.88,70.63) .. (619.88,68.57) -- cycle ;
				%Shape: Ellipse [id:dp1704381012410563] 
				\draw  [fill={rgb, 255:red, 74; green, 144; blue, 226 }  ,fill opacity=1 ] (629.38,68.57) .. controls (629.38,66.5) and (631.06,64.83) .. (633.12,64.83) .. controls (635.18,64.83) and (636.86,66.5) .. (636.86,68.57) .. controls (636.86,70.63) and (635.18,72.31) .. (633.12,72.31) .. controls (631.06,72.31) and (629.38,70.63) .. (629.38,68.57) -- cycle ;
				%Shape: Ellipse [id:dp2529943997626998] 
				\draw  [fill={rgb, 255:red, 74; green, 144; blue, 226 }  ,fill opacity=0.6 ] (624.31,55.9) .. controls (624.31,53.83) and (625.99,52.16) .. (628.05,52.16) .. controls (630.12,52.16) and (631.79,53.83) .. (631.79,55.9) .. controls (631.79,57.96) and (630.12,59.63) .. (628.05,59.63) .. controls (625.99,59.63) and (624.31,57.96) .. (624.31,55.9) -- cycle ;
				%Straight Lines [id:da8335152557203693] 
				\draw    (628.05,59.63) -- (633.12,64.83) ;
				%Straight Lines [id:da02651495645950086] 
				\draw    (628.05,59.63) -- (623.62,64.83) ;
				%Straight Lines [id:da3473294429751561] 
				\draw    (605.24,52.16) -- (612.91,45.13) ;
				%Straight Lines [id:da6397248496761334] 
				\draw    (620.39,45.13) -- (628.05,52.16) ;
				%Shape: Ellipse [id:dp18278668350276184] 
				\draw  [fill={rgb, 255:red, 248; green, 231; blue, 28 }  ,fill opacity=0.5 ] (511.65,108.3) .. controls (511.65,95.28) and (522.2,84.72) .. (535.22,84.72) .. controls (548.24,84.72) and (558.8,95.28) .. (558.8,108.3) .. controls (558.8,121.32) and (548.24,131.88) .. (535.22,131.88) .. controls (522.2,131.88) and (511.65,121.32) .. (511.65,108.3) -- cycle ;
				%Shape: Ellipse [id:dp8125065998432488] 
				\draw  [fill={rgb, 255:red, 74; green, 144; blue, 226 }  ,fill opacity=1 ] (515.47,118.24) .. controls (515.47,116.17) and (517.14,114.5) .. (519.2,114.5) .. controls (521.27,114.5) and (522.94,116.17) .. (522.94,118.24) .. controls (522.94,120.3) and (521.27,121.97) .. (519.2,121.97) .. controls (517.14,121.97) and (515.47,120.3) .. (515.47,118.24) -- cycle ;
				%Shape: Ellipse [id:dp4407553142488989] 
				\draw  [fill={rgb, 255:red, 74; green, 144; blue, 226 }  ,fill opacity=1 ] (524.97,118.24) .. controls (524.97,116.17) and (526.64,114.5) .. (528.71,114.5) .. controls (530.77,114.5) and (532.44,116.17) .. (532.44,118.24) .. controls (532.44,120.3) and (530.77,121.97) .. (528.71,121.97) .. controls (526.64,121.97) and (524.97,120.3) .. (524.97,118.24) -- cycle ;
				%Shape: Ellipse [id:dp6879706460962463] 
				\draw  [fill={rgb, 255:red, 74; green, 144; blue, 226 }  ,fill opacity=0.6 ] (519.9,105.57) .. controls (519.9,103.5) and (521.57,101.83) .. (523.64,101.83) .. controls (525.7,101.83) and (527.38,103.5) .. (527.38,105.57) .. controls (527.38,107.63) and (525.7,109.3) .. (523.64,109.3) .. controls (521.57,109.3) and (519.9,107.63) .. (519.9,105.57) -- cycle ;
				%Shape: Ellipse [id:dp1451533911225249] 
				\draw  [fill={rgb, 255:red, 74; green, 144; blue, 226 }  ,fill opacity=0.4 ] (531.3,94.8) .. controls (531.3,92.73) and (532.98,91.06) .. (535.04,91.06) .. controls (537.11,91.06) and (538.78,92.73) .. (538.78,94.8) .. controls (538.78,96.86) and (537.11,98.53) .. (535.04,98.53) .. controls (532.98,98.53) and (531.3,96.86) .. (531.3,94.8) -- cycle ;
				%Straight Lines [id:da8313108989507183] 
				\draw    (523.64,109.3) -- (528.71,114.5) ;
				%Straight Lines [id:da10897027427027917] 
				\draw    (523.64,109.3) -- (519.2,114.5) ;
				%Shape: Ellipse [id:dp6109367997042796] 
				\draw  [fill={rgb, 255:red, 74; green, 144; blue, 226 }  ,fill opacity=1 ] (538.27,118.24) .. controls (538.27,116.17) and (539.95,114.5) .. (542.01,114.5) .. controls (544.07,114.5) and (545.75,116.17) .. (545.75,118.24) .. controls (545.75,120.3) and (544.07,121.97) .. (542.01,121.97) .. controls (539.95,121.97) and (538.27,120.3) .. (538.27,118.24) -- cycle ;
				%Shape: Ellipse [id:dp9895943975233079] 
				\draw  [fill={rgb, 255:red, 74; green, 144; blue, 226 }  ,fill opacity=1 ] (547.78,118.24) .. controls (547.78,116.17) and (549.45,114.5) .. (551.51,114.5) .. controls (553.58,114.5) and (555.25,116.17) .. (555.25,118.24) .. controls (555.25,120.3) and (553.58,121.97) .. (551.51,121.97) .. controls (549.45,121.97) and (547.78,120.3) .. (547.78,118.24) -- cycle ;
				%Shape: Ellipse [id:dp4979014537641204] 
				\draw  [fill={rgb, 255:red, 74; green, 144; blue, 226 }  ,fill opacity=0.6 ] (542.71,105.57) .. controls (542.71,103.5) and (544.38,101.83) .. (546.45,101.83) .. controls (548.51,101.83) and (550.18,103.5) .. (550.18,105.57) .. controls (550.18,107.63) and (548.51,109.3) .. (546.45,109.3) .. controls (544.38,109.3) and (542.71,107.63) .. (542.71,105.57) -- cycle ;
				%Straight Lines [id:da41491292793861634] 
				\draw    (546.45,109.3) -- (551.51,114.5) ;
				%Straight Lines [id:da7089612038102051] 
				\draw    (546.45,109.3) -- (542.01,114.5) ;
				%Straight Lines [id:da8616156596025852] 
				\draw    (523.64,101.83) -- (531.3,94.8) ;
				%Straight Lines [id:da4783730598178382] 
				\draw    (538.78,94.8) -- (546.45,101.83) ;
				%Shape: Ellipse [id:dp36226927970993983] 
				\draw  [fill={rgb, 255:red, 248; green, 231; blue, 28 }  ,fill opacity=1 ] (512.54,165.87) .. controls (512.54,152.85) and (523.1,142.3) .. (536.12,142.3) .. controls (549.14,142.3) and (559.7,152.85) .. (559.7,165.87) .. controls (559.7,178.89) and (549.14,189.45) .. (536.12,189.45) .. controls (523.1,189.45) and (512.54,178.89) .. (512.54,165.87) -- cycle ;
				%Shape: Ellipse [id:dp007526058307440353] 
				\draw  [fill={rgb, 255:red, 74; green, 144; blue, 226 }  ,fill opacity=1 ] (516.11,175.81) .. controls (516.11,173.75) and (517.78,172.07) .. (519.85,172.07) .. controls (521.91,172.07) and (523.59,173.75) .. (523.59,175.81) .. controls (523.59,177.87) and (521.91,179.55) .. (519.85,179.55) .. controls (517.78,179.55) and (516.11,177.87) .. (516.11,175.81) -- cycle ;
				%Shape: Ellipse [id:dp0005069214121367072] 
				\draw  [fill={rgb, 255:red, 74; green, 144; blue, 226 }  ,fill opacity=1 ] (525.61,175.81) .. controls (525.61,173.75) and (527.29,172.07) .. (529.35,172.07) .. controls (531.42,172.07) and (533.09,173.75) .. (533.09,175.81) .. controls (533.09,177.87) and (531.42,179.55) .. (529.35,179.55) .. controls (527.29,179.55) and (525.61,177.87) .. (525.61,175.81) -- cycle ;
				%Shape: Ellipse [id:dp45447998753808583] 
				\draw  [fill={rgb, 255:red, 74; green, 144; blue, 226 }  ,fill opacity=0.6 ] (520.55,163.14) .. controls (520.55,161.07) and (522.22,159.4) .. (524.28,159.4) .. controls (526.35,159.4) and (528.02,161.07) .. (528.02,163.14) .. controls (528.02,165.2) and (526.35,166.88) .. (524.28,166.88) .. controls (522.22,166.88) and (520.55,165.2) .. (520.55,163.14) -- cycle ;
				%Shape: Ellipse [id:dp17770413729848222] 
				\draw  [fill={rgb, 255:red, 74; green, 144; blue, 226 }  ,fill opacity=0.4 ] (531.95,152.37) .. controls (531.95,150.3) and (533.62,148.63) .. (535.69,148.63) .. controls (537.75,148.63) and (539.42,150.3) .. (539.42,152.37) .. controls (539.42,154.43) and (537.75,156.11) .. (535.69,156.11) .. controls (533.62,156.11) and (531.95,154.43) .. (531.95,152.37) -- cycle ;
				%Straight Lines [id:da3052225220125486] 
				\draw    (524.28,166.88) -- (529.35,172.07) ;
				%Straight Lines [id:da7672987762790939] 
				\draw    (524.28,166.88) -- (519.85,172.07) ;
				%Shape: Ellipse [id:dp14801689221811754] 
				\draw  [fill={rgb, 255:red, 74; green, 144; blue, 226 }  ,fill opacity=1 ] (538.92,175.81) .. controls (538.92,173.75) and (540.59,172.07) .. (542.66,172.07) .. controls (544.72,172.07) and (546.39,173.75) .. (546.39,175.81) .. controls (546.39,177.87) and (544.72,179.55) .. (542.66,179.55) .. controls (540.59,179.55) and (538.92,177.87) .. (538.92,175.81) -- cycle ;
				%Shape: Ellipse [id:dp290714940008723] 
				\draw  [fill={rgb, 255:red, 74; green, 144; blue, 226 }  ,fill opacity=1 ] (548.42,175.81) .. controls (548.42,173.75) and (550.09,172.07) .. (552.16,172.07) .. controls (554.22,172.07) and (555.9,173.75) .. (555.9,175.81) .. controls (555.9,177.87) and (554.22,179.55) .. (552.16,179.55) .. controls (550.09,179.55) and (548.42,177.87) .. (548.42,175.81) -- cycle ;
				%Shape: Ellipse [id:dp6463167420535021] 
				\draw  [fill={rgb, 255:red, 74; green, 144; blue, 226 }  ,fill opacity=0.6 ] (543.35,163.14) .. controls (543.35,161.07) and (545.03,159.4) .. (547.09,159.4) .. controls (549.15,159.4) and (550.83,161.07) .. (550.83,163.14) .. controls (550.83,165.2) and (549.15,166.88) .. (547.09,166.88) .. controls (545.03,166.88) and (543.35,165.2) .. (543.35,163.14) -- cycle ;
				%Straight Lines [id:da4826652503521436] 
				\draw    (547.09,166.88) -- (552.16,172.07) ;
				%Straight Lines [id:da9304892051988369] 
				\draw    (547.09,166.88) -- (542.66,172.07) ;
				%Straight Lines [id:da7822885951011576] 
				\draw    (524.28,159.4) -- (531.95,152.37) ;
				%Straight Lines [id:da5897642004024468] 
				\draw    (539.42,152.37) -- (547.09,159.4) ;
				%Shape: Ellipse [id:dp3100683772323869] 
				\draw  [fill={rgb, 255:red, 248; green, 231; blue, 28 }  ,fill opacity=1 ] (562.14,165.49) .. controls (562.14,152.47) and (572.7,141.92) .. (585.72,141.92) .. controls (598.74,141.92) and (609.29,152.47) .. (609.29,165.49) .. controls (609.29,178.51) and (598.74,189.07) .. (585.72,189.07) .. controls (572.7,189.07) and (562.14,178.51) .. (562.14,165.49) -- cycle ;
				%Shape: Ellipse [id:dp7611332066656022] 
				\draw  [fill={rgb, 255:red, 74; green, 144; blue, 226 }  ,fill opacity=1 ] (565.71,175.43) .. controls (565.71,173.37) and (567.38,171.69) .. (569.45,171.69) .. controls (571.51,171.69) and (573.18,173.37) .. (573.18,175.43) .. controls (573.18,177.49) and (571.51,179.17) .. (569.45,179.17) .. controls (567.38,179.17) and (565.71,177.49) .. (565.71,175.43) -- cycle ;
				%Shape: Ellipse [id:dp26335499380815985] 
				\draw  [fill={rgb, 255:red, 74; green, 144; blue, 226 }  ,fill opacity=1 ] (575.21,175.43) .. controls (575.21,173.37) and (576.88,171.69) .. (578.95,171.69) .. controls (581.01,171.69) and (582.69,173.37) .. (582.69,175.43) .. controls (582.69,177.49) and (581.01,179.17) .. (578.95,179.17) .. controls (576.88,179.17) and (575.21,177.49) .. (575.21,175.43) -- cycle ;
				%Shape: Ellipse [id:dp7364042090305751] 
				\draw  [fill={rgb, 255:red, 74; green, 144; blue, 226 }  ,fill opacity=0.6 ] (570.14,162.76) .. controls (570.14,160.69) and (571.82,159.02) .. (573.88,159.02) .. controls (575.94,159.02) and (577.62,160.69) .. (577.62,162.76) .. controls (577.62,164.82) and (575.94,166.5) .. (573.88,166.5) .. controls (571.82,166.5) and (570.14,164.82) .. (570.14,162.76) -- cycle ;
				%Shape: Ellipse [id:dp9974435571199238] 
				\draw  [fill={rgb, 255:red, 74; green, 144; blue, 226 }  ,fill opacity=0.4 ] (581.55,151.99) .. controls (581.55,149.92) and (583.22,148.25) .. (585.28,148.25) .. controls (587.35,148.25) and (589.02,149.92) .. (589.02,151.99) .. controls (589.02,154.05) and (587.35,155.73) .. (585.28,155.73) .. controls (583.22,155.73) and (581.55,154.05) .. (581.55,151.99) -- cycle ;
				%Straight Lines [id:da21033328306803334] 
				\draw    (573.88,166.5) -- (578.95,171.69) ;
				%Straight Lines [id:da8023310936512733] 
				\draw    (573.88,166.5) -- (569.45,171.69) ;
				%Shape: Ellipse [id:dp12579653809685865] 
				\draw  [fill={rgb, 255:red, 74; green, 144; blue, 226 }  ,fill opacity=1 ] (588.52,175.43) .. controls (588.52,173.37) and (590.19,171.69) .. (592.25,171.69) .. controls (594.32,171.69) and (595.99,173.37) .. (595.99,175.43) .. controls (595.99,177.49) and (594.32,179.17) .. (592.25,179.17) .. controls (590.19,179.17) and (588.52,177.49) .. (588.52,175.43) -- cycle ;
				%Shape: Ellipse [id:dp8390814343581312] 
				\draw  [fill={rgb, 255:red, 74; green, 144; blue, 226 }  ,fill opacity=1 ] (598.02,175.43) .. controls (598.02,173.37) and (599.69,171.69) .. (601.76,171.69) .. controls (603.82,171.69) and (605.49,173.37) .. (605.49,175.43) .. controls (605.49,177.49) and (603.82,179.17) .. (601.76,179.17) .. controls (599.69,179.17) and (598.02,177.49) .. (598.02,175.43) -- cycle ;
				%Shape: Ellipse [id:dp914814366397001] 
				\draw  [fill={rgb, 255:red, 74; green, 144; blue, 226 }  ,fill opacity=0.6 ] (592.95,162.76) .. controls (592.95,160.69) and (594.62,159.02) .. (596.69,159.02) .. controls (598.75,159.02) and (600.42,160.69) .. (600.42,162.76) .. controls (600.42,164.82) and (598.75,166.5) .. (596.69,166.5) .. controls (594.62,166.5) and (592.95,164.82) .. (592.95,162.76) -- cycle ;
				%Straight Lines [id:da7662411894838497] 
				\draw    (596.69,166.5) -- (601.76,171.69) ;
				%Straight Lines [id:da34402793182202696] 
				\draw    (596.69,166.5) -- (592.25,171.69) ;
				%Straight Lines [id:da3686780642523684] 
				\draw    (573.88,159.02) -- (581.55,151.99) ;
				%Straight Lines [id:da27383890767274544] 
				\draw    (589.02,151.99) -- (596.69,159.02) ;
				%Shape: Ellipse [id:dp30060652855787495] 
				\draw  [fill={rgb, 255:red, 248; green, 231; blue, 28 }  ,fill opacity=1 ] (650.25,165.87) .. controls (650.25,152.85) and (660.81,142.3) .. (673.83,142.3) .. controls (686.85,142.3) and (697.4,152.85) .. (697.4,165.87) .. controls (697.4,178.89) and (686.85,189.45) .. (673.83,189.45) .. controls (660.81,189.45) and (650.25,178.89) .. (650.25,165.87) -- cycle ;
				%Shape: Ellipse [id:dp4662771810312143] 
				\draw  [fill={rgb, 255:red, 74; green, 144; blue, 226 }  ,fill opacity=1 ] (653.82,175.81) .. controls (653.82,173.75) and (655.49,172.07) .. (657.55,172.07) .. controls (659.62,172.07) and (661.29,173.75) .. (661.29,175.81) .. controls (661.29,177.87) and (659.62,179.55) .. (657.55,179.55) .. controls (655.49,179.55) and (653.82,177.87) .. (653.82,175.81) -- cycle ;
				%Shape: Ellipse [id:dp506056753055524] 
				\draw  [fill={rgb, 255:red, 74; green, 144; blue, 226 }  ,fill opacity=1 ] (663.32,175.81) .. controls (663.32,173.75) and (664.99,172.07) .. (667.06,172.07) .. controls (669.12,172.07) and (670.8,173.75) .. (670.8,175.81) .. controls (670.8,177.87) and (669.12,179.55) .. (667.06,179.55) .. controls (664.99,179.55) and (663.32,177.87) .. (663.32,175.81) -- cycle ;
				%Shape: Ellipse [id:dp1923590637435757] 
				\draw  [fill={rgb, 255:red, 74; green, 144; blue, 226 }  ,fill opacity=0.6 ] (658.25,163.14) .. controls (658.25,161.07) and (659.93,159.4) .. (661.99,159.4) .. controls (664.05,159.4) and (665.73,161.07) .. (665.73,163.14) .. controls (665.73,165.2) and (664.05,166.88) .. (661.99,166.88) .. controls (659.93,166.88) and (658.25,165.2) .. (658.25,163.14) -- cycle ;
				%Shape: Ellipse [id:dp9827800146990613] 
				\draw  [fill={rgb, 255:red, 74; green, 144; blue, 226 }  ,fill opacity=0.4 ] (669.66,152.37) .. controls (669.66,150.3) and (671.33,148.63) .. (673.39,148.63) .. controls (675.46,148.63) and (677.13,150.3) .. (677.13,152.37) .. controls (677.13,154.43) and (675.46,156.11) .. (673.39,156.11) .. controls (671.33,156.11) and (669.66,154.43) .. (669.66,152.37) -- cycle ;
				%Straight Lines [id:da7287743614171558] 
				\draw    (661.99,166.88) -- (667.06,172.07) ;
				%Straight Lines [id:da368250026953425] 
				\draw    (661.99,166.88) -- (657.55,172.07) ;
				%Shape: Ellipse [id:dp8260525724272922] 
				\draw  [fill={rgb, 255:red, 74; green, 144; blue, 226 }  ,fill opacity=1 ] (676.62,175.81) .. controls (676.62,173.75) and (678.3,172.07) .. (680.36,172.07) .. controls (682.43,172.07) and (684.1,173.75) .. (684.1,175.81) .. controls (684.1,177.87) and (682.43,179.55) .. (680.36,179.55) .. controls (678.3,179.55) and (676.62,177.87) .. (676.62,175.81) -- cycle ;
				%Shape: Ellipse [id:dp49275528783800193] 
				\draw  [fill={rgb, 255:red, 74; green, 144; blue, 226 }  ,fill opacity=1 ] (686.13,175.81) .. controls (686.13,173.75) and (687.8,172.07) .. (689.86,172.07) .. controls (691.93,172.07) and (693.6,173.75) .. (693.6,175.81) .. controls (693.6,177.87) and (691.93,179.55) .. (689.86,179.55) .. controls (687.8,179.55) and (686.13,177.87) .. (686.13,175.81) -- cycle ;
				%Shape: Ellipse [id:dp5591783684123859] 
				\draw  [fill={rgb, 255:red, 74; green, 144; blue, 226 }  ,fill opacity=0.6 ] (681.06,163.14) .. controls (681.06,161.07) and (682.73,159.4) .. (684.8,159.4) .. controls (686.86,159.4) and (688.53,161.07) .. (688.53,163.14) .. controls (688.53,165.2) and (686.86,166.88) .. (684.8,166.88) .. controls (682.73,166.88) and (681.06,165.2) .. (681.06,163.14) -- cycle ;
				%Straight Lines [id:da33455594100525077] 
				\draw    (684.8,166.88) -- (689.86,172.07) ;
				%Straight Lines [id:da08935580234411411] 
				\draw    (684.8,166.88) -- (680.36,172.07) ;
				%Straight Lines [id:da9732375374056514] 
				\draw    (661.99,159.4) -- (669.66,152.37) ;
				%Straight Lines [id:da05635172945864109] 
				\draw    (677.13,152.37) -- (684.8,159.4) ;
				%Shape: Ellipse [id:dp3341999534287774] 
				\draw  [fill={rgb, 255:red, 248; green, 231; blue, 28 }  ,fill opacity=0.5 ] (672.86,107.79) .. controls (672.86,94.77) and (683.41,84.22) .. (696.44,84.22) .. controls (709.46,84.22) and (720.01,94.77) .. (720.01,107.79) .. controls (720.01,120.81) and (709.46,131.37) .. (696.44,131.37) .. controls (683.41,131.37) and (672.86,120.81) .. (672.86,107.79) -- cycle ;
				%Shape: Ellipse [id:dp821905762563998] 
				\draw  [fill={rgb, 255:red, 74; green, 144; blue, 226 }  ,fill opacity=1 ] (676.43,117.73) .. controls (676.43,115.67) and (678.1,113.99) .. (680.16,113.99) .. controls (682.23,113.99) and (683.9,115.67) .. (683.9,117.73) .. controls (683.9,119.79) and (682.23,121.47) .. (680.16,121.47) .. controls (678.1,121.47) and (676.43,119.79) .. (676.43,117.73) -- cycle ;
				%Shape: Ellipse [id:dp5386111801570164] 
				\draw  [fill={rgb, 255:red, 74; green, 144; blue, 226 }  ,fill opacity=1 ] (685.93,117.73) .. controls (685.93,115.67) and (687.6,113.99) .. (689.67,113.99) .. controls (691.73,113.99) and (693.4,115.67) .. (693.4,117.73) .. controls (693.4,119.79) and (691.73,121.47) .. (689.67,121.47) .. controls (687.6,121.47) and (685.93,119.79) .. (685.93,117.73) -- cycle ;
				%Shape: Ellipse [id:dp01992075078974942] 
				\draw  [fill={rgb, 255:red, 74; green, 144; blue, 226 }  ,fill opacity=0.6 ] (680.86,105.06) .. controls (680.86,103) and (682.53,101.32) .. (684.6,101.32) .. controls (686.66,101.32) and (688.34,103) .. (688.34,105.06) .. controls (688.34,107.12) and (686.66,108.8) .. (684.6,108.8) .. controls (682.53,108.8) and (680.86,107.12) .. (680.86,105.06) -- cycle ;
				%Shape: Ellipse [id:dp09291286542895638] 
				\draw  [fill={rgb, 255:red, 74; green, 144; blue, 226 }  ,fill opacity=0.4 ] (692.26,94.29) .. controls (692.26,92.23) and (693.94,90.55) .. (696,90.55) .. controls (698.07,90.55) and (699.74,92.23) .. (699.74,94.29) .. controls (699.74,96.35) and (698.07,98.03) .. (696,98.03) .. controls (693.94,98.03) and (692.26,96.35) .. (692.26,94.29) -- cycle ;
				%Straight Lines [id:da599785988850726] 
				\draw    (684.6,108.8) -- (689.67,113.99) ;
				%Straight Lines [id:da8448980255230618] 
				\draw    (684.6,108.8) -- (680.16,113.99) ;
				%Shape: Ellipse [id:dp6198582497591532] 
				\draw  [fill={rgb, 255:red, 74; green, 144; blue, 226 }  ,fill opacity=1 ] (699.23,117.73) .. controls (699.23,115.67) and (700.91,113.99) .. (702.97,113.99) .. controls (705.03,113.99) and (706.71,115.67) .. (706.71,117.73) .. controls (706.71,119.79) and (705.03,121.47) .. (702.97,121.47) .. controls (700.91,121.47) and (699.23,119.79) .. (699.23,117.73) -- cycle ;
				%Shape: Ellipse [id:dp37176383743635055] 
				\draw  [fill={rgb, 255:red, 74; green, 144; blue, 226 }  ,fill opacity=1 ] (708.74,117.73) .. controls (708.74,115.67) and (710.41,113.99) .. (712.47,113.99) .. controls (714.54,113.99) and (716.21,115.67) .. (716.21,117.73) .. controls (716.21,119.79) and (714.54,121.47) .. (712.47,121.47) .. controls (710.41,121.47) and (708.74,119.79) .. (708.74,117.73) -- cycle ;
				%Shape: Ellipse [id:dp516156944283138] 
				\draw  [fill={rgb, 255:red, 74; green, 144; blue, 226 }  ,fill opacity=0.6 ] (703.67,105.06) .. controls (703.67,103) and (705.34,101.32) .. (707.4,101.32) .. controls (709.47,101.32) and (711.14,103) .. (711.14,105.06) .. controls (711.14,107.12) and (709.47,108.8) .. (707.4,108.8) .. controls (705.34,108.8) and (703.67,107.12) .. (703.67,105.06) -- cycle ;
				%Straight Lines [id:da8767569342062373] 
				\draw    (707.4,108.8) -- (712.47,113.99) ;
				%Straight Lines [id:da6314007965605795] 
				\draw    (707.4,108.8) -- (702.97,113.99) ;
				%Straight Lines [id:da7871981301972124] 
				\draw    (684.6,101.32) -- (692.26,94.29) ;
				%Straight Lines [id:da80362446867511] 
				\draw    (699.74,94.29) -- (707.4,101.32) ;
				%Straight Lines [id:da38533334741568637] 
				\draw    (535.22,131.88) -- (585.72,141.92) ;
				%Straight Lines [id:da41083035161701953] 
				\draw    (486.92,141.9) -- (535.22,131.88) ;
				%Straight Lines [id:da9919409707195785] 
				\draw    (696.44,131.37) -- (723.42,141.92) ;
				%Straight Lines [id:da2617734483213039] 
				\draw    (669.83,142.3) -- (696.44,131.37) ;
				%Straight Lines [id:da7179150623948458] 
				\draw    (593.51,58.63) -- (535.22,84.72) ;
				%Shape: Ellipse [id:dp6161795165415112] 
				\draw  [fill={rgb, 255:red, 248; green, 231; blue, 28 }  ,fill opacity=1 ] (463.34,165.48) .. controls (463.34,152.46) and (473.9,141.9) .. (486.92,141.9) .. controls (499.94,141.9) and (510.5,152.46) .. (510.5,165.48) .. controls (510.5,178.5) and (499.94,189.05) .. (486.92,189.05) .. controls (473.9,189.05) and (463.34,178.5) .. (463.34,165.48) -- cycle ;
				%Shape: Ellipse [id:dp43818785898237433] 
				\draw  [fill={rgb, 255:red, 74; green, 144; blue, 226 }  ,fill opacity=1 ] (466.91,175.41) .. controls (466.91,173.35) and (468.58,171.68) .. (470.65,171.68) .. controls (472.71,171.68) and (474.39,173.35) .. (474.39,175.41) .. controls (474.39,177.48) and (472.71,179.15) .. (470.65,179.15) .. controls (468.58,179.15) and (466.91,177.48) .. (466.91,175.41) -- cycle ;
				%Shape: Ellipse [id:dp1219997206264789] 
				\draw  [fill={rgb, 255:red, 74; green, 144; blue, 226 }  ,fill opacity=1 ] (476.41,175.41) .. controls (476.41,173.35) and (478.09,171.68) .. (480.15,171.68) .. controls (482.22,171.68) and (483.89,173.35) .. (483.89,175.41) .. controls (483.89,177.48) and (482.22,179.15) .. (480.15,179.15) .. controls (478.09,179.15) and (476.41,177.48) .. (476.41,175.41) -- cycle ;
				%Shape: Ellipse [id:dp3604365126205369] 
				\draw  [fill={rgb, 255:red, 74; green, 144; blue, 226 }  ,fill opacity=0.6 ] (471.35,162.74) .. controls (471.35,160.68) and (473.02,159.01) .. (475.08,159.01) .. controls (477.15,159.01) and (478.82,160.68) .. (478.82,162.74) .. controls (478.82,164.81) and (477.15,166.48) .. (475.08,166.48) .. controls (473.02,166.48) and (471.35,164.81) .. (471.35,162.74) -- cycle ;
				%Shape: Ellipse [id:dp8198679350501424] 
				\draw  [fill={rgb, 255:red, 74; green, 144; blue, 226 }  ,fill opacity=0.4 ] (482.75,151.97) .. controls (482.75,149.91) and (484.42,148.24) .. (486.49,148.24) .. controls (488.55,148.24) and (490.22,149.91) .. (490.22,151.97) .. controls (490.22,154.04) and (488.55,155.71) .. (486.49,155.71) .. controls (484.42,155.71) and (482.75,154.04) .. (482.75,151.97) -- cycle ;
				%Straight Lines [id:da8451186759496354] 
				\draw    (475.08,166.48) -- (480.15,171.68) ;
				%Straight Lines [id:da678763593553671] 
				\draw    (475.08,166.48) -- (470.65,171.68) ;
				%Shape: Ellipse [id:dp7594984710545862] 
				\draw  [fill={rgb, 255:red, 74; green, 144; blue, 226 }  ,fill opacity=1 ] (489.72,175.41) .. controls (489.72,173.35) and (491.39,171.68) .. (493.46,171.68) .. controls (495.52,171.68) and (497.19,173.35) .. (497.19,175.41) .. controls (497.19,177.48) and (495.52,179.15) .. (493.46,179.15) .. controls (491.39,179.15) and (489.72,177.48) .. (489.72,175.41) -- cycle ;
				%Shape: Ellipse [id:dp38464802829768885] 
				\draw  [fill={rgb, 255:red, 74; green, 144; blue, 226 }  ,fill opacity=1 ] (499.22,175.41) .. controls (499.22,173.35) and (500.89,171.68) .. (502.96,171.68) .. controls (505.02,171.68) and (506.7,173.35) .. (506.7,175.41) .. controls (506.7,177.48) and (505.02,179.15) .. (502.96,179.15) .. controls (500.89,179.15) and (499.22,177.48) .. (499.22,175.41) -- cycle ;
				%Shape: Ellipse [id:dp6095130346426982] 
				\draw  [fill={rgb, 255:red, 74; green, 144; blue, 226 }  ,fill opacity=0.6 ] (494.15,162.74) .. controls (494.15,160.68) and (495.83,159.01) .. (497.89,159.01) .. controls (499.95,159.01) and (501.63,160.68) .. (501.63,162.74) .. controls (501.63,164.81) and (499.95,166.48) .. (497.89,166.48) .. controls (495.83,166.48) and (494.15,164.81) .. (494.15,162.74) -- cycle ;
				%Straight Lines [id:da7185271791500384] 
				\draw    (497.89,166.48) -- (502.96,171.68) ;
				%Straight Lines [id:da9625147431459744] 
				\draw    (497.89,166.48) -- (493.46,171.68) ;
				%Straight Lines [id:da12887414475548586] 
				\draw    (475.08,159.01) -- (482.75,151.97) ;
				%Straight Lines [id:da04465170743039848] 
				\draw    (490.22,151.97) -- (497.89,159.01) ;
				%Straight Lines [id:da23467829819207964] 
				\draw    (535.22,131.88) -- (536.12,142.3) ;
				%Shape: Ellipse [id:dp12590399997489998] 
				\draw  [fill={rgb, 255:red, 248; green, 231; blue, 28 }  ,fill opacity=1 ] (699.85,165.87) .. controls (699.85,152.85) and (710.4,142.3) .. (723.42,142.3) .. controls (736.44,142.3) and (747,152.85) .. (747,165.87) .. controls (747,178.89) and (736.44,189.45) .. (723.42,189.45) .. controls (710.4,189.45) and (699.85,178.89) .. (699.85,165.87) -- cycle ;
				%Shape: Ellipse [id:dp7182101317154163] 
				\draw  [fill={rgb, 255:red, 74; green, 144; blue, 226 }  ,fill opacity=1 ] (703.41,175.81) .. controls (703.41,173.75) and (705.09,172.07) .. (707.15,172.07) .. controls (709.22,172.07) and (710.89,173.75) .. (710.89,175.81) .. controls (710.89,177.87) and (709.22,179.55) .. (707.15,179.55) .. controls (705.09,179.55) and (703.41,177.87) .. (703.41,175.81) -- cycle ;
				%Shape: Ellipse [id:dp7880820626536978] 
				\draw  [fill={rgb, 255:red, 74; green, 144; blue, 226 }  ,fill opacity=1 ] (712.92,175.81) .. controls (712.92,173.75) and (714.59,172.07) .. (716.65,172.07) .. controls (718.72,172.07) and (720.39,173.75) .. (720.39,175.81) .. controls (720.39,177.87) and (718.72,179.55) .. (716.65,179.55) .. controls (714.59,179.55) and (712.92,177.87) .. (712.92,175.81) -- cycle ;
				%Shape: Ellipse [id:dp3292445105759051] 
				\draw  [fill={rgb, 255:red, 74; green, 144; blue, 226 }  ,fill opacity=0.6 ] (707.85,163.14) .. controls (707.85,161.07) and (709.52,159.4) .. (711.59,159.4) .. controls (713.65,159.4) and (715.32,161.07) .. (715.32,163.14) .. controls (715.32,165.2) and (713.65,166.88) .. (711.59,166.88) .. controls (709.52,166.88) and (707.85,165.2) .. (707.85,163.14) -- cycle ;
				%Shape: Ellipse [id:dp9181979254909683] 
				\draw  [fill={rgb, 255:red, 74; green, 144; blue, 226 }  ,fill opacity=0.4 ] (719.25,152.37) .. controls (719.25,150.3) and (720.93,148.63) .. (722.99,148.63) .. controls (725.05,148.63) and (726.73,150.3) .. (726.73,152.37) .. controls (726.73,154.43) and (725.05,156.11) .. (722.99,156.11) .. controls (720.93,156.11) and (719.25,154.43) .. (719.25,152.37) -- cycle ;
				%Straight Lines [id:da51008178080684] 
				\draw    (711.59,166.88) -- (716.65,172.07) ;
				%Straight Lines [id:da9072148533306574] 
				\draw    (711.59,166.88) -- (707.15,172.07) ;
				%Shape: Ellipse [id:dp25628240420961024] 
				\draw  [fill={rgb, 255:red, 74; green, 144; blue, 226 }  ,fill opacity=1 ] (726.22,175.81) .. controls (726.22,173.75) and (727.89,172.07) .. (729.96,172.07) .. controls (732.02,172.07) and (733.7,173.75) .. (733.7,175.81) .. controls (733.7,177.87) and (732.02,179.55) .. (729.96,179.55) .. controls (727.89,179.55) and (726.22,177.87) .. (726.22,175.81) -- cycle ;
				%Shape: Ellipse [id:dp6871350431861706] 
				\draw  [fill={rgb, 255:red, 74; green, 144; blue, 226 }  ,fill opacity=1 ] (735.72,175.81) .. controls (735.72,173.75) and (737.4,172.07) .. (739.46,172.07) .. controls (741.53,172.07) and (743.2,173.75) .. (743.2,175.81) .. controls (743.2,177.87) and (741.53,179.55) .. (739.46,179.55) .. controls (737.4,179.55) and (735.72,177.87) .. (735.72,175.81) -- cycle ;
				%Shape: Ellipse [id:dp81479545086639] 
				\draw  [fill={rgb, 255:red, 74; green, 144; blue, 226 }  ,fill opacity=0.6 ] (730.66,163.14) .. controls (730.66,161.07) and (732.33,159.4) .. (734.39,159.4) .. controls (736.46,159.4) and (738.13,161.07) .. (738.13,163.14) .. controls (738.13,165.2) and (736.46,166.88) .. (734.39,166.88) .. controls (732.33,166.88) and (730.66,165.2) .. (730.66,163.14) -- cycle ;
				%Straight Lines [id:da7402537220434001] 
				\draw    (734.39,166.88) -- (739.46,172.07) ;
				%Straight Lines [id:da31480058672911193] 
				\draw    (734.39,166.88) -- (729.96,172.07) ;
				%Straight Lines [id:da3716303324207695] 
				\draw    (711.59,159.4) -- (719.25,152.37) ;
				%Straight Lines [id:da56478016583621] 
				\draw    (726.73,152.37) -- (734.39,159.4) ;
				%Straight Lines [id:da500215414720777] 
				\draw    (696.44,84.22) -- (640.66,58.63) ;
				
				% Text Node
				\draw (260.95,162.87) node [anchor=north west][inner sep=0.75pt]  [font=\footnotesize]  {$x_{i,1}^{[ 1]}$};
				% Text Node
				\draw (302.75,162.25) node [anchor=north west][inner sep=0.75pt]  [font=\footnotesize]  {$x_{i,2}^{[ 1]}$};
				% Text Node
				\draw (281.39,107.13) node [anchor=north west][inner sep=0.75pt]  [font=\footnotesize]  {$x_{i,1}^{[ 2]}$};
				% Text Node
				\draw (340.11,52.01) node [anchor=north west][inner sep=0.75pt]  [font=\footnotesize]  {$x_{i,1}^{[ 34]}$};
				% Text Node
				\draw (345.82,111.46) node [anchor=north west][inner sep=0.75pt]   [align=left] {...};
				% Text Node
				\draw (345.82,167.51) node [anchor=north west][inner sep=0.75pt]   [align=left] {...};
				% Text Node
				\draw (228.46,138.24) node [anchor=north west][inner sep=0.75pt]   [align=left] {{\footnotesize 30-min}};
				% Text Node
				\draw (247.92,86.1) node [anchor=north west][inner sep=0.75pt]   [align=left] {{\footnotesize hourly}};
				% Text Node
				\draw (311.04,32) node [anchor=north west][inner sep=0.75pt]   [align=left] {{\footnotesize daily}};
				% Text Node
				\draw (376.93,163.64) node [anchor=north west][inner sep=0.75pt]  [font=\footnotesize]  {$x_{i,33}^{[ 1]}$};
				% Text Node
				\draw (418.73,162.56) node [anchor=north west][inner sep=0.75pt]  [font=\footnotesize]  {$x_{i,34}^{[ 1]}$};
				% Text Node
				\draw (397.37,107.44) node [anchor=north west][inner sep=0.75pt]  [font=\footnotesize]  {$x_{i,17}^{[ 2]}$};
				% Text Node
				\draw (27.86,162.36) node [anchor=north west][inner sep=0.75pt]  [font=\footnotesize]  {$A_{1}$};
				% Text Node
				\draw (54.48,110.02) node [anchor=north west][inner sep=0.75pt]  [font=\footnotesize]  {$Z_{118}$};
				% Text Node
				\draw (117.54,51.37) node [anchor=north west][inner sep=0.75pt]  [font=\footnotesize]  {$M_{136}$};
				% Text Node
				\draw (129.35,114.25) node [anchor=north west][inner sep=0.75pt]   [align=left] {...};
				% Text Node
				\draw (130.27,167.3) node [anchor=north west][inner sep=0.75pt]   [align=left] {...};
				% Text Node
				\draw (54.8,162.73) node [anchor=north west][inner sep=0.75pt]  [font=\footnotesize]  {$A_{2}$};
				% Text Node
				\draw (82.29,162.73) node [anchor=north west][inner sep=0.75pt]  [font=\footnotesize]  {$A_{3}$};
				% Text Node
				\draw (184.53,110.33) node [anchor=north west][inner sep=0.75pt]  [font=\footnotesize]  {$Z_{135}$};
				% Text Node
				\draw (169.64,162.73) node [anchor=north west][inner sep=0.75pt]  [font=\footnotesize]  {$A_{116}$};
				% Text Node
				\draw (197.6,162.73) node [anchor=north west][inner sep=0.75pt]  [font=\footnotesize]  {$A_{117}$};
				% Text Node
				\draw (1.97,142.16) node [anchor=north west][inner sep=0.75pt]   [align=left] {{\footnotesize Area}};
				% Text Node
				\draw (26.98,88.11) node [anchor=north west][inner sep=0.75pt]   [align=left] {{\footnotesize Zone}};
				% Text Node
				\draw (70.79,31.97) node [anchor=north west][inner sep=0.75pt]   [align=left] {{\footnotesize Market}};
				% Text Node
				\draw (285.2,7.4) node [anchor=north west][inner sep=0.75pt]  [font=\footnotesize] [align=left] {\textbf{temporal hierarchy}};
				% Text Node
				\draw (45,7.4) node [anchor=north west][inner sep=0.75pt]  [font=\footnotesize] [align=left] {\textbf{cross-sectional hierarchy}};
				% Text Node
				\draw (618,100) node [anchor=north west][inner sep=0.75pt]   [align=left] {\textbf{{\large ...}}};
				% Text Node
				\draw (618,160) node [anchor=north west][inner sep=0.75pt]   [align=left] {\textbf{{\large ...}}};
				% Text Node
				\draw (455.48,128.38) node [anchor=north west][inner sep=0.75pt]  [font=\footnotesize] [align=left] {Area};
				% Text Node
				\draw (495.57,75.23) node [anchor=north west][inner sep=0.75pt]  [font=\footnotesize] [align=left] {Zone};
				% Text Node
				\draw (556.08,32) node [anchor=north west][inner sep=0.75pt]  [font=\footnotesize] [align=left] {Market};
				% Text Node
				\draw (520.8,7.4) node [anchor=north west][inner sep=0.75pt]  [font=\footnotesize] [align=left] {\textbf{cross-temporal hierarchy}};

			\end{tikzpicture}
	\end{minipage}}
	\caption{Cross-sectional  (left), temporal  (middle) and cross-temporal (right) hierarchy in the platform application.} \label{hierarchy-platform}
\end{figure}
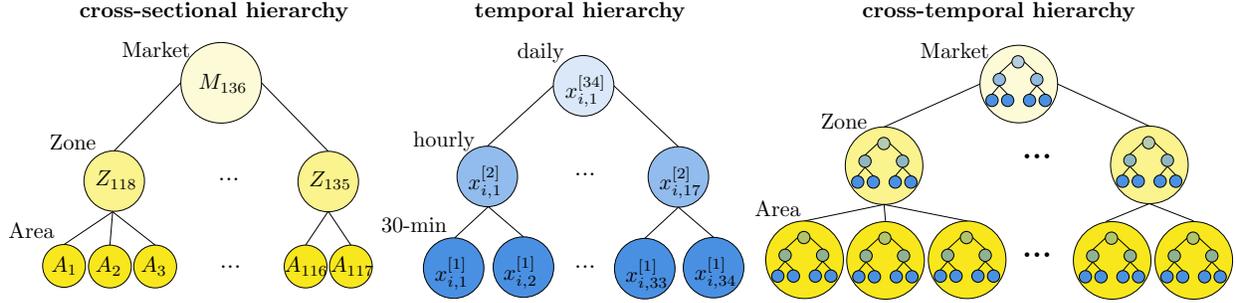

In our London platform application, regarding the cross-sectional hierarchy the bottom-level consists of $n_b=117$ delivery areas that are aggregated into the upper levels consisting of 18 zones (middle) and finally one market (top level), as visualized in the left panel of Figure \ref{hierarchy-platform}. The total number of variables is thus $n = 117 + 18 +1 = 136$. 
Regarding the temporal hierarchy, we have three frequencies $p=3$, (30-min - hourly - daily), as visualized in the middle panel of Figure \ref{hierarchy-platform},
$m=34$ and $k_1 = 1,\ k_2 = 2$ and $k_{3} = 34$
so that $\boldsymbol{x}_{i,\tau}^{[1]}$  ($\boldsymbol{x}_{i,\tau}^{[2]}$) collects the $m_1 = 34$  ($m_2 = 17$) 30-min (hourly) demand observations on day $\tau$, and  $x_{i,\tau}^{[34]}$ corresponds to the single daily observation for variable $i$. 
Finally, the combined cross-temporal hierarchy is displayed in the right panel of Figure \ref{hierarchy-platform}. 
In our Manchester platform application, the temporal hierarchy is the same as for London but the cross-sectional hierarchy consists of $n_b=24$ bottom-level series that are aggregated in a single upper-level series for the market.
Finally, for the Citi Bike application, the cross-section consists of $n_b=6$ bottom level series that are aggregated in a single upper-level series, whereas for the temporal 
hierarchy, the set of $p=10$ temporal aggregation orders is given by $K=\{k_1 = 1, 2, 3, 4, 6, 8, 12, 16, 24, k_p = 48\}$ with $m=48$.

\subsection{Forecast Reconciliation via Machine Learning} \label{subsec:MLreconciliation-for-CT}
We are now ready to introduce our machine learning-based forecast reconciliation approach for cross-temporal hierarchies, thereby extending the work of  \cite{spiliotis2021hierarchical} for cross-sectional hierarchies.

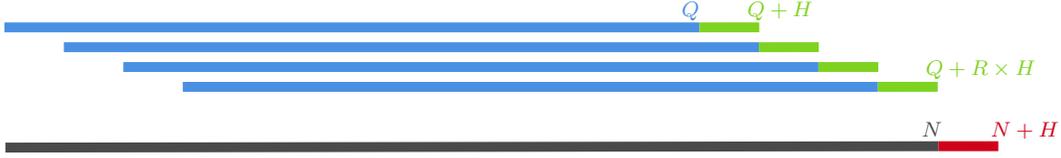
\begin{figure}[t]
	\centering
	\tikzset{every picture/.style={line width=0.75pt}} 
	
	\begin{tikzpicture}[x=0.75pt,y=0.75pt,yscale=-1,xscale=1]
		
		%Straight Lines [id:da6826880536751605] 
		\draw [color={rgb, 255:red, 74; green, 144; blue, 226 }  ,draw opacity=1 ][line width=3.75]    (59.71,81.86) -- (410.29,81.86) ;
		%Straight Lines [id:da5818702050703384] 
		\draw [color={rgb, 255:red, 126; green, 211; blue, 33 }  ,draw opacity=1 ][line width=3.75]    (410.29,81.86) -- (440.45,81.84) ;
		%Straight Lines [id:da12633158005803558] 
		\draw [color={rgb, 255:red, 74; green, 144; blue, 226 }  ,draw opacity=1 ][line width=3.75]    (89.71,91.86) -- (440.29,91.86) ;
		%Straight Lines [id:da6725621337816694] 
		\draw [color={rgb, 255:red, 126; green, 211; blue, 33 }  ,draw opacity=1 ][line width=3.75]    (440.29,91.86) -- (470.45,91.84) ;
		%Straight Lines [id:da6258471218417176] 
		\draw [color={rgb, 255:red, 74; green, 144; blue, 226 }  ,draw opacity=1 ][line width=3.75]    (119.71,101.86) -- (470.29,101.86) ;
		%Straight Lines [id:da9498445983982271] 
		\draw [color={rgb, 255:red, 126; green, 211; blue, 33 }  ,draw opacity=1 ][line width=3.75]    (470.29,101.86) -- (500.45,101.84) ;
		%Straight Lines [id:da2132347102890637] 
		\draw [color={rgb, 255:red, 74; green, 144; blue, 226 }  ,draw opacity=1 ][line width=3.75]    (149.71,111.86) -- (500.29,111.86) ;
		%Straight Lines [id:da6579846655045232] 
		\draw [color={rgb, 255:red, 126; green, 211; blue, 33 }  ,draw opacity=1 ][line width=3.75]    (500.29,111.86) -- (530.45,111.84) ;
		%Straight Lines [id:da009413434453608982] 
		\draw [color={rgb, 255:red, 74; green, 74; blue, 74 }  ,draw opacity=1 ][line width=3.75]    (60.21,142.52) -- (530.83,141.83) ;
		%Straight Lines [id:da1043731796692351] 
		\draw [color={rgb, 255:red, 208; green, 2; blue, 27 }  ,draw opacity=1 ][line width=3.75]    (530.83,141.83) -- (560.99,141.82) ;
		
		% Text Node
		\draw (521,128.2) node [anchor=north west][inner sep=0.75pt]  [font=\scriptsize,color={rgb, 255:red, 74; green, 74; blue, 74 }  ,opacity=1 ]  {$N$};
		% Text Node
		\draw (399.8,67.2) node [anchor=north west][inner sep=0.75pt]  [font=\scriptsize]  {$\textcolor[rgb]{0.29,0.56,0.89}{Q}$};
		% Text Node
		\draw (522.9,97.2) node [anchor=north west][inner sep=0.75pt]  [font=\scriptsize,color={rgb, 255:red, 126; green, 211; blue, 33 }  ,opacity=1 ]  {$Q+R\times H$};
		% Text Node
		\draw (433,67.2) node [anchor=north west][inner sep=0.75pt]  [font=\scriptsize,color={rgb, 255:red, 126; green, 211; blue, 33 }  ,opacity=1 ]  {$Q+H$};
		% Text Node
		\draw (555.8,128.2) node [anchor=north west][inner sep=0.75pt]  [font=\scriptsize,color={rgb, 255:red, 208; green, 2; blue, 27 }  ,opacity=1 ]  {$N+H$};

	\end{tikzpicture}
	\caption{Rolling-window procedure for machine-learning based forecast reconciliation.
	}
	\label{fig:tikz_rollingwindow}
\end{figure}

The general idea behind  ML-based reconciliation is to model the relationship between each bottom-level time series $\boldsymbol{x}_{b,\tau}^{[1]}$ ($b=1, \ldots, n_b$)  of the cross-temporal hierarchy and the  base forecasts across the entire cross-temporal tree $\widehat{\boldsymbol{x}}_{i,\tau}^{[k]}$ ($i=1, \ldots, n$ and $k=k_1, \ldots, k_p$). 
We use a time series rolling-window procedure to construct these base forecasts as visualized in  Figure \ref{fig:tikz_rollingwindow}. The rolling-window size of the estimation sample is denoted by $Q$ (length of the blue bars) and the validation sample with  forecast horizon by $H$ (length of the green bars), both quantities are expressed in number of observations for the top temporal aggregation level (i.e.\ the daily time series in our applications). 
Then, for each cross-sectional series at temporal aggregation level $k$ and rolling window iteration $r=1, \ldots, R$ (i.e.\ $R=4$ in Figure \ref{fig:tikz_rollingwindow} as used in our last-mile logistics application), we estimate base forecast models on the estimation sample, consisting of $m_k\times Q$ observations, and obtain 
$m_k \times H$ multistep forecasts, with step-ahead-periods ranging from $1$ to $m_k \times H$.

\definecolor{myyellow}{RGB}{248,231,28}
\definecolor{myblue}{RGB}{74,144,226}

Next, from the observations in the validation sample $\mathcal{V}$ consisting of observations  $\mathcal{V}=\{Q+1, \ldots, Q+R\times H\}$, we construct the response vector and features matrix  for the ML-model which can be generally written as
\begin{align}
	\boldsymbol{y}_{b,\mathcal{V}} = f_{b}(\widehat{\boldsymbol{X}}_{b,\mathcal{V}}) + \boldsymbol\varepsilon_{b,\mathcal{V}}, \label{ML-model}
\end{align}
where $f_b(\cdot)$ denotes the forecast function to be trained for bottom-level series $b$, see Section \ref{subsec:baseforecasts-MLmethods} for specifications, and $\boldsymbol\varepsilon_{b, \mathcal{V}}$ represents the error term.  Note that every bottom-level series in the cross-temporal hierarchy (i.e.\ 30-min series 
in our applications) is predicted by a separate ML model, thereby allowing the forecast reconciliation approach to adapt to different patterns in each series.
The response variable $\boldsymbol{y}_{b,\mathcal{V}}$ in \eqref{ML-model} is simply given by the $m_1\times R\times H$ observations of the bottom-level series for variable $b$; hence $${\boldsymbol{y}}_{b,\mathcal{V}} = 
\left[{\boldsymbol{x}_{b,Q+1}^{[1]}}^\top,\dots, {\boldsymbol{x}_{b,Q+RH} ^{[1]}}^\top\right]^\top. 
$$
As features matrix to the ML-model, the most natural generalization of \cite{spiliotis2021hierarchical} to the cross-temporal framework would be to use the same features matrix for each bottom-level series $b$, namely $\widehat{\boldsymbol{X}}_{b,\mathcal{V}}=\widehat{\boldsymbol{X}}_{\mathcal{V}}=$
\begin{small}
	\begin{equation}
		\begin{bmatrix}
			\hat{\boldsymbol{x}}_{1,Q+1}^{[ 1]} & \dotsc    & \hat{\boldsymbol{x}}_{n,Q+1}^{[ 1]}  & \hat{\boldsymbol{x}}_{1,Q+1}^{[ k_{2}]} \otimes \boldsymbol{1}_{k_{2}} & \dotsc  & \hat{\boldsymbol{x}}_{n,Q+1}^{[ k_{2}]} \otimes \boldsymbol{1}_{k_{2}} & \dotsc & \hat{x}_{1,Q+1}^{[ k_{p}]} \otimes \boldsymbol{1}_{k_{p}} & \dotsc & \hat{x}_{n,Q+1}^{[ k_{p}]} \otimes \boldsymbol{1}_{k_{p}}\\
			\vdots  & \ddots    & \vdots    & \vdots  & \ddots  & \vdots & \ddots  & \vdots & \ddots & \vdots \\
			\hat{\boldsymbol{x}}_{1,Q+RH}^{[ 1]} & \dotsc  & \hat{\boldsymbol{x}}_{n,Q+RH}^{[ 1]} & \hat{\boldsymbol{x}}_{1,Q+RH}^{[ k_{2}]} \otimes \boldsymbol{1}_{k_{2}} & \cdots &\hat{\boldsymbol{x}}_{n,Q+RH}^{[ k_{2}]} \otimes \boldsymbol{1}_{k_{2}} & \dotsc & \hat{x}_{1,Q+RH}^{[ k_{p}]} \otimes \boldsymbol{1}_{k_{p}} & \dotsc & \hat{x}_{n,Q+RH}^{[ k_{p}]} \otimes \boldsymbol{1}_{k_{p}}
		\end{bmatrix}, \nonumber
	\end{equation}
\end{small}
which contains all the base forecasts for all the series in the cross-temporal hierarchy on the validation sample, where $\otimes$ denotes the Kronecker product and $\boldsymbol{1}$ is a column vector of ones. Due to the frequency mismatch, each of the lower-frequency series 
at level $k_{a}$ for $a=2, \ldots, p$ 
are repeated respectively $k_a$ times. Note, however, that $\widehat{\boldsymbol{X}}_{\mathcal{V}}$ contains $p\times n$ features in its columns which may quickly become large and this curse of dimensionality may lead to a serious computational burden in practice. To solve this issue, we suggest a more compact bottom-series specific features matrix as alternative, namely
$\widehat{\boldsymbol{X}}_{b,\mathcal{V}}=$
\noindent
\begin{adjustwidth*}{-6.8em}{-7em}
	\begin{tikzpicture}[x=1pt,y=1pt,yscale=-1,xscale=1]
		%uncomment if require: \path (0,300); %set diagram left start at 0, and has height of 300
		
		%Shape: Rectangle [id:dp03652222876668265] 
		\draw  [color={rgb, 255:red, 74; green, 144; blue, 226 }  ,draw opacity=1 ][fill={rgb, 255:red, 74; green, 144; blue, 226 }  ,fill opacity=0.5 ][line width=1.5]  (265,4.3) -- (478,4.3) -- (478,66.7) -- (265,66.7) -- cycle ;
		%Shape: Rectangle [id:dp6437348097703011] 
		\draw  [color={rgb, 255:red, 248; green, 231; blue, 28 }  ,draw opacity=1 ][fill={rgb, 255:red, 248; green, 231; blue, 28 }  ,fill opacity=0.5 ][line width=1.5]  (2.67,6) -- (311.8,6) -- (311.8,65) -- (2.67,65) -- cycle ;
		
		% Text Node
		\draw (-81,40.4) node [anchor=north west][inner sep=0.75pt]    {$\ \ \ $};
		% Text Node
		\draw (-2.17,3.57) node [anchor=north west][inner sep=0.75pt]  [font=\footnotesize]  {$\begin{bmatrix}
				\hat{\boldsymbol{x}}_{1,Q+1}^{[ 1]} & \dotsc  & \hat{\boldsymbol{x}}_{b-1,Q+1}^{[ 1]} & \hat{\boldsymbol{x}}_{b+1,Q+1}^{[ 1]} & \dotsc  & \hat{\boldsymbol{x}}_{n,Q+1}^{[ 1]} & \hat{\boldsymbol{x}}_{b,Q+1}^{[ 1]} & \hat{\boldsymbol{x}}_{b,Q+1}^{[ k_{2}]} \otimes \boldsymbol{1}_{k_{2}} & \dotsc  & \hat{x}_{b,Q+1}^{[ k_{p}]} \otimes \boldsymbol{1}_{k_{p}}\\
				\vdots  & \ddots  & \vdots  & \vdots  & \ddots  & \vdots  & \vdots  & \vdots  & \ddots  & \vdots \\
				\hat{\boldsymbol{x}}_{1,Q+RH}^{[ 1]} & \dotsc  & \hat{\boldsymbol{x}}_{b-1,Q+RH}^{[ 1]} & \hat{\boldsymbol{x}}_{b+1,Q+RH}^{[ 1]} & \dotsc  & \hat{\boldsymbol{x}}_{n,Q+RH}^{[ 1]} & \hat{\boldsymbol{x}}_{b,Q+RH}^{[ 1]} & \hat{\boldsymbol{x}}_{b,Q+RH}^{[ k_{2}]} \otimes \boldsymbol{1}_{k_{2}} & \dotsc  & \hat{x}_{b,Q+RH}^{[ k_{p}]} \otimes \boldsymbol{1}_{k_{p}}
			\end{bmatrix},$};
	\end{tikzpicture}
\end{adjustwidth*}
which contains a selection of base forecasts on the validation sample. 
More specifically,
the first set of predictors, highlighted in yellow, exploits information in the cross-sectional hierarchy and contains the $n_b$ highest-frequency (i.e.\ 30-min in our applications) 
base forecasts for all bottom-level variables, 
as well as the $n_a$ highest-frequency base forecasts for the upper level series in the cross-temporal hierarchy 
(i.e.\ zone and market for London; market for Manchester and Citi Bike).
The second set of predictors highlighted in blue exploits information in the temporal hierarchy and contains all temporal 
base forecasts for the  bottom-level variable $b$ for which we are constructing our ML-model.
The number of features in the columns of the matrix $\widehat{\boldsymbol{X}}_{b,\mathcal{V}}$ is now substantially reduced to  $n+p-1$ features, but does rest on
the assumption that the temporal gain in reconciling forecasts mainly comes from the focal variable's own temporally aggregated series to keep this second set of predictors compact.
Finally, note the central role for the bottom level variable $b$ itself (highlighted in yellow and blue) whose base forecast we aim to improve upon in this reconciliation step.
In the remainder of the paper, to keep the computational burden feasible, we use this more compact features matrix when displaying the main results.\footnote{We ran some small-scale experiments for the last-mile delivery applications and did not find a substantial gain in out-of-sample forecast performance from using the complete features matrix $\widehat{\boldsymbol{X}}_{\mathcal{V}}$  instead of the more compact features matrix $\widehat{\boldsymbol{X}}_{b,\mathcal{V}}$, thereby providing empirical support for choosing the latter in these applications. 
	For the Citi Bike dataset, we find an improvement in out-of-sample forecast performance from using $\widehat{\boldsymbol{X}}_{\mathcal{V}}$  instead of $\widehat{\boldsymbol{X}}_{b,\mathcal{V}}$ for the H3 cells but not for the market, see Section \ref{subsec:sensitivity-Citibike} for further details.}

Finally, to reconcile the actual out-of-sample demand forecasts of horizon $h = 1,\ldots,H$ for forecast evaluation purposes, we use the complete sample up to $N=Q+R\times H$ to obtain the $m_k\times H$ multistep forecasts (with step-ahead-periods ranging from $1$ to $m_k \times H$)
for each cross-sectional series at temporal aggregation level $k$. This is visualized in the bottom part of Figure \ref{fig:tikz_rollingwindow}, where the  gray bar represents the training sample (of length $N$) on which the base forecast models are estimated and the red bar (of length $H$) represents the test set on which we evaluate out-of-sample forecast performance.  
The out-of-sample base forecasts 
$\widehat{\boldsymbol{x}}_{i, N+h}^{[1]}$ ($i=1, \ldots, n$ and $h=1, \ldots, H$), and $\widehat{\boldsymbol{x}}_{b, N+h}^{[k]}$ ($b=1, \ldots, n_{b}, \ h=1, \ldots, H$, and $k=k_{2}, \ldots, k_{p}$)
are  used as inputs to the previously estimated  ML models (one for each bottom-level series) to provide the revised bottom-level forecasts 
$\widetilde{\boldsymbol{x}}_{b,N+h}^{[1]}$ ($b=1, \ldots, n_b$ and $h=1,\ldots, H$) 
of the cross-temporal hierarchy.
As a last step, these revised bottom-level forecasts are then aggregated  as explained in Section \ref{subsec:CThierarchy} to obtain coherent forecasts across the complete hierarchy.

\section{Design of the Forecast Study  \label{sec:setup}}
We start by discussing in Section \ref{subsec:forecast:setup} the forecast set-up we use to 
compare the forecast reconciliation methods. Section \ref{subsec:forecast-evaluation}  discusses the metrics we use to evaluate out-of-sample forecast performance.
Section \ref{subsec:baseforecasts-MLmethods} summarizes the base forecast models all forecast reconciliation methods rely on, together with the machine learning methods used by our procedure to obtain the reconciled forecasts.
Finally, Section \ref{subsec:benchmarks} outlines the benchmark forecast methods against which we compare the performance of our procedure.

\subsection{Forecast Set-up} \label{subsec:forecast:setup}
To compare the forecast performance of our forecast reconciliation method against its benchmarks, we use a standard rolling-window approach.
We first describe the set-up used for London and Manchester, then summarize the differences in set-up for Citi Bike at the end.

\textit{London and Manchester.} We use the period February 20, 2023 to September 24, 2023 
(31 weeks) as test sample. 
In each rolling window iteration, we use the most recent 
24 weeks of data to estimate the base models on the estimation set (first 
20 weeks), and  the most recent  
4 weeks of data to estimate the ML model on the validation set.
Hence, in line with the notation used in Figure \ref{fig:tikz_rollingwindow}, we take $Q=140$ days, $H=7$ days and $R=4$.
By taking $H=7$, it should be noted that we combine base forecasts with different forecast uncertainty, as the latter typically increases with the forecast horizon.
Finally, since the demand data in our platform applications consists of non-negative integers we ensure this by rounding the base forecasts prior to using them as inputs in the ML model, i.e. $\hat{x}_{i,t}^{{[k]}^\star} = \text{max}(0, \lfloor \hat{x}_{i,t}^{[k]} \rceil)$, we similarly round  the bottom-level reconciled forecasts before aggregating them. 

In a next step, we combine the estimation and validation sample ($N=168$ days) to estimate the base models and obtain final out-of-sample base forecasts as well as reconciled forecasts on the test set. 
As forecast horizon, we consider one week of data 
(i.e.\ we obtain $m_{k} \times H$ multistep forecasts where the step-ahead periods range from 1 to $m_k \times H$ 
for temporal aggregation level $k$) 
as this corresponds to the planning horizon for the compensation schemes of couriers through dynamic pricing which crucially relies on demand forecasts as inputs.
Hence, for the 30-min series, we obtain $34\times 7  = 238$ forecasts; for the hourly series we obtain $17\times 7 = 119$ forecasts and for the daily series we obtain $H=7$ forecasts.
This concludes one iteration in the outer rolling window used for forecast evaluation purposes. In the next iteration, we move one-week ahead and repeat the same steps. We roll forward until we reach the end of our sample, thereby resulting in 31 outer rolling windows.  

Figure \ref{fig:tikz_iterations} in Appendix \ref{app:rolling-window}
provides a visualization of iteration $r$ and $r+1$ in the outer rolling window. 
We hereby highlight the estimation (blue bars) and validation (green bars) sets for the inner rolling window used to construct the inputs to the ML model (see Section \ref{subsec:MLreconciliation-for-CT}). Due to our inner rolling-window set-up, 
the training set (blue bar) and forecast horizon (green bar) of the last three (inner) rolling windows in iteration $r$ overlap with the first three (inner) rolling windows in iteration $r+1$, as indicated with the shading. This offers a 
clear computational advantage as beyond iteration $r=1$, we can re-use the base forecasts for the first three weeks from the previous iteration and need to fit the forecasting model only once to obtain the base forecasts for the fourth week to construct the validation set.

\textit{Citi Bike.} We use the period June 18, 2023 to December 31, 2023 as test sample and take $Q=140$ days, $H=1$ day, $R=28$ days and $N=168$ days. 
As forecast horizon, we thus consider the natural choice $H=1$ day which is of  great interest in many other fields where forecast reconciliation has been applied. In each iteration of the outer-rolling window we then roll one day forward, resulting in 197 outer rolling windows.

\subsection{Evaluating Forecast Performance} \label{subsec:forecast-evaluation}
To evaluate the forecast performance of the various forecast methods on the three datasets, we use the Weighted Absolute Percentage Error (WAPE), alternatively referred to as the Weighted Average Percentage Error, and known as normalized version of the mean absolute error (MAE), aka as MAD/Mean ratio as firstly considered by \cite{hoover2006measuring}, where MAD is a synonym for MAE. 
The WAPE accuracy index for each temporal factor $k$ and cross-sectional time series $i=1, \ldots, n$ is given by 
$$
\text{WAPE}^{[k]}_i = \frac{\sum_{j=1}^{T_{\text{test}}/k} |A_{i,j}^{[k]} - F_{i,j}^{[k]}|}{\sum_{j=1}^{T_{\text{test}}/k} A_{i,j}^{[k]}},
$$
which measures the overall deviation of the forecasted values $F_{i,j}^{[k]}$ from the actual values $A_{i,j}^{[k]}$, for all time points $j$ in the test set of size $T_{\text{test}}/k$ where
$T_{\text{test}}$ denotes the number of high-frequency (i.e.\ 30-min) observations in the test set.
Note that the WAPE can be multiplied by 100 to get a percentage value.
We then  average these WAPEs across all cross-sectional units belonging to the same cross-temporal level; for instance the overall WAPE for the bottom-level series (i.e.\ 30-min for the $n_b$ areas) is then given by 
$\text{WAPE}^{[1]} = (1/n_b) \sum_{b=1}^{n_b} \text{WAPE}^{[1]}_b$. 

The WAPE is a popular forecast metric to evaluate performance at delivery platforms 
(i) because of its scale-independence thereby facilitating comparisons across heterogeneous areas as well as different time/geographical aggregation levels, (ii) it is agnostic to zeros as it is very unlikely that the denominator is zero, thereby making it a suitable metric even when the demand in an area is low or intermittent, and (iii) it emphasizes accuracy for items with larger demand, thereby prioritizing accurate forecasting of demand during busy periods. 

Note that, as a robustness check, we also report results for the Mean Absolute Scaled Error (MASE), defined as
\begin{equation}
	\text{MASE}_{i}^{[k]} =   \frac{\frac{1}{m_k H} \sum\limits_{j=m_kN+1}^{m_k(N+H)} |A_{i,j}^{[k]} - F^{[k]}_{i,j}|}{
		\frac{1}{m_k (N-7)} \sum\limits_{j=7m_k+1}^{m_kN} |A_{i,j}^{[k]} - A_{i,j - 7m_k}^{[k]}|
	}, \nonumber
\end{equation}
for each temporal factor $k$ and cross-sectional time series $i=1, \ldots, n$.
Here the absolute (out-of-sample) errors are scaled based on the in-sample MAE
from a ``naive" forecast method 
in which the forecast for
each future  time slot of next week is the actual value for the same specific time slot and day of this week.
While the WAPE assumes that the mean is stable over time, making it sensitive to trend, seasonality or other patterns in the data, the MASE is unaffected by such patterns, for example  the seasonal pattern that repeats itself every week for a specific hour of the day and day of the week in the datasets we consider.
We refer the interested reader to \cite{hyndman2006anotherlook}, \cite{hyndman2006} for a more elaborate discussion.

\subsection{Base Forecasts and Machine Learning Methods} \label{subsec:baseforecasts-MLmethods} 
Our procedure allows practitioners to choose their preferred base forecast models as well as their preferred ML method to perform the ML-based reconciliation.
In this paper, we consider four base forecast models and three popular ML methods to this end.
\bigskip

\noindent
{\bf Base Forecasts.} We consider one popular industry forecast model for platform data, two all-round time series models for forecasting univariate time series and a forecast combination of the former three as ensemble, base forecast model.
The first base forecast model is a simple yet often-used 
model in the platform industry which we will label as ``Naive" since the forecast for a specific  slot (for instance, 30-min, hourly or daily) of next week is simply the value observed in the same slot and day of the previous week. 
The following two  base forecast models are oftentimes used in the hierarchical forecasting literature to obtain base forecasts, namely the class of SARIMA and exponential smoothing models.
Full details on the base forecast methods are available in Appendix \ref{app:base-forecasts}.

\bigskip
\noindent
{\bf Machine Learning Methods.} We consider three popular ML models, namely random forest, XGBoost and LightGBM.
The former two are also considered in \cite{spiliotis2021hierarchical} whereas the latter has recently shown great success on the M5 competition \citep{makridakis2022m5accuracyresults}
consisting of an application with hierarchical retail sales time series, thereby making  all suitable candidates for our platform data.
More details on the ML models are available in Appendix \ref{app:ML-methods}. 

For all three ML methods, we report main results based on their standard implementations with default tuning parameters (see Appendix \ref{app:ML-methods}).  
We do this, first to avoid excessive computing times on our streaming platform data, but second, and importantly, also because off-the-shelf implementations of tree-based methods often attain excellent performance across a variety of settings (e.g, \citealp{januschowski2022forecasting}). 
In Section \ref{subsec:result:tuning}, we investigate the improvements in forecast performance one can obtain by tuning these ML models.

Finally, when reporting our main results, we use the sum of squared errors as loss function to train all ML models.
However, our framework allows practitioners to use another loss function as they see suited for the data at hand. 
We also investigated the performance of the ML-based reconciliation methods when using the Tweedie loss function instead of the standard squared loss, since the former showed superior performance in the M5 competition (\citealp{januschowski2022forecasting, makridakis2022m5accuracyresults}) on forecasting hierarchical time series as it is specifically geared towards sparse (zero-inflated) target data, which some of the considered delivery areas also display (see Figure \ref{fig:map_London_totalareademand}).
However, using a Tweedie loss instead of the regular squared loss function did not improve forecast results of our proposed ML reconciliation methods on the considered data, and are hence omitted but available from the authors upon request.

\subsection{Benchmark Reconciliation Methods} \label{subsec:benchmarks}
We compare our new ML forecast reconciliation procedure against the base forecasts and five state-of-the-art linear reconciliation benchmarks for cross-temporal hierarchies, as summarized in Table \ref{tabel:linearbenchmarks}  and implemented in the  \texttt{FoReco} package \citep{FoReco} in \verb|R| \citep{Rcoreteam}. 
Apart from the Bottom Up method, the reconciliation step requires specifying how to estimate the covariance matrix of the in-sample fit errors since the benchmark methods are geared towards minimizing reconciliation errors as opposed to our procedure which is geared towards minimizing forecast combination errors. 
For the tcs, cst and ite benchmark methods, the temporal reconciliation covariance matrix estimator is the series variance scaling matrix (thf\_comb = ``wlsv") which is the diagonal matrix that contains the estimated variances of the in-sample residuals across each level.
The cross-sectional reconciliation covariance matrix estimator is a shrinkage covariance matrix  of the in-sample residuals (hts\_comb = ``shr"), where the off-diagonal elements are shrunk towards zero. 
For the oct method, the cross-temporal based covariance matrix estimator is simply containing the in-sample residual variance on the main diagonal (csts\_comb = ``wlsv").
Full details on the cross-temporal benchmark methods are available in \cite{difonzo2023cross}.\footnote{There exist two different implementations of the iterative cross-temporal forecast reconciliation ``ite". The first reconciliation step in each iteration can either start from the temporal or from the cross-sectional reconciliation. We estimate both procedures, but only report upon the latter since it is the most competitive on the considered datasets.}  
Finally, note that possible negative reconciled values for the linear reconciliation benchmarks are dealt with in the same way as for the ML-based reconciliation methods.

\begin{table}[t]
	\begin{center}
		\caption{Overview of cross-temporal forecast reconciliation benchmarks.   \label{tabel:linearbenchmarks}}
		\begin{tabular}{p{3cm}p{12.5cm}}
			\midrule
			Name & Description  \\
			\midrule
			Bottom Up & Simple aggregation of bottom level forecasts for any cross-temporal hierarchy  \\[0.8cm]
			tcs & First-temporal-then-cross-sectional cross-temporal forecast reconciliation - \cite{kourentzes2019cross} \\[0.8cm]
			cst & First-cross-sectional-then-temporal cross-temporal forecast reconciliation - \cite{difonzo2023cross}\\[0.8cm]
			ite & Iterative cross-temporal forecast reconciliation starting with cross-sectional reconciliation - \cite{difonzo2023cross}  \\[0.8cm]
			oct & Optimal combination cross-temporal forecast reconciliation in a single reconciliation step - \cite{difonzo2023cross} \\
			\midrule
		\end{tabular} 
	\end{center}
	\raggedright
	\footnotesize
	$\;$\\[-0.6cm]
	Notes: This table summarizes the benchmark cross-temporal forecast reconciliation methods.   \\ 
\end{table}

\section{Results  \label{sec:results}}
This section is divided in two parts. 
We first present the results for the last mile delivery platform and then for Citi Bike.

\subsection{Results for Last-Mile Logistics Platform  Applications} \label{sec:Stuart-results}
In Section \ref{subsec:result:overall}, we first discuss the overall forecast performance of the forecast reconciliation methods for London and Manchester.
We then zoom into the London dataset  and present 
detailed insights on our
ML-based reconciliation approach (Section \ref{subsec:result:MLinsights}), and 
discuss the trade-off between computing time and forecast accuracy when incorporating hyperparameter tuning into our ML-based reconciliation proposal (Section \ref{subsec:result:tuning}).

\subsubsection{Overall Forecast Performance}  \label{subsec:result:overall}
We start by discussing the forecast performance on the London dataset and then highlight how shifts in the demand data streams affect the performance of the reconciliation methods on the Manchester dataset.

\begin{table}[htbp]
	\centering
	\caption{
		London forecast reconciliation results.}
	\resizebox{\textwidth}{!}{ \begin{tabular}{ll c c c|ccc|ccc}
			\hline
			\textbf{Base} & \textbf{Forecast} & \multicolumn{3}{c}{\textbf{Delivery Area}} & \multicolumn{3}{c}{\textbf{Zone}} & \multicolumn{3}{c}{\textbf{Market}} \\
			\textbf{Forecasts} & \textbf{Method} \\
			\hline
			&& \textbf{30min} & \textbf{hour} & \textbf{day} & \textbf{30min} & \textbf{hour} & \textbf{day} &\textbf{30min} & \textbf{hour} & \textbf{day}  \\
			\\
			Naive & Base & 0.3576 & 0.2703 & 0.1017 & 0.1475 & 0.1196 & 0.0638 & 0.0793 & 0.0756 & 0.0559 \\ 
			& Random forest & \cellcolor{lightgray}0.2762 & \cellcolor{lightgray}0.2162 & \cellcolor{lightgray}0.0976 & \cellcolor{lightgray}0.1271 & \cellcolor{lightgray}0.1081 & \cellcolor{lightgray}0.0636 & \cellcolor{lightgray}0.0819 & \cellcolor{lightgray}0.0782 & \cellcolor{lightgray}0.0576 \\ 
			& XGBoost & 0.3017 & 0.2338 & 0.0993 & 0.1352 & 0.1132 & 0.0654 & 0.0841 & 0.0796 & 0.0588 \\ 
			& LightGBM & 0.2898 & 0.2252 & 0.0991 & 0.1321 & 0.1110 & 0.0650 & 0.0842 & 0.0796 & 0.0587 \\ 
			\\
			\\
			ETS & Base & 0.3180 & 0.2654 & 0.0837 & 0.1935 & 0.1769 & 0.0539 & 0.1621 & 0.1521 & 0.0471 \\ 
			& Bottom Up & 0.3180 & 0.2682 & 0.1755 & 0.1842 & 0.1716 & 0.1378 & 0.1523 & 0.1501 & 0.1313 \\ 
			& tcs & 0.3105 & 0.2591 & 0.1587 & 0.1769 & 0.1641 & 0.1290 & 0.1440 & 0.1420 & 0.1229 \\ 
			& cst & 0.3070 & 0.2557 & 0.1536 & 0.1699 & 0.1570 & 0.1191 & 0.1349 & 0.1331 & 0.1117 \\ 
			& ite & 0.3112 & 0.2606 & 0.1617 & 0.1782 & 0.1659 & 0.1316 & 0.1454 & 0.1439 & 0.1251 \\ 
			& oct & 0.3099 & 0.2577 & 0.1545 & 0.1765 & 0.1633 & 0.1267 & 0.1435 & 0.1411 & 0.1206 \\ 
			& Random forest & \cellcolor{lightgray}0.2809 & \cellcolor{lightgray}0.2229 & \cellcolor{lightgray}0.1064 & \cellcolor{lightgray}0.1360 & \cellcolor{lightgray}0.1192 & \cellcolor{lightgray}0.0729 & 0.0959 & 0.0933 & 0.0673 \\ 
			& XGBoost & 0.3161 & 0.2545 & 0.1189 & 0.1439 & 0.1249 & 0.0735 & \cellcolor{lightgray}0.0920 & \cellcolor{lightgray}0.0889 & \cellcolor{lightgray}0.0634 \\ 
			& LightGBM & 0.2960 & 0.2374 & 0.1164 & 0.1387 & 0.1209 & 0.0737 & 0.0926 & 0.0896 & 0.0645 \\ 
			\\
			\\
			SARIMA & Base & 0.3118 & 0.2584 & 0.0894 & 0.1793 & 0.1656 & 0.0568 & 0.1490 & 0.1445 & 0.0526 \\ 
			& Bottom Up & 0.3118 & 0.2605 & 0.1637 & 0.1808 & 0.1682 & 0.1339 & 0.1511 & 0.1489 & 0.1286 \\ 
			& tcs & 0.3049 & 0.2524 & 0.1486 & 0.1742 & 0.1612 & 0.1253 & 0.1436 & 0.1416 & 0.1201 \\ 
			& cst & 0.3013 & 0.2488 & 0.1424 & 0.1679 & 0.1549 & 0.1164 & 0.1351 & 0.1334 & 0.1100 \\  
			& ite & 0.3051 & 0.2533 & 0.1508 & 0.1750 & 0.1627 & 0.1276 & 0.1446 & 0.1431 & 0.1221 \\ 
			& oct & 0.3043 & 0.2510 & 0.1443 & 0.1718 & 0.1583 & 0.1210 & 0.1408 & 0.1385 & 0.1164 \\ 
			& Random forest & \cellcolor{lightgray}0.2801 & \cellcolor{lightgray}0.2206 & \cellcolor{lightgray}0.1002 & \cellcolor{lightgray}0.1305 & \cellcolor{lightgray}0.1132 & 0.0657 & 0.0883 & 0.0861 & 0.0586 \\ 
			& XGBoost & 0.3195 & 0.2528 & 0.1109 & 0.1385 & 0.1182 & 0.0659 & 0.0845 & 0.0815 & 0.0551 \\ 
			& LightGBM & 0.2933 & 0.2327 & 0.1067 & 0.1320 & 0.1134 & \cellcolor{lightgray}0.0641 & \cellcolor{lightgray}0.0832 & \cellcolor{lightgray}0.0806 & \cellcolor{lightgray}0.0538 \\ 
			\\
			\\
			Forecast & Base & 0.2950 & 0.2365 & 0.0850 & 0.1502 & 0.1333 & 0.0538 & 0.1134 & 0.1089 & 0.0485 \\ 
			Combination & Bottom Up & 0.2950 & 0.2358 & 0.1289 & 0.1493 & 0.1333 & 0.0991 & 0.1120 & 0.1097 & 0.0938 \\ 
			& tcs & 0.2925 & 0.2349 & 0.1215 & 0.1470 & 0.1310 & 0.0962 & 0.1093 & 0.1073 & 0.0913 \\ 
			& cst & 0.2909 & 0.2331 & 0.1182 & 0.1438 & 0.1277 & 0.0908 & 0.1046 & 0.1027 & 0.0853 \\ 
			& ite & 0.2931 & 0.2359 & 0.1238 & 0.1483 & 0.1328 & 0.0984 & 0.1111 & 0.1095 & 0.0933 \\ 
			& oct & 0.2917 & 0.2336 & 0.1182 & 0.1454 & 0.1290 & 0.0928 & 0.1070 & 0.1046 & 0.0879 \\ 
			& Random forest & \cellcolor{lightgray}0.2758 & \cellcolor{lightgray}0.2161 & 0.1018 & \cellcolor{lightgray}0.1285 & 0.1108 & 0.0717 & 0.0871 & 0.0844 & 0.0666 \\ 
			& XGBoost & 0.3026 & 0.2355 & 0.1011 & 0.1336 & 0.1129 & 0.0671 & 0.0836 & 0.0803 & 0.0601 \\ 
			& LightGBM & 0.2900 & 0.2260 & \cellcolor{lightgray}0.0999 & 0.1301 & \cellcolor{lightgray}0.1103 & \cellcolor{lightgray}0.0664 & \cellcolor{lightgray}0.0831 & \cellcolor{lightgray}0.0799 & \cellcolor{lightgray}0.0593 \\ 
			
			\midrule
	\end{tabular}}
	\label{tab:London_FR_results}
	\raggedright
	\footnotesize
	$\;$\\[-0.1cm]
	Notes: This table shows forecast accuracy measured in weighted absolute percentage error (WAPE).  
	The best forecast reconciliation results for each base forecast (Naive, ETS, SARIMA, Forecast Combination) are highlighted in gray.
	The linear benchmark methods are defined in Table \ref{tabel:linearbenchmarks}.  
	\\[0.1cm]
\end{table}

\textit{London.} 
Table \ref{tab:London_FR_results} 
summarizes the forecast accuracy, in terms of WAPE,  across the complete cross-temporal hierarchy for the different combinations of  base forecasts and reconciliation methods.\footnote{Note that for the Naive base method, the base and bottom up forecasts are identical, and no results are reported for the linear forecast reconciliation methods since the original base forecasts are already reconciled. Indeed, the standard linear reconciliation methods reconcile forecasts by minimizing the reconciliation error between the base and reconciled forecasts, making this step redundant in this case. In contrast, the non-linear ML-based reconciliation methods are  geared towards minimizing forecast combination errors to improve out-of-sample forecast performance. Then, there is value in starting off from  the naive base forecasts since they might not be ``inherently reconciled" in the ``optimal way" (in terms of out-of-sample forecast performance), yet are  attractive to use in a production environment where computational speed is key since they do not require estimation time, unlike SARIMA and ETS.
} 
Overall, all methods find that forecasting is easier higher up in the hierarchy, i.e.\ lower temporal frequency and larger geographic areas, which confirms the general findings in the literature. For the set of considered base forecasts and reconciliation methods, WAPE falls from around 30\% to 5\% for respectively 30-minute delivery areas and the daily London hierarchy levels.

The ML-based forecast reconciliation results using random forest  are consistently outperforming all benchmarks at the delivery area and zone levels at all frequencies. For example, using 30-minute frequency SARIMA base forecasts, we obtain a WAPE of 0.3013 for the cst method (i.e.\ the best linear forecast reconciliation method) compared to 0.2801 for random forest based reconciliation (i.e.\ the best non-linear forecast reconciliation method). However, ML-based forecast reconciliation does not systematically dominate linear methods as shown for instance by XGBoost with a WAPE of 0.3195. 
Furthermore, ML-based forecast reconciliation seems to be rather insensitive to the choice of the base forecasts. In fact, the Naive, ETS and SARIMA results are qualitatively the same. For example in the case of random forest, the respective WAPEs are very close with 0.2762, 0.2809 and 0.2801 respectively. 

\begin{figure}[t]
	\centering
	\begin{subfigure}{0.4\textwidth}
		\includegraphics[scale = 0.40]{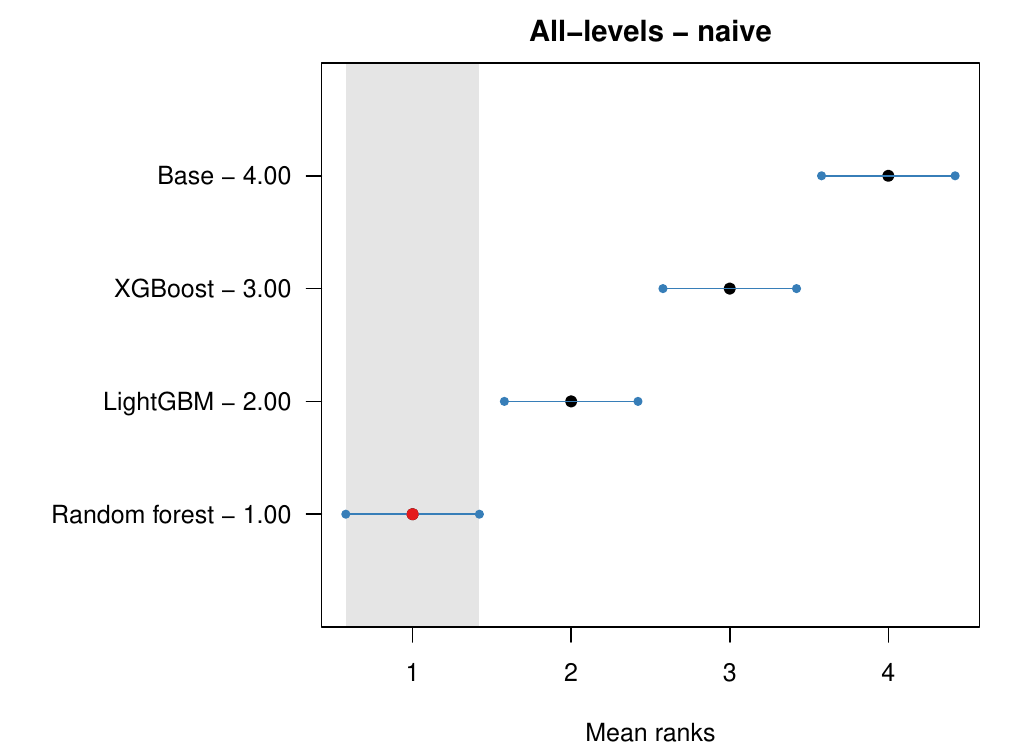}
	\end{subfigure} 
	\begin{subfigure}{0.4\textwidth}
		\includegraphics[scale = 0.4]{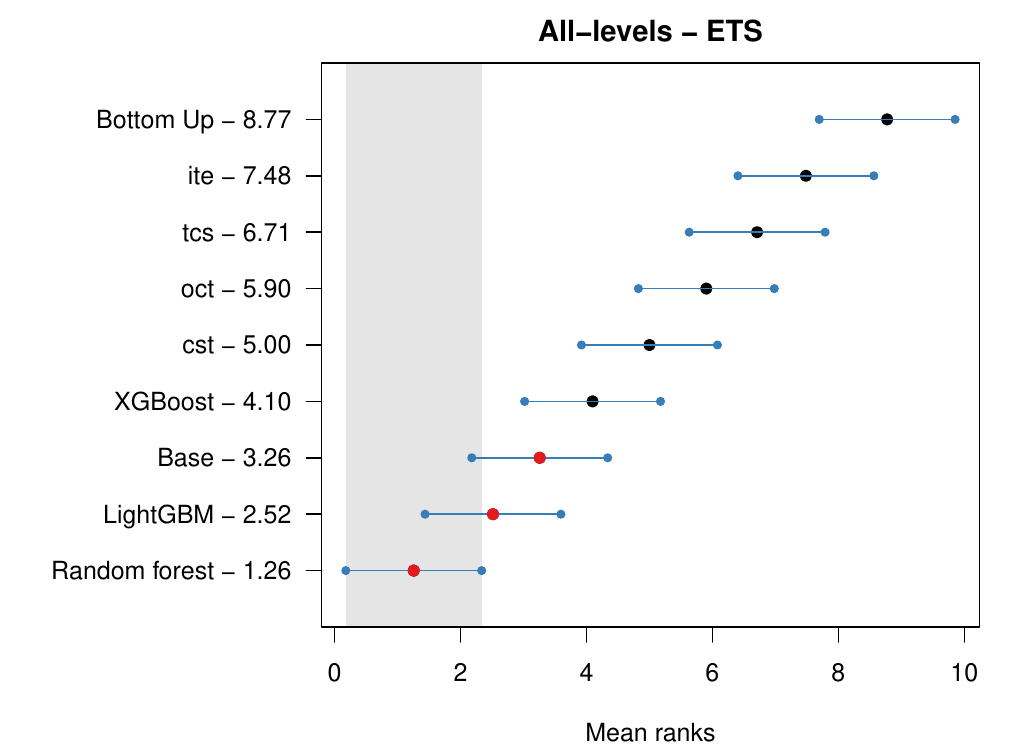}
	\end{subfigure} 
	\begin{subfigure}{0.4\textwidth}
		\includegraphics[scale = 0.4]{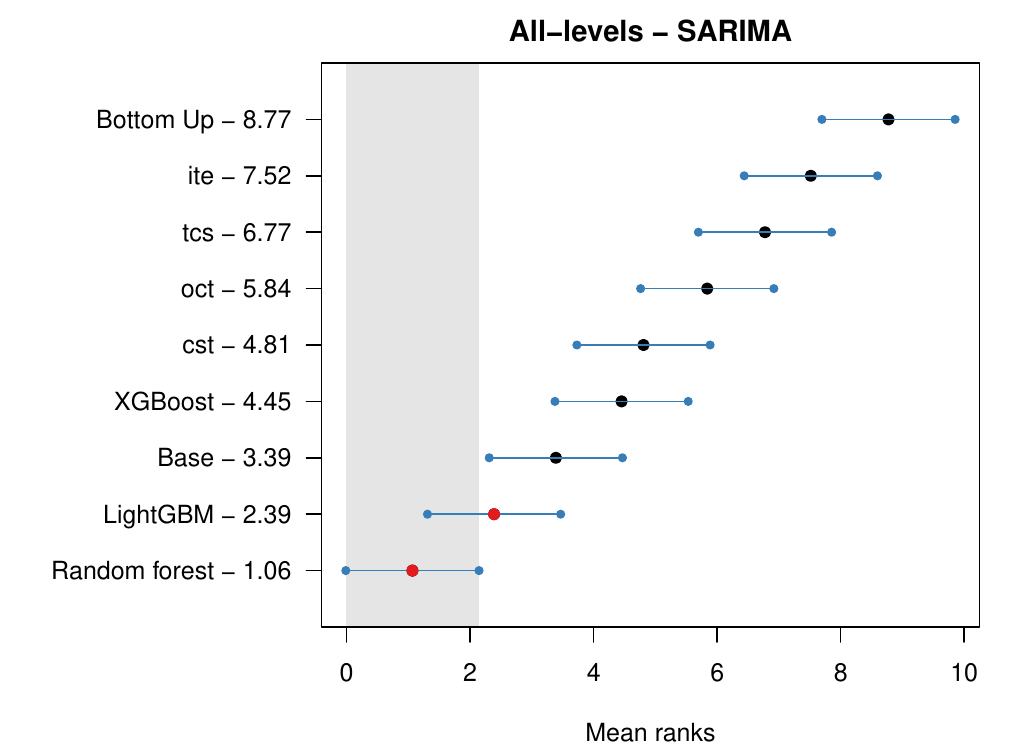}
	\end{subfigure} 
	\begin{subfigure}{0.4\textwidth}
		\includegraphics[scale = 0.4]{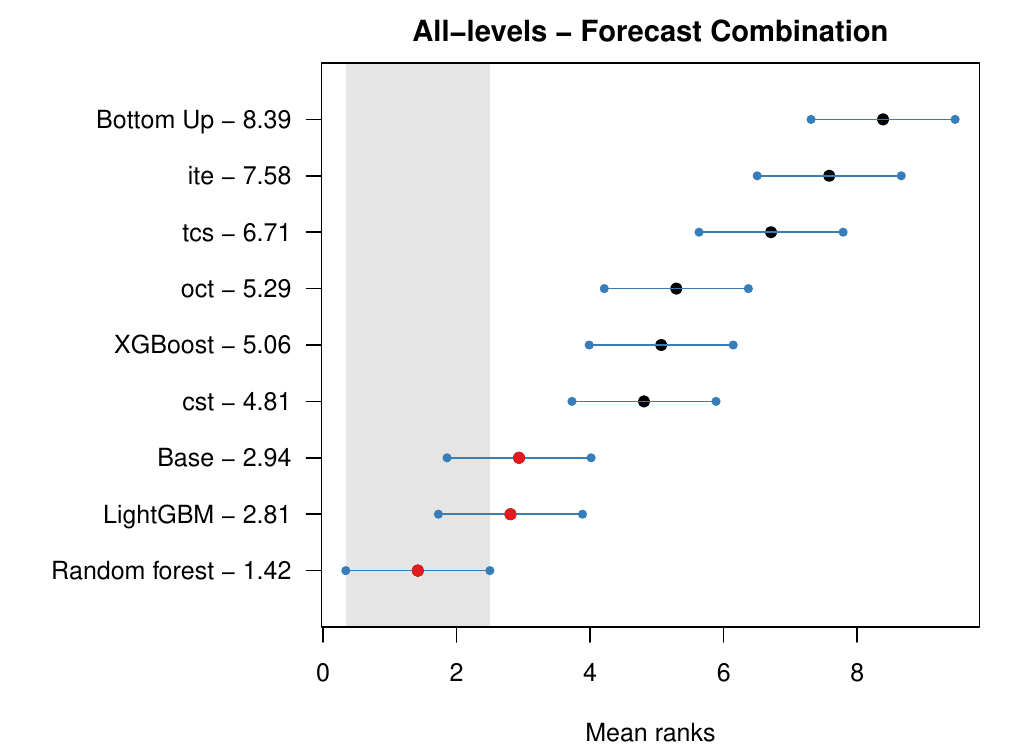}
	\end{subfigure} 
	\caption{
		London: MCB test with confidence level 0.95, using all levels in the hierarchy.   Methods that do not overlap with the gray zone indicating the confidence interval of the best method are significantly worse than the best.}
	\label{fig:mcb-all-levels}
\end{figure}

To assess the significance of our findings, we report in Figure \ref{fig:mcb-all-levels} the results of the accuracy ranking based multiple comparisons with the best (MCB) test, a methodology popularized in  the forecasting literature since \cite{KONING2005397}. We compute the MCB test across all levels in the cross-temporal hierarchy and report results for each of the four base forecast methods. 
To this end, we use the \texttt{R}-package \texttt{tsutils} \citep{tsutils-Rpackage} with function \texttt{nemenyi()} to compute the MCB test. The test ranks the performance of the examined methods across the series being forecast comparing their average ranks by considering a critical difference, determined through a confidence interval.
For each base forecast panel  in Figure \ref{fig:mcb-all-levels}, the methods that do not overlap with the best ranked procedure (as highlighted by the gray zone)  are significantly worse than the best. Visual inspection of the results confirms our overall finding that random forest reconciliation dominates the other procedures.

The important business question is whether ML-based forecast reconciliation helps improving the base forecast methods for high-frequency (i.e.\ 30-minute and hourly)  delivery area levels, because these forecasts are direct inputs for automated decision making. The answer is yes  and the forecast gains are between 10 and 15\%. As an illustration  for random forest reconciliation with ETS base forecasts, the WAPE is 0.2809 compared to 0.3180 for non-reconciled base forecasts.
It also turns out that these high-frequency delivery area level gains become even stronger at the zone and market levels. In fact, the forecast gains mount to even 40\% at the market level. For example, the London 30-minute WAPEs decrease from 0.1621 to 0.0959 with ETS base forecasts and random forest reconciliation. Here, the linear reconciliation methods only deliver marginal accuracy gains, with the best (cst) WAPE being 0.1349. However, we also note that at the London level, the overall smallest WAPEs are obtained using Naive base forecasts  for 30-minute and hourly frequencies. 

Finally, in the bottom panel of Table \ref{tab:London_FR_results}, we provide forecast combination results, that is by combining with equal weights the base forecasts. We find this strategy to slightly pay off compared to the best performing base methods. 
For example,  at the 30-min delivery area level, 
the random forest ML method for the base forecasts are respectively 0.2762, 0.2809 and 0.2801 while the forecast combination yields 0.2758. 
The gains are overall stronger for the linear reconciliation methods compared to the non-linear ones, yet the former are, overall, outperformed by the ML-based reconciliation methods.
Still in practice, we generally do not know a priori which base forecasts method will perform best out-of-sample, and moreover, this  likely varies across the different levels of the cross-temporal hierarchy. Then, a forecast combination base forecast approach forms an effective practice to obtain accurate and robust base forecasts, which can still be further improved upon via ML-based reconciliation to obtain coherent forecasts.

\textit{Manchester.}
Table \ref{tab:Manchester_reconiliation_WAPE_results} summarizes the forecast accuracy for the cross-temporal hierarchy for SARIMA  base forecasts and all reconciliation methods; the other base forecast results are similar and available on request. We report results for the period before the data shift (i.e.\ prior to June 14), the period after the data shift (i.e.\ after July 14), and the one month period in between characterizing the data shift. 
There are several interesting findings for the delivery area level of the hierarchy. 
First, for  ``Before shift" period  we see that random forest reconciliation for delivery areas outperforms other methods with the same relative magnitudes as found in the London market where no data shifts occur. 
For example, random forest and cst WAPEs are 0.3226 and 0.3435, respectively. 
Second, we see that the volatile ``During shift" period WAPEs at the delivery area level all more than double compared to the stable period before. The LightGBM ML reconciliation method performs best with a WAPE of 0.7516. 
Third, the  ``After shift" period is generally characterized by better accuracies than during the shift, but the size of the gains are much stronger for the ML-based reconciliation methods. To illustrate the magnitude of this, consider cst with WAPEs of 0.7986 and 0.7023 and random forest WAPEs of 0.8004 and 0.4206, respectively for the ``During" and ``After shift" periods.
Fourth, the un-reconciled ``After shift" SARIMA base forecasts are of low quality with WAPEs still double the size compared to the ``Before shift" period. This can somehow be expected given that parameter estimates are based on historical data subject to the shift. However, our ML-based reconciliation method is able to transform these low quality projections into forecasts that yield WAPEs much closer to the period prior to the shift. 
At the market level part of the hierarchy, we observe only small differences in terms of forecasting accuracy over the three periods. This is expected since the aggregation of delivery areas not all being subject to a data shift, results in a smoother market level time series, see Figure \ref{fig:map_Manchester_shift}, bottom panel. The LightGBM method has consistently the smallest WAPEs at the 30-minute and hourly frequency levels.

\begin{table}[htbp]
	\caption{
		Manchester forecast reconciliation results.}
	\centering
	\footnotesize
	%\resizebox{\textwidth}{!}{ 
		\begin{tabular}{ll c c c |ccc}
			\hline
			\textbf{Period} & \textbf{Forecast} & \multicolumn{3}{c}{\textbf{Delivery Area}} &  \multicolumn{3}{c}{\textbf{Market}} \\
			\textbf{} & \textbf{Method} \\
			\hline
			&& \textbf{30min} & \textbf{hour} & \textbf{day} &\textbf{30min} & \textbf{hour} & \textbf{day} \\
			\\
			
			Before shift & Base & 0.3549 & 0.2922 & 0.1100 & 0.1884 & 0.1832 & 0.0815 \\ 
			& Bottom Up & 0.3549 & 0.2945 & 0.1879 & 0.1949 & 0.1905 & 0.1630 \\ 
			& tcs & 0.3459 & 0.2837 & 0.1734 & 0.1847 & 0.1794 & 0.1533 \\ 
			& cst & 0.3435 & 0.2807 & 0.1680 & 0.1794 & 0.1735 & 0.1460 \\ 
			&  ite &  0.3464 &  0.2840 &  0.1743 &  0.1855 &  0.1800 &  0.1549 \\ 
			& oct & 0.3470 & 0.2846 & 0.1731 & 0.1848 & 0.1797 & 0.1520 \\ 
			& Random forest & \cellcolor{lightgray}0.3226 & \cellcolor{lightgray}0.2523 & \cellcolor{lightgray}0.1146 & 0.1216 & 0.1141 & 0.0764 \\ 
			& XGBoost & 0.3650 & 0.2885 & 0.1315 & 0.1205 & 0.1121 & \cellcolor{lightgray}0.0743 \\ 
			& LightGBM & 0.3418 & 0.2706 & 0.1300 & \cellcolor{lightgray}0.1196 & \cellcolor{lightgray}0.1118 & 0.0759 \\ 
			\\
			\\
			During shift & Base & 0.8124 & 0.7687 & 0.4634 & 0.2091 & 0.2034 & 0.0615 \\ 
			& Bottom Up & 0.8124 & 0.7587 & 0.5923 & 0.2092 & 0.2059 & 0.1805 \\ 
			& tcs & 0.7991 & 0.7683 & 0.5926 & 0.1965 & 0.1926 & 0.1642 \\ 
			& cst & 0.7986 & 0.7651 & 0.5747 & 0.1885 & 0.1841 & 0.1520 \\ 
			&  ite &  0.8107 &  0.7710 &  0.5971 &  0.1997 &  0.1956 &  0.1678 \\ 
			& oct & 0.7885 & 0.7606 & 0.5789 & 0.1910 & 0.1875 & 0.1581 \\ 
			& Random forest & 0.8004 & 0.7346 & 0.5885 & 0.1110 & 0.1022 & 0.0646 \\ 
			& XGBoost & 0.8330 & 0.7534 & 0.5683 & 0.1128 & 0.1017 & \cellcolor{lightgray}0.0633 \\ 
			& LightGBM & \cellcolor{lightgray}0.7516 & \cellcolor{lightgray}0.6818 & \cellcolor{lightgray}0.5230 & \cellcolor{lightgray}0.1101 & \cellcolor{lightgray}0.1002 & 0.0658 \\ 
			\\
			\\
			After shift & Base & 0.7166 & 0.6402 & 0.1610 & 0.1776 & 0.1740 & 0.0524 \\ 
			& Bottom Up & 0.7166 & 0.6464 & 0.4506 & 0.1825 & 0.1786 & 0.1526 \\ 
			& tcs & 0.6969 & 0.6446 & 0.3694 & 0.1682 & 0.1631 & 0.1364 \\ 
			& cst & 0.7023 & 0.6484 & 0.3381 & 0.1615 & 0.1558 & 0.1256 \\ 
			&  ite &  0.7049 & 0.6521 &  0.3553 &  0.1699 &  0.1646 &  0.1381 \\ 
			& oct & 0.6925 & 0.6402 & 0.3755 & 0.1650 & 0.1601 & 0.1322 \\ 
			& Random forest & \cellcolor{lightgray}0.4206 & \cellcolor{lightgray}0.3300 & \cellcolor{lightgray}0.1457 & 0.0992 & 0.0904 & \cellcolor{lightgray}0.0465 \\ 
			& XGBoost & 0.4863 & 0.3874 & 0.1464 & 0.1007 & 0.0896 & 0.0485 \\ 
			& LightGBM & 0.4386 & 0.3488 & 0.1551 & \cellcolor{lightgray}0.0980 & \cellcolor{lightgray}0.0880 & 0.0474 \\

			\midrule
		\end{tabular}%}
	\label{tab:Manchester_reconiliation_WAPE_results}
	% \footnotesize
	$\;$\\[0.2cm]
	\raggedright
	Notes: This table shows forecast accuracy measured in weighted absolute percentage error (WAPE) for SARIMA base forecasts.  
	The best forecast reconciliation results for the periods Before shift (prior to June 14, 2023), After shift (after July 14, 2023), and During shift (period in between) are highlighted in gray. 
	The linear benchmark methods are defined in Table \ref{tabel:linearbenchmarks}.  \\[0.1cm]
\end{table}

To sum up, random forest forecast reconciliation performs particularly well but not all ML methods are doing better than linear methods. The strongest gains are obtained for the 30-minute and hourly frequencies. Forecast combination, i.e. averaging the base forecasts, is (marginally) beneficial.
These main findings are, overall, qualitatively the same when we measure forecast accuracy in terms of the MASE, as can be seen from Tables \ref{tab:London_FR_results_mase} and \ref{tab:Manchester_reconiliation_MASE_results} 
in Appendix \ref{app:platformapp}. 
Random forest  remains the top performing reconciliation method for the bottom levels of the cross-temporal hierarchy across the London and Manchester datasets.
LightGBM becomes even more favorable  to use-- compared to random forest --for the upper levels of the cross-temporal hierarchy when using the MASE accuracy index instead of WAPE.
For Manchester, in the ``After shift" period the margin by which the ML-based reconciliation methods outperform the linear ones is less outspoken in terms of MASE compared to WAPE.

\subsubsection{ML-Based Reconciliation Insights} \label{subsec:result:MLinsights}
From the previous results, we find that random forest based forecast reconciliation outperforms the considered benchmarks, especially at the more granular levels in the hierarchy. To gain more insights,  we investigate the link between forecast accuracy and size of the delivery areas on the London dataset.
In addition, we investigate what drives forecast performance by computing SHAP values which are considered to be the current state-of-the-art method for interpreting ML models, see \cite{molnar2022interpretable} for an introduction. 

\begin{figure}[t]
	\includegraphics[width=\textwidth]{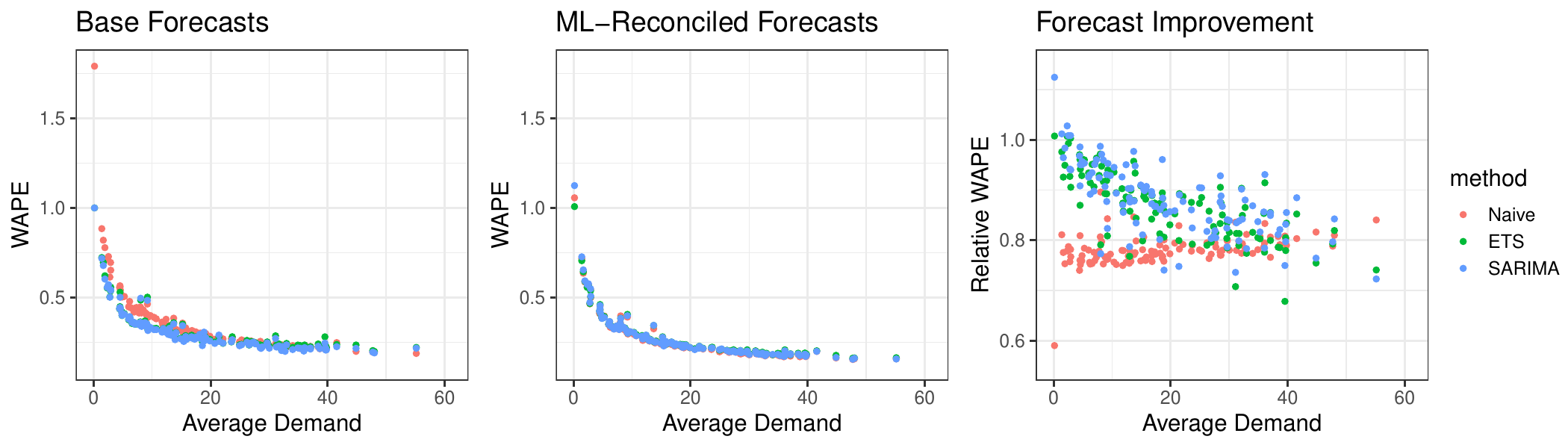}
	\caption{ Forecast performance measured in WAPE for base forecasts (left), for ML-reconciled forecasts via random forest (middle) and relative  improvement in WAPE of the latter over the former versus average 30-min demand for the different delivery areas (dots) in London. }
	\label{fig:banana_insights_plots}
\end{figure}

Figure \ref{fig:banana_insights_plots} plots forecast performance (WAPE, vertical axis) of each delivery area (dot) based on its average demand size (horizontal axis). The left panel shows this relationship using un-reconciled base forecasts, and we find a smirk shape pattern with many WAPEs above 0.5. The Naive base forecasts are, overall, worse than ETS and SARIMA when average demand is below 20. The middle figure shows the random forest forecast reconciliation results and we see that the smirk flattens out, with hardly any noticeable difference between the base forecast methods, and few areas having WAPEs above 0.5. The right figure highlights that random forest based forecast reconciliation has a capacity to ``correct" bad quality base forecasts as the relative WAPEs of the ML-reconciled forecasts over the base forecasts are mostly below 1.

\begin{figure}[t]
	\centering
	\includegraphics[width=\textwidth]{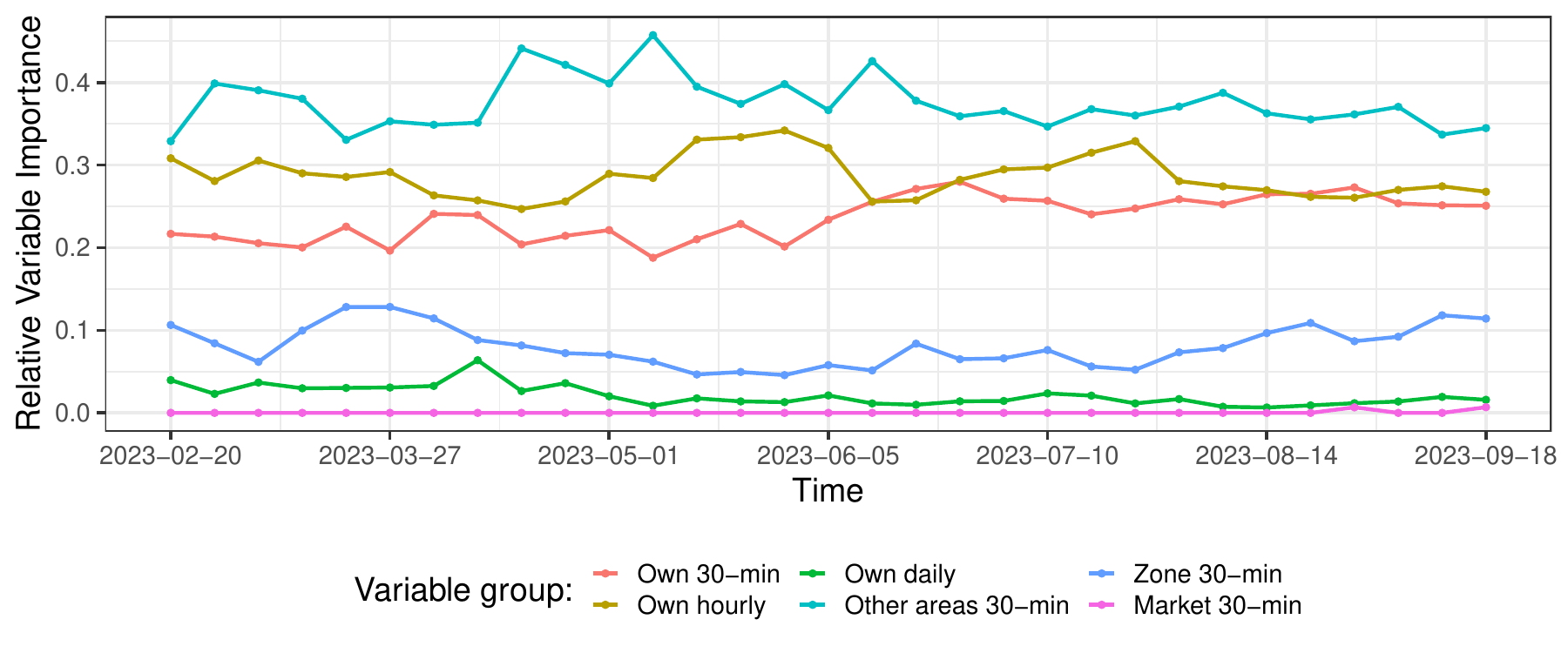}
	\caption{
		Variable importance in random forest ML model with Naive base forecasts for the delivery area City of London.}
	\label{fig:shapley_CentralLondon_best36}
\end{figure}

Next, for each random forest model with Naive base forecasts in the (outer) rolling-window setup, we compute SHAP values \citep{shapley1953value, lundberg2017unified} for all observations in the test set. To this end, we use  the package \texttt{treeshap} \citep{treeshapR} in \texttt{R}, a fast implementation for tree ensemble models \citep{lundberg2018consistent}. Figure \ref{fig:shapley_CentralLondon_best36} shows the resulting relative variable importance  plots by variable group for City of London as an illustration.
Given the large number of features used in the ML-model, we focus on the 36 highest ranked features in terms of variable importance (which together account for more than 80\% of the overall variable importance).
Overall, the proportions are stable over the test sample. While City of London's  own 30-min and hourly streams count for a sizable portion in the 20-30\% range, the other areas 30-minute streams count together for almost 40\%. Zone and market level, or own daily information contributes only marginally.

\subsubsection{Computational Aspects and Hyperparameter Tuning}  \label{subsec:result:tuning}
Our main results above are obtained by training the ML algorithms with their default hyperparameters as provided by the software packages. By doing so, we save a substantial amount of computing time which is an important consideration when deploying forecast reconciliation in a production environment. However, 
in this section, we  investigate whether the performance of the ML-based forecast reconciliation methods can be further improved by hyperparameter tuning. 
To this end, we  use the common practice of grid search to tune the hyperparameters of the random forest algorithm, since there are only few hyperparameters, whereas we use Bayesian optimization to tune the hyperparameters of the XGBoost and LightGBM algorithms as the latter two have a large amount of hyperparameters.
Details on the tuning process are provided in Appendix \ref{app:ML-methods}.

Table \ref{tab:London_FR_results_tuned} reports results for the ML-based reconciliation methods-- default and tuned --with SARIMA base forecasts on the London dataset. The random forest algorithm with default tuning parameters, these were best performing in most of our previous results, performs equally well as its tuned version. This is a great advantage when implementing the forecasting procedure in a high-dimensional streaming data environment with frequent updates. In contrast to random forest, the XGBoost algorithm greatly benefits from tuning  at the high-frequency delivery area level. Here, the loss decreases from 0.3196 to 0.2790, a reduction of about 13\% and XGBoost is now at the same accuracy level as random forest. For the zone and market hierarchies, the differences between default and tuning are marginal. Finally, the LightGBM algorithm is only mildly impacted, with small tuning improvements at the delivery area level.

\begin{table}[t]
	\centering
	\caption{  
		London forecast reconciliation results with hyperparameter tuning.}
	\resizebox{0.995\textwidth}{!}{ \begin{tabular}{lll c c c|ccc|ccc}
			\hline
			\multicolumn{2}{l}{\textbf{Forecast Method}}  &  \textbf{Time}& \multicolumn{3}{c}{\textbf{Delivery Area}} & \multicolumn{3}{c}{\textbf{Zone}} & \multicolumn{3}{c}{\textbf{Market}} \\
			&  & \textbf{(seconds)}  \\
			\hline
			& && \textbf{30min} & \textbf{hour} & \textbf{day} & \textbf{30min} & \textbf{hour} & \textbf{day} &\textbf{30min} & \textbf{hour} & \textbf{day}  \\
			\\
			Random forest & default &  7&  0.2801 & 0.2206 & 0.1002 & 0.1305 & 0.1132 &  0.0657 & 0.0883 & 0.0861 & 0.0586 \\ 
			Random forest & tuned &  400 &  0.2784 & 0.2198 & 0.1033 & 0.1324 & 0.1155 & 0.0692 & 0.0911 & 0.0889 & 0.0621 \\ 
			XGBoost & default &  1 & 0.3195 & 0.2528 & 0.1109 & 0.1385 & 0.1182 & 0.0659 & 0.0845 & 0.0815 & 0.0551 \\ 
			XGBoost & tuned &  600 & 0.2790 & 0.2209 & 0.1032 & 0.1314 & 0.1144 & 0.0652 & 0.0886 & 0.0864 & 0.0568 \\ 
			LightGBM &default &  1 & 0.2933 & 0.2327 & 0.1067 & 0.1320 & 0.1134 & 0.0641 & 0.0832 & 0.0806 & 0.0538 \\ 
			LightGBM &tuned &  900 & 0.2801 & 0.2221 & 0.1034 & 0.1334 & 0.1165 & 0.0668 & 0.0917 & 0.0895 & 0.0589 \\

			\midrule
	\end{tabular}}
	\label{tab:London_FR_results_tuned}
	\raggedright
	\footnotesize
	$\;$\\[-0.1cm]
	Notes: This table shows forecast accuracy measured in weighted absolute percentage error (WAPE) for SARIMA base forecasts. 
	Computing times are approximate, measured in seconds and recorded on a Windows 10 Enterprise LTSC machine with Intel Core i5 2.4GHz processor.
	\\[0.1cm]
\end{table}

While hyperparameter tuning does result, overall, in a marginal gain in forecast accuracy, it is important to investigate the computational burden that comes along with it. 
Table \ref{tab:London_FR_results_tuned} reports in the third column the approximate computing times (in seconds) for one ML fit with fixed tuning parameters as well as one round of hyperparameter tuning, and this for random forest, XGBoost and LightGBM. 
As well known, both the XGBoost and LightGBM algorithms are fast with default tuning parameters, requiring only about one second to train. Random forest is in comparison relatively slow with seven seconds to train. However, the prohibitive cost of tuning is appalling. All algorithms need large multiples of computing time to tune the hyperparameters. In particular the XGBoost and LightGBM algorithms with many tuning hyperparameters are long to optimize.\footnote{We use Bayesian optimization to tune the hyperparameters of XGBoost and LightGBM. Other methods or only subsets of parameters could be tuned. This would make the tuning faster.}  

To conclude, among the considered ML algorithms, the random forest algorithm is preferred because the default tuning parameters yield good accuracy levels close to the tuned ones. This is a great advantage when faced with computational constraints such as given by the platform environments we consider in this paper.

\subsection{ Results for the Bicycle Sharing Platform Application} \label{subsec:citibike-results}
Section \ref{subsubsec:citibike-results} discusses the overall performance of the forecast reconciliation methods on the Citi Bike dataset, Section \ref{subsec:sensitivity-Citibike} details two sensitivity analyses on the choice of temporal aggregation orders in the reconciliation step.

\subsubsection{Overall Forecast Performance} \label{subsubsec:citibike-results}
Table \ref{tab:Citibike_results_areas} summarizes the WAPE forecast accuracy for the H3 cells on the complete temporal hierarchy for the different combinations of base forecasts and reconciliation methods, the results for the New York City market are displayed in Table \ref{tab:Citibike_results_market}.\footnote{The ML-based forecast reconciliation results are reported for fixed choices of tuning parameters. For random forest and LightGBM, tuning does, overall, not result in improved forecast performance, for XGBoost it leads to a minor improvement of  1.86\% on average.}

\begin{table}[htbp]
	\centering
	\caption{
		Citi Bike forecast reconciliation results for the H3 cells.}
	\resizebox{\textwidth}{!}{ \begin{tabular}{ll cccccccccc}
			\midrule
			\textbf{Base} & \textbf{Forecast} & \multicolumn{10}{c}{\textbf{Temporal Frequency}} \\
			\textbf{Forecasts} & \textbf{Method} \\
			\midrule
			&& \textbf{30min} & \textbf{1h} & \textbf{1.5h} & \textbf{2h} & \textbf{3h} & \textbf{4h} &\textbf{6h} & \textbf{8h} & \textbf{12h} & \textbf{24h} \\
			\\
			Naive & Base & 0.2883 & 0.2675 & 0.2583 & 0.2525 & 0.2434 & 0.2365 & 0.2286 & 0.2199 & 0.2132 & 0.1981 \\ 
			& Random forest & \cellcolor{lightgray}0.2682 & \cellcolor{lightgray}0.2530 & \cellcolor{lightgray}0.2456 & \cellcolor{lightgray}0.2400 & \cellcolor{lightgray}0.2317 & \cellcolor{lightgray}0.2241 & \cellcolor{lightgray}0.2136 & \cellcolor{lightgray}0.2014 & \cellcolor{lightgray}0.1962 & \cellcolor{lightgray}0.1775 \\ 
			& XGBoost & 0.2951 & 0.2775 & 0.2682 & 0.2620 & 0.2513 & 0.2426 & 0.2305 & 0.2157 & 0.2099 & 0.1879 \\ 
			& LightGBM & 0.2882 & 0.2723 & 0.2639 & 0.2583 & 0.2484 & 0.2408 & 0.2278 & 0.2154 & 0.2089 & 0.1869 \\ 
			\\
			\\
			ETS & Base & 0.3699 & 0.3382 & 0.3314 & 0.3246 & 0.3089 & 0.2677 & 0.2482 & 0.2383 & 0.2237 & 0.1552 \\ 
			& Bottom Up & 0.3699 & 0.3593 & 0.3542 & 0.3498 & 0.3408 & 0.3305 & 0.3183 & 0.3052 & 0.2958 & 0.2555 \\ 
			& tcs & 0.3453 & 0.3338 & 0.3283 & 0.3237 & 0.3136 & 0.3016 & 0.2902 & 0.2734 & 0.2620 & 0.2180 \\ 
			& cst & 0.3549 & 0.3445 & 0.3391 & 0.3348 & 0.3262 & 0.3149 & 0.3042 & 0.2894 & 0.2795 & 0.2375 \\ 
			& ite & 0.3552 & 0.3445 & 0.3391 & 0.3347 & 0.3258 & 0.3146 & 0.3037 & 0.2892 & 0.2795 & 0.2367 \\ 
			& oct & 0.3481 & 0.3366 & 0.3311 & 0.3264 & 0.3162 & 0.3042 & 0.2926 & 0.2765 & 0.2653 & 0.2210 \\ 
			& Random forest & \cellcolor{lightgray}0.2606 & \cellcolor{lightgray}0.2446 & \cellcolor{lightgray}0.2367 & \cellcolor{lightgray}0.2306 & \cellcolor{lightgray}0.2213 & \cellcolor{lightgray}0.2115 & \cellcolor{lightgray}0.1989 & \cellcolor{lightgray}0.1908 & \cellcolor{lightgray}0.1827 & \cellcolor{lightgray}0.1654 \\ 
			& XGBoost & 0.2822 & 0.2642 & 0.2541 & 0.2475 & 0.2369 & 0.2271 & 0.2125 & 0.2034 & 0.1941 & 0.1738 \\ 
			& LightGBM & 0.2742 & 0.2580 & 0.2489 & 0.2430 & 0.2328 & 0.2240 & 0.2096 & 0.1994 & 0.1924 & 0.1718 \\ 
			\\
			\\
			SARIMA & Base & 0.3467 & 0.3110 & 0.3022 & 0.2982 & 0.2706 & 0.2321 & 0.2110 & 0.2004 & 0.1700 & 0.1546 \\ 
			& Bottom Up & 0.3467 & 0.3353 & 0.3295 & 0.3248 & 0.3137 & 0.3019 & 0.2868 & 0.2728 & 0.2583 & 0.2169 \\ 
			& tcs & 0.3176 & 0.3045 & 0.2977 & 0.2920 & 0.2784 & 0.2636 & 0.2487 & 0.2334 & 0.2171 & 0.1756 \\ 
			& cst & 0.3197 & 0.3068 & 0.3000 & 0.2939 & 0.2804 & 0.2652 & 0.2496 & 0.2353 & 0.2185 & 0.1766 \\  
			& ite & 0.3182 & 0.3051 & 0.2981 & 0.2920 & 0.2782 & 0.2628 & 0.2470 & 0.2323 & 0.2158 & 0.1738 \\ 
			& oct & 0.3174 & 0.3043 & 0.2976 & 0.2920 & 0.2785 & 0.2637 & 0.2488 & 0.2335 & 0.2173 & 0.1760 \\ 
			& Random forest & \cellcolor{lightgray}0.2605 & \cellcolor{lightgray}0.2445 & \cellcolor{lightgray}0.2361 & \cellcolor{lightgray}0.2303 & \cellcolor{lightgray}0.2201 &\cellcolor{lightgray}0.2086 & \cellcolor{lightgray}0.1947 & \cellcolor{lightgray}0.1866 & \cellcolor{lightgray}0.1773 & \cellcolor{lightgray}0.1589 \\ 
			& XGBoost & 0.2801 & 0.2624 & 0.2518 & 0.2453 & 0.2332 & 0.2232 & 0.2077 & 0.1984 & 0.1903 & 0.1680 \\ 
			& LightGBM & 0.2715 & 0.2550 & 0.2459 & 0.2398 & 0.2287 & 0.2188 & 0.2037 & 0.1948 & 0.1861 & 0.1648 \\ 
			\\
			\\
			Forecast  &Base & 0.2942 & 0.2660 & 0.2578 & 0.2532 & 0.2364 & 0.2109 & 0.2017 & 0.1911 & 0.1758 & 0.1527 \\ 
			Combination & Bottom Up & 0.2942 & 0.2820 & 0.2759 & 0.2719 & 0.2641 & 0.2544 & 0.2443 & 0.2336 & 0.2253 & 0.1951 \\ 
			& tcs & 0.2799 & 0.2667 & 0.2602 & 0.2557 & 0.2470 & 0.2363 & 0.2264 & 0.2150 & 0.2051 & 0.1743 \\ 
			& cst & 0.2853 & 0.2728 & 0.2665 & 0.2624 & 0.2540 & 0.2437 & 0.2338 & 0.2230 & 0.2132 & 0.1830 \\ 
			& ite & 0.2869 & 0.2743 & 0.2680 & 0.2638 & 0.2554 & 0.2453 & 0.2352 & 0.2245 & 0.2150 & 0.1848 \\ 
			& oct & 0.2793 & 0.2661 & 0.2596 & 0.2549 & 0.2461 & 0.2352 & 0.2253 & 0.2138 & 0.2037 & 0.1726 \\ 
			& Random forest & \cellcolor{lightgray}0.2521 & \cellcolor{lightgray}0.2366 & \cellcolor{lightgray}0.2288 & \cellcolor{lightgray}0.2237 & \cellcolor{lightgray}0.2156 & \cellcolor{lightgray}0.2076 & \cellcolor{lightgray}0.1964 & \cellcolor{lightgray}0.1894 & \cellcolor{lightgray}0.1817 & \cellcolor{lightgray}0.1647 \\ 
			& XGBoost & 0.2756 & 0.2580 & 0.2482 & 0.2419 & 0.2316 & 0.2240 & 0.2113 & 0.2033 & 0.1953 & 0.1763 \\ 
			& LightGBM & 0.2692 & 0.2535 & 0.2443 & 0.2389 & 0.2295 & 0.2222 & 0.2094 & 0.2023 & 0.1948 & 0.1752 \\ 
			
			\midrule
	\end{tabular}}
	\label{tab:Citibike_results_areas}
	\raggedright
	\footnotesize
	$\;$\\[-0.1cm]
	Notes: This table shows forecast accuracy measured in weighted absolute percentage error (WAPE). 
	The best forecast reconciliation results for each base forecast (Naive, ETS, SARIMA, Forecast Combination) are highlighted in gray.
	The linear benchmark methods are defined in Table \ref{tabel:linearbenchmarks}.  
	\\[0.1cm]
\end{table}

\begin{table}[htbp]
	\centering
	\caption{
		Citi Bike forecast reconciliation results for the market.}
	\resizebox{\textwidth}{!}{ \begin{tabular}{ll cccccccccc}
			\midrule
			\textbf{Base} & \textbf{Forecast} & \multicolumn{10}{c}{\textbf{Temporal Frequency}} \\
			\textbf{Forecasts} & \textbf{Method} \\
			\midrule
			&& \textbf{30min} & \textbf{1h} & \textbf{1.5h} & \textbf{2h} & \textbf{3h} & \textbf{4h} &\textbf{6h} & \textbf{8h} & \textbf{12h} & \textbf{24h} \\
			\\
			Naive & Base & 0.2270 & 0.2243 & 0.2226 & 0.2213 & 0.2179 & 0.2150 & 0.2097 & 0.2043 & 0.1999 & 0.1891 \\ 
			& Randomforest & \cellcolor{lightgray}0.2225 & \cellcolor{lightgray}0.2190 & \cellcolor{lightgray}0.2166 & \cellcolor{lightgray}0.2143 & \cellcolor{lightgray}0.2102 & \cellcolor{lightgray}0.2057 & \cellcolor{lightgray}0.1976 & \cellcolor{lightgray}0.1875 & \cellcolor{lightgray}0.1832 & \cellcolor{lightgray}0.1695 \\ 
			& XGBoost & 0.2338 & 0.2295 & 0.2263 & 0.2237 & 0.2183 & 0.2133 & 0.2052 & 0.1943 & 0.1885 & 0.1732 \\ 
			& LightGBM & 0.2356 & 0.2315 & 0.2287 & 0.2266 & 0.2215 & 0.2176 & 0.2080 & 0.1971 & 0.1914 & 0.1751 \\    
			\\
			\\
			ETS & Base & 0.3839 & 0.3044 & 0.2978 & 0.2928 & 0.2751 & 0.2386 & 0.2166 & 0.2115 & 0.1954 & 0.1469 \\ 
			& Bottom Up & 0.3474 & 0.3446 & 0.3430 & 0.3409 & 0.3328 & 0.3260 & 0.3150 & 0.3031 & 0.2993 & 0.2737 \\ 
			& tcs & 0.3116 & 0.3084 & 0.3063 & 0.3039 & 0.2950 & 0.2862 & 0.2760 & 0.2578 & 0.2507 & 0.2241 \\ 
			& cst & 0.3144 & 0.3114 & 0.3091 & 0.3066 & 0.2980 & 0.2890 & 0.2791 & 0.2611 & 0.2553 & 0.2265 \\  
			& ite & 0.3218 & 0.3188 & 0.3166 & 0.3142 & 0.3057 & 0.2972 & 0.2869 & 0.2708 & 0.2659 & 0.2373 \\ 
			& oct & 0.3176 & 0.3145 & 0.3125 & 0.3100 & 0.3013 & 0.2929 & 0.2826 & 0.2654 & 0.2600 & 0.2319 \\ 
			& Random forest & \cellcolor{lightgray}0.2154 & \cellcolor{lightgray}0.2104 & \cellcolor{lightgray}0.2075 & \cellcolor{lightgray}0.2044 & \cellcolor{lightgray}0.1979 & \cellcolor{lightgray}0.1912 & \cellcolor{lightgray}0.1804 & \cellcolor{lightgray}0.1724 & \cellcolor{lightgray}0.1689 & \cellcolor{lightgray}0.1562 \\ 
			& XGBoost & 0.2231 & 0.2180 & 0.2146 & 0.2115 & 0.2048 & 0.1989 & 0.1876 & 0.1790 & 0.1759 & 0.1609 \\ 
			& LightGBM & 0.2201 & 0.2153 & 0.2120 & 0.2097 & 0.2026 & 0.1975 & 0.1856 & 0.1779 & 0.1750 & 0.1622 \\ 
			\\
			\\
			SARIMA & Base & 0.2915 & 0.2796 & 0.2775 & 0.2779 & 0.2451 & 0.2100 & 0.1988 & 0.1909 & 0.1596 & 0.1573 \\ 
			& Bottom Up & 0.2870 & 0.2830 & 0.2802 & 0.2774 & 0.2661 & 0.2541 & 0.2423 & 0.2224 & 0.2097 & 0.1817 \\ 
			& tcs & 0.2747 & 0.2705 & 0.2673 & 0.2640 & 0.2520 & 0.2392 & 0.2262 & 0.2090 & 0.1939 & 0.1648 \\ 
			& cst & 0.2716 & 0.2672 & 0.2639 & 0.2601 & 0.2475 & 0.2342 & 0.2206 & 0.2058 & 0.1889 & 0.1591 \\ 
			& ite& 0.2748 & 0.2704 & 0.2670 & 0.2634 & 0.2509 & 0.2378 & 0.2239 & 0.2086 & 0.1920 & 0.1627 \\ 
			& oct & 0.2749 & 0.2707 & 0.2675 & 0.2642 & 0.2523 & 0.2395 & 0.2266 & 0.2093 & 0.1944 & 0.1651 \\ 
			& Random forest & \cellcolor{lightgray}0.2143 & \cellcolor{lightgray}0.2087 & \cellcolor{lightgray}0.2054 & \cellcolor{lightgray}0.2023 & \cellcolor{lightgray}0.1947 & \cellcolor{lightgray}0.1870 & \cellcolor{lightgray}0.1751 & \cellcolor{lightgray}0.1683 & \cellcolor{lightgray}0.1619 & \cellcolor{lightgray}0.1490 \\ 
			& XGBoost & 0.2155 & 0.2099 & 0.2058 & 0.2027 & 0.1959 & 0.1901 & 0.1782 & 0.1706 & 0.1664 & 0.1510 \\ 
			& LightGBM & 0.2133 & 0.2080 & 0.2041 & 0.2007 & 0.1938 & 0.1876 & 0.1764 & 0.1693 & 0.1643 & 0.1499 \\ 
			\\
			\\
			Forecast & Base & 0.2604 & 0.2339 & 0.2304 & 0.2291 & 0.2113 & 0.1912 & 0.1844 & 0.1768 & 0.1613 & 0.1498 \\ 
			Combination & Bottom Up & 0.2522 & 0.2497 & 0.2474 & 0.2460 & 0.2400 & 0.2333 & 0.2239 & 0.2128 & 0.2076 & 0.1889 \\ 
			& tcs & 0.2402 & 0.2374 & 0.2351 & 0.2335 & 0.2272 & 0.2190 & 0.2107 & 0.1985 & 0.1911 & 0.1712 \\ 
			& cst & 0.2434 & 0.2407 & 0.2384 & 0.2368 & 0.2306 & 0.2225 & 0.2140 & 0.2024 & 0.1950 & 0.1742 \\ 
			& ite & 0.2489 & 0.2464 & 0.2442 & 0.2427 & 0.2366 & 0.2287 & 0.2198 & 0.2089 & 0.2021 & 0.1814 \\ 
			& oct & 0.2393 & 0.2364 & 0.2342 & 0.2324 & 0.2261 & 0.2178 & 0.2098 & 0.1972 & 0.1895 & 0.1695 \\ 
			& Randomforest & \cellcolor{lightgray}0.2065 & \cellcolor{lightgray}0.2023 & \cellcolor{lightgray}\cellcolor{lightgray}0.1998 & \cellcolor{lightgray}0.1979 & \cellcolor{lightgray}0.1931 & \cellcolor{lightgray}0.1874 & \cellcolor{lightgray}0.1785 & \cellcolor{lightgray}0.1742 & \cellcolor{lightgray}0.1677 & \cellcolor{lightgray}0.1580 \\ 
			& XGBoost & 0.2135 & 0.2090 & 0.2063 & 0.2040 & 0.1991 & 0.1930 & 0.1842 & 0.1791 & 0.1729 & 0.1622 \\ 
			& LightGBM & 0.2124 & 0.2082 & 0.2051 & 0.2039 & 0.1983 & 0.1935 & 0.1835 & 0.1790 & 0.1740 & 0.1620 \\ 
			
			\midrule
	\end{tabular}}
	\label{tab:Citibike_results_market}
	\raggedright
	\footnotesize
	$\;$\\[-0.1cm]
	Notes: This table shows forecast accuracy measured in weighted absolute percentage error (WAPE). 
	The best forecast reconciliation results for each base forecast (Naive, ETS, SARIMA, Forecast Combination) are highlighted in gray.
	The linear benchmark methods are defined in Table \ref{tabel:linearbenchmarks}.  
	\\[0.1cm]
\end{table}

The ML-based forecast reconciliation methods generally result in substantial forecast accuracy improvements compared to the linear reconciliation methods with the largest improvement for ETS base forecasts. 
Furthermore, unlike the linear reconciliation methods, the ML-based forecast reconciliation methods are considerably more stable for different base forecasts.
Similar to the other platform applications, ML-based reconciliation with random forest performs  best overall, and is in fact the only reconciliation method that offers a consistent improvement over the (already coherent) naive base forecasts.  For example, the H3 cells random forest WAPEs with SARIMA base forecasts for the highest and lowest temporal frequencies are respectively 0.2605 and 0.1589, compared to 0.3174 and 0.1760 respectively for oct linear reconciliation. These results carry over for the market level results with random forest 0.2143 and 0.1490, and oct 0.2749 and 0.1651 respectively. 
Combining the base forecasts yields sizable gains only for the linear reconciliation methods.
Changing the metric to the MASE forecast accuracy index instead of the WAPE, similar conclusions can be drawn, see Tables \ref{tab:Citibike_results_areas_MASE} and  \ref{tab:Citibike_results_market_MASE} in Appendix \ref{app:platformapp} for details. 

\subsubsection{Sensitivity Analyses on Temporal Aggregation Orders} \label{subsec:sensitivity-Citibike}
The main results for Citi Bike are obtained when using the compact features matrix $\widehat{\boldsymbol{X}}_{b,\mathcal{V}}$  across all ML models, thereby using all  $p=10$ temporal frequencies but only for the focal bottom-level series for which the ML-model is trained.
This entails two direct questions:
(1) Is there a gain in forecast accuracy when using the complete features matrix $\widehat{\boldsymbol{X}}_{\mathcal{V}}$ instead of the compact one $\widehat{\boldsymbol{X}}_{b,\mathcal{V}}$?
(2) Is there a gain in forecast accuracy for  certain temporal frequencies of interest, for instance 30-min, 1-hour and 24-hour, when using all ten temporal frequencies for reconciliation instead of only those of three of interest?
We next address these questions  through sensitivity analyses.

Table \ref{tab:sensitivity_bigX_citibike_results_areas} in Appendix \ref{app:citibike} addresses the first question and presents relative WAPEs when using the complete features matrix instead of the compact one.
A value below one indicates better performance when using the complete features matrix.
For the H3 cells, see top panel of Table \ref{tab:sensitivity_bigX_citibike_results_areas}, usage of the complete features matrix improves forecast accuracy for all base forecasts and ML forecast methods.  
In contrast, when turning to the results for the whole market, see bottom panel of Table \ref{tab:sensitivity_bigX_citibike_results_areas}, the results remain either virtually unaffected or deteriorate. 
Hence, the relative performance of the ML-based reconciliation method with the complete versus compact features matrix is application-specific.

Table \ref{tab:sensitvity_3freq_citibike_results} provides an answer to the second question and reports relative WAPEs for the temporal frequencies of interest (30-min, 1-hour and 24-hour) when  using all ten temporal frequencies for reconciliation compared to using only those three.
A value below one  indicates better performance when using the full set rather than the reduced set of temporal granularities.
We consistently find forecast performance to improve when using the full set for both the linear and the ML-based reconciliation methods. 
The improvement is typically larger for the ML-based reconciliation methods than for the linear ones, at least for the bottom-level in the cross-temporal hierarchy (30-min H3 cells).
Note that the base forecasts and the bottom-up reconciliation method remain unaffected under this choice but are  reported for completeness.
This finding provides  support for using intermediate temporal frequencies in the forecast reconciliation procedure beyond the frequencies of interest, though it comes with the cost of additional computing time for the ML-based reconciliation methods.

\section{Conclusion  \label{sec:conclusion}}
Forecast reconciliation based on linear methods has been shown to be very useful and successful in a wide variety of applications. Recently, ML-based forecast reconciliation methods are proposed for cross-sectional time series hierarchies. We extend the latter approach to  cross-temporal hierarchies. We provide a general description of the methodology and applications to unique platform datasets from an on-demand logistics company and a bicycle sharing system. In particular, the datasets consist of demand streams that constitute a rich hierarchy in spatial and time dimensions, with bottom level time series defined 
at 30-minute frequency.

Our key empirical finding is that ML-based forecast reconciliation for our platform datasets can result in substantial forecast accuracy improvements compared to existing linear reconciliation methods. However, unless one can face the huge computational cost of hyperparameter tuning, ML-based reconciliation is not uniformly superior to linear methods. 
Another key finding is that, as platform data streams are potentially impacted by data shifts, our approach is able to react swiftly to such instability when compared with linear approaches.

Several extensions should be considered. 
First, applications to other datasets consisting of different cross-temporal hierarchies. 
Second, our ML algorithms in the platform application consist of tree-based methods only, though our approach can be  used with any ML algorithm. Taking into account computing time, it would be interesting to test alternative neural network based methods using our approach. 
Third, we currently retrain the ML algorithms at every iteration of the rolling window, it would be interesting to investigate if online ML methods that incrementally update the forecast function as new data becomes available can lead to further computational efficiencies.

\bibliographystyle{apalike}
\bibliography{references}

\begin{thebibliography}{}

\bibitem[Abolghasemi et~al., 2024]{abolghasemi2022machine}
Abolghasemi, M., Tarr, G., and Bergmeir, C. (2024).
\newblock Machine learning applications in hierarchical time series
  forecasting: Investigating the impact of promotions.
\newblock {\em International Journal of Forecasting}, 40(2):597--615.

\bibitem[Anderer and Li, 2022]{anderer2022hierarchical}
Anderer, M. and Li, F. (2022).
\newblock Hierarchical forecasting with a top-down alignment of
  independent-level forecasts.
\newblock {\em International Journal of Forecasting}, 38(4):1405--1414.

\bibitem[Athanasopoulos et~al., 2009]{Hyndman_2009_INTFOR}
Athanasopoulos, G., Ahmed, R.~A., and Hyndman, R.~J. (2009).
\newblock Hierarchical forecasts for {A}ustralian domestic tourism.
\newblock {\em International Journal of Forecasting}, 25(1):146--166.

\bibitem[Athanasopoulos et~al., 2024a]{ATHANASOPOULOS2024}
Athanasopoulos, G., Hyndman, R.~J., Kourentzes, N., and Panagiotelis, A.
  (2024a).
\newblock Editorial: Innovations in hierarchical forecasting.
\newblock {\em International Journal of Forecasting}, 40(2):427--429.

\bibitem[Athanasopoulos et~al., 2024b]{athanasopoulos2023review}
Athanasopoulos, G., Hyndman, R.~J., Kourentzes, N., and Panagiotelis, A.
  (2024b).
\newblock Forecast reconciliation: A review.
\newblock {\em International Journal of Forecasting}, 40(2):430--456.

\bibitem[Athanasopoulos et~al., 2017]{athanasopoulos2017temporal}
Athanasopoulos, G., Hyndman, R.~J., Kourentzes, N., and Petropoulos, F. (2017).
\newblock Forecasting with temporal hierarchies.
\newblock {\em European Journal of Operational Research}, 262(1):60--74.

\bibitem[Bischl et~al., 2016]{mlr}
Bischl, B., Lang, M., Kotthoff, L., Schiffner, J., Richter, J., Studerus, E.,
  Casalicchio, G., and Jones, Z.~M. (2016).
\newblock {mlr}: Machine learning in {R}.
\newblock {\em Journal of Machine Learning Research}, 17(170):1--5.

\bibitem[Breiman, 2001]{Breiman2001}
Breiman, L. (2001).
\newblock Random forests.
\newblock {\em Machine Learning}, 45:5--32.

\bibitem[Caporin et~al., 2023]{caporin2023exploiting}
Caporin, M., Di~Fonzo, T., and Girolimetto, D. (2023).
\newblock Exploiting intraday decompositions in realized volatility
  forecasting: A forecast reconciliation approach.
\newblock {\em arXiv preprint arXiv:2306.02952}.

\bibitem[Chen and Guestrin, 2016]{Chen_XGBoost_2016}
Chen, T. and Guestrin, P.~C. (2016).
\newblock {XGB}oost: A scalable tree boosting system.
\newblock {\em in Proceedings of the 22nd ACM SIGKDD International Conference
  on Knowledge Discovery and Data Mining. New York: ACM}, 0(0):785--794.

\bibitem[Chen et~al., 2024]{chen2015xgboost}
Chen, T., He, T., Benesty, M., Khotilovich, V., Tang, Y., Cho, H., Chen, K.,
  Mitchell, R., Cano, I., Zhou, T., Li, M., Xie, J., Lin, M., Geng, Y., Li, Y.,
  and Yuan, J. (2024).
\newblock {\em xgboost: Extreme Gradient Boosting}.
\newblock {R} package version 1.7.7.1.

\bibitem[Di~Fonzo and Girolimetto, 2023a]{difonzo2023cross}
Di~Fonzo, T. and Girolimetto, D. (2023a).
\newblock Cross-temporal forecast reconciliation: Optimal combination method
  and heuristic alternatives.
\newblock {\em International Journal of Forecasting}, 39(1):39--57.

\bibitem[Di~Fonzo and Girolimetto, 2023b]{difonzo2023spatio}
Di~Fonzo, T. and Girolimetto, D. (2023b).
\newblock Spatio-temporal reconciliation of solar forecasts.
\newblock {\em Solar Energy}, 251:13--29.

\bibitem[{Di Fonzo} and Girolimetto, 2024]{DIFONZO2024490}
{Di Fonzo}, T. and Girolimetto, D. (2024).
\newblock Forecast combination-based forecast reconciliation: Insights and
  extensions.
\newblock {\em International Journal of Forecasting}, 40(2):490--514.

\bibitem[Efron and Hastie, 2016]{Efron_Hastie_2016}
Efron, B. and Hastie, T. (2016).
\newblock {\em Computer Age Statistical Inference}.
\newblock Cambridge University Press, Cambridge.

\bibitem[Elliott and Timmermann, 2016]{Elliott16}
Elliott, G. and Timmermann, A. (2016).
\newblock {\em Economic Forecasting}.
\newblock Princeton University Press, Princeton, New Jersey.

\bibitem[Friedman, 2001]{Friedman_AOS_2001}
Friedman, J.~H. (2001).
\newblock Greedy function approximation: A gradient boosting machine.
\newblock {\em The Annals of Statistics}, 29(5):1189--1232.

\bibitem[Girolimetto et~al., 2023]{GIROLIMETTO2023}
Girolimetto, D., Athanasopoulos, G., {Di Fonzo}, T., and Hyndman, R.~J. (2023).
\newblock Cross-temporal probabilistic forecast reconciliation: Methodological
  and practical issues.
\newblock {\em International Journal of Forecasting}, forthcoming.

\bibitem[Girolimetto and {Di Fonzo}, 2023]{FoReco}
Girolimetto, D. and {Di Fonzo}, T. (2023).
\newblock {\em FoReco: Point Forecast Reconciliation}.
\newblock {R} package version 0.2.6.

\bibitem[Girolimetto and Di~Fonzo, 2023]{girolimetto2023point}
Girolimetto, D. and Di~Fonzo, T. (2023).
\newblock Point and probabilistic forecast reconciliation for general linearly
  constrained multiple time series.
\newblock {\em Statistical Methods \& Applications}, forthcoming.

\bibitem[Hollyman et~al., 2021]{HOLLYMAN_EJOR_2021149}
Hollyman, R., Petropoulos, F., and Tipping, M.~E. (2021).
\newblock Understanding forecast reconciliation.
\newblock {\em European Journal of Operational Research}, 294(1):149--160.

\bibitem[Hoover, 2006]{hoover2006measuring}
Hoover, J. (2006).
\newblock Measuring forecast accuracy: Omissions in today's forecasting engines
  and demand-planning software.
\newblock {\em Foresight: The International Journal of Applied Forecasting},
  4:32--35.

\bibitem[Hyndman et~al., 2024]{Rforecast-ref1}
Hyndman, R., Athanasopoulos, G., Bergmeir, C., Caceres, G., Chhay, L.,
  Kuroptev, K., O'Hara-Wild, M., Petropoulos, F., Razbash, S., Wang, E.,
  Yasmeen, F., Garza, F., Girolimetto, D., Ihaka, R., {R Core Team}, Reid, D.,
  Shaub, D., Tang, Y., Wang, X., and Zhou, Z. (2024).
\newblock {\em {forecast}: Forecasting functions for time series and linear
  models}.
\newblock {R} package version 8.22.0.

\bibitem[Hyndman, 2006]{hyndman2006anotherlook}
Hyndman, R.~J. (2006).
\newblock Another look at forecast-accuracy metrics for intermittent demand.
\newblock {\em Foresight: The International Journal of Applied Forecasting},
  4(4):43--46.

\bibitem[Hyndman et~al., 2011]{Hyndman_2011_CSDA}
Hyndman, R.~J., Ahmed, R.~A., Athanasopoulos, G., and Shang, H.~L. (2011).
\newblock Optimal combination forecasts for hierarchical time series.
\newblock {\em Computational Statistics \& Data Analysis}, 55(9):2579--2589.

\bibitem[Hyndman and Athanasopoulos, 2014]{hyndman2014optimally}
Hyndman, R.~J. and Athanasopoulos, G. (2014).
\newblock Optimally reconciling forecasts in a hierarchy.
\newblock {\em Foresight: The International Journal of Applied Forecasting},
  35:42--48.

\bibitem[Hyndman and Athanasopoulos, 2021]{hyndman2021forecasting}
Hyndman, R.~J. and Athanasopoulos, G. (2021).
\newblock {\em Forecasting: principles and practice}.
\newblock 3rd. Melbourne, Australia: OTexts.

\bibitem[Hyndman and Khandakar, 2008]{Rforecast-ref2}
Hyndman, R.~J. and Khandakar, Y. (2008).
\newblock Automatic time series forecasting: The forecast package for {R}.
\newblock {\em Journal of Statistical Software}, 26(3):1--22.

\bibitem[Hyndman and Koehler, 2006]{hyndman2006}
Hyndman, R.~J. and Koehler, A.~B. (2006).
\newblock Another look at measures of forecast accuracy.
\newblock {\em International Journal of Forecasting}, 22(4):679--688.

\bibitem[Hyndman et~al., 2002]{hyndman2002state}
Hyndman, R.~J., Koehler, A.~B., Snyder, R.~D., and Grose, S. (2002).
\newblock A state space framework for automatic forecasting using exponential
  smoothing methods.
\newblock {\em International Journal of forecasting}, 18(3):439--454.

\bibitem[Hyndman et~al., 2016]{hyndman2016fast}
Hyndman, R.~J., Lee, A.~J., and Wang, E. (2016).
\newblock Fast computation of reconciled forecasts for hierarchical and grouped
  time series.
\newblock {\em Computational Statistics \& Data Analysis}, 97:16--32.

\bibitem[Januschowski et~al., 2022]{januschowski2022forecasting}
Januschowski, T., Wang, Y., Torkkola, K., Erkkil{\"a}, T., Hasson, H., and
  Gasthaus, J. (2022).
\newblock Forecasting with trees.
\newblock {\em International Journal of Forecasting}, 38(4):1473--1481.

\bibitem[Komisarczyk et~al., 2024]{treeshapR}
Komisarczyk, K., Kozminski, P., Maksymiuk, S., Kapsner, L.~A., Spytek, M.,
  Krzyzinski, M., and Biecek, P. (2024).
\newblock {\em treeshap: Compute SHAP values for your tree-based models using
  the `TreeSHAP' algorithm}.
\newblock {R} package version 0.3.1.

\bibitem[Koning et~al., 2005]{KONING2005397}
Koning, A.~J., Franses, P.~H., Hibon, M., and Stekler, H. (2005).
\newblock The {M3} competition: Statistical tests of the results.
\newblock {\em International Journal of Forecasting}, 21(3):397--409.

\bibitem[Kourentzes and Athanasopoulos, 2019]{kourentzes2019cross}
Kourentzes, N. and Athanasopoulos, G. (2019).
\newblock Cross-temporal coherent forecasts for {A}ustralian tourism.
\newblock {\em Annals of Tourism Research}, 75:393--409.

\bibitem[Kourentzes et~al., 2023]{tsutils-Rpackage}
Kourentzes, N., Svetunkov, I., and Schaer, O. (2023).
\newblock {\em tsutils: Time Series Exploration, Modelling and Forecasting}.
\newblock {R} package version 0.9.4.

\bibitem[Liaw and Wiener, 2002]{randomForest}
Liaw, A. and Wiener, M. (2002).
\newblock Classification and regression by randomforest.
\newblock {\em R News}, 2(3):18--22.

\bibitem[Lundberg et~al., 2018]{lundberg2018consistent}
Lundberg, S.~M., Erion, G.~G., and Lee, S.-I. (2018).
\newblock Consistent individualized feature attribution for tree ensembles.
\newblock {\em arXiv preprint arXiv:1802.03888}.

\bibitem[Lundberg and Lee, 2017]{lundberg2017unified}
Lundberg, S.~M. and Lee, S.-I. (2017).
\newblock A unified approach to interpreting model predictions.
\newblock In Guyon, I., Luxburg, U.~V., Bengio, S., Wallach, H., Fergus, R.,
  Vishwanathan, S., and Garnett, R., editors, {\em Advances in Neural
  Information Processing Systems, vol. 30}, volume~30 of {\em NIPS'17}, pages
  4765--4774. Curran Associates, Inc.

\bibitem[Makridakis et~al., 2020]{makridakis2020m4}
Makridakis, S., Spiliotis, E., and Assimakopoulos, V. (2020).
\newblock The {M4} competition: 100,000 time series and 61 forecasting methods.
\newblock {\em International Journal of Forecasting}, 36(1):54--74.

\bibitem[Makridakis et~al., 2022a]{makridakis2022m5accuracyresults}
Makridakis, S., Spiliotis, E., and Assimakopoulos, V. (2022a).
\newblock M5 accuracy competition: Results, findings, and conclusions.
\newblock {\em International Journal of Forecasting}, 38(4):1346--1364.

\bibitem[Makridakis et~al., 2022b]{makridakis2022m5}
Makridakis, S., Spiliotis, E., and Assimakopoulos, V. (2022b).
\newblock The {M5} competition: Background, organization, and implementation.
\newblock {\em International Journal of Forecasting}, 38(4):1325--1336.

\bibitem[Makridakis et~al., 2022c]{makridakis2022m5hypothesizing}
Makridakis, S., Spiliotis, E., and Assimakopoulos, V. (2022c).
\newblock Predicting/hypothesizing the findings of the {M5} competition.
\newblock {\em International Journal of Forecasting}, 38(4):1337--1345.

\bibitem[Medeiros et~al., 2021]{Medeiros_2021}
Medeiros, M.~C., Vasconcelos, G. F.~R., Álvaro Veiga, and Zilberman, E.
  (2021).
\newblock Forecasting inflation in a data-rich environment: The benefits of
  machine learning methods.
\newblock {\em Journal of Business \& Economic Statistics}, 39(1):98--119.

\bibitem[Molnar, 2023]{molnar2022interpretable}
Molnar, C. (2023).
\newblock {\em Interpretable Machine Learning}.
\newblock 2 edition.

\bibitem[Panagiotelis et~al., 2021]{panagiotelis2021forecast}
Panagiotelis, A., Athanasopoulos, G., Gamakumara, P., and Hyndman, R.~J.
  (2021).
\newblock Forecast reconciliation: A geometric view with new insights on bias
  correction.
\newblock {\em International Journal of Forecasting}, 37(1):343--359.

\bibitem[{R Core Team}, 2023]{Rcoreteam}
{R Core Team} (2023).
\newblock {\em R: A Language and Environment for Statistical Computing. R
  Foundation for Statistical Computing, Vienna, Austria.}

\bibitem[Rostami-Tabar and Hyndman, 2024]{Hyndman_Rostami_2024}
Rostami-Tabar, B. and Hyndman, R.~J. (2024).
\newblock Hierarchical time series forecasting in emergency medical services.
\newblock {\em Journal of Service Research}, forthcoming.

\bibitem[Shapley et~al., 1953]{shapley1953value}
Shapley, L.~S. et~al. (1953).
\newblock A value for n-person games.
\newblock {\em Contributions to the theory of games}, 2:307--317.

\bibitem[Shi et~al., 2024]{lightgbm-Rpackage}
Shi, Y., Ke, G., Soukhavong, D., Lamb, J., Meng, Q., Finley, T., Wang, T.,
  Chen, W., Ma, W., Ye, Q., Liu, T.-Y., Titov, N., Yan, Y., {Microsoft
  Corporation}, {Dropbox, Inc.}, Ferreira, A., Lemire, D., Zverovich, V., {IBM
  Corporation}, Cortes, D., and Mayer, M. (2024).
\newblock {\em lightgbm: Light Gradient Boosting Machine}.
\newblock {R} package version 4.3.0.

\bibitem[Spiliotis et~al., 2021]{spiliotis2021hierarchical}
Spiliotis, E., Abolghasemi, M., Hyndman, R.~J., Petropoulos, F., and
  Assimakopoulos, V. (2021).
\newblock Hierarchical forecast reconciliation with machine learning.
\newblock {\em Applied Soft Computing}, 112:107756.

\bibitem[Theodosiou and Kourentzes, 2021]{theodosiou2021forecasting}
Theodosiou, F. and Kourentzes, N. (2021).
\newblock Forecasting with deep temporal hierarchies.
\newblock {\em Available at SSRN 3918315}.

\bibitem[Van~Erven and Cugliari, 2015]{van2015game}
Van~Erven, T. and Cugliari, J. (2015).
\newblock Game-theoretically optimal reconciliation of contemporaneous
  hierarchical time series forecasts.
\newblock In Antoniadis, A., Poggi, J.-M., and Brossat, X., editors, {\em
  Modeling and stochastic learning for forecasting in high dimensions}, pages
  297--317. Springer.

\bibitem[Wang et~al., 2022]{wang2022end}
Wang, S., Zhou, F., Sun, Y., Ma, L., Zhang, J., and Zheng, Y. (2022).
\newblock End-to-end modeling of hierarchical time series using autoregressive
  transformer and conditional normalizing flow-based reconciliation.
\newblock In {\em 2022 IEEE International Conference on Data Mining Workshops
  (ICDMW)}, pages 1087--1094. IEEE.

\bibitem[Wickramasuriya et~al., 2019]{wickramasuriya2019optimal}
Wickramasuriya, S.~L., Athanasopoulos, G., and Hyndman, R.~J. (2019).
\newblock Optimal forecast reconciliation for hierarchical and grouped time
  series through trace minimization.
\newblock {\em Journal of the American Statistical Association},
  114(526):804--819.

\bibitem[Yan, 2021]{rBayesianOptimization-Rpackage}
Yan, Y. (2021).
\newblock {\em rBayesianOptimization: Bayesian Optimization of
  Hyperparameters}.
\newblock {R} package version 1.2.0.

\end{thebibliography}

\appendix

\newpage

\renewcommand{\thetable}{A.\arabic{table}}
\setcounter{table}{0}

\renewcommand{\thefigure}{A.\arabic{figure}}
\setcounter{figure}{0}

\begin{appendices}

\clearpage
\newpage

\section{Design of the Forecast Study} \label{app:setup}

\subsection{Rolling-Window Set-Up} \label{app:rolling-window}

\begin{figure}[h]
    \centering
    \begin{tikzpicture}[x=0.75pt,y=0.75pt,yscale=-1,xscale=1]
%uncomment if require: \path (0,300); %set diagram left start at 0, and has height of 300

%Straight Lines [id:da7094050589759195] 
\draw [color={rgb, 255:red, 74; green, 144; blue, 226 }  ,draw opacity=1 ][line width=3.75]    (39.71,57.86) -- (390.29,57.86) ;
%Straight Lines [id:da5190560545509719] 
\draw [color={rgb, 255:red, 126; green, 211; blue, 33 }  ,draw opacity=1 ][line width=3.75]    (390.29,57.86) -- (420.45,57.84) ;
%Straight Lines [id:da22900312763039987] 
\draw [color={rgb, 255:red, 74; green, 144; blue, 226 }  ,draw opacity=1 ][line width=3.75]    (69.71,67.86) -- (420.29,67.86) ;
%Straight Lines [id:da41714262093587173] 
\draw [color={rgb, 255:red, 126; green, 211; blue, 33 }  ,draw opacity=1 ][line width=3.75]    (420.29,67.86) -- (450.45,67.84) ;
%Straight Lines [id:da6540008095092569] 
\draw [color={rgb, 255:red, 74; green, 144; blue, 226 }  ,draw opacity=1 ][line width=3.75]    (99.71,77.86) -- (450.29,77.86) ;
%Straight Lines [id:da8639789751919968] 
\draw [color={rgb, 255:red, 126; green, 211; blue, 33 }  ,draw opacity=1 ][line width=3.75]    (450.29,77.86) -- (480.45,77.84) ;
%Straight Lines [id:da8086869884096664] 
\draw [color={rgb, 255:red, 74; green, 144; blue, 226 }  ,draw opacity=1 ][line width=3.75]    (129.71,87.86) -- (480.29,87.86) ;
%Straight Lines [id:da8773178251305596] 
\draw [color={rgb, 255:red, 126; green, 211; blue, 33 }  ,draw opacity=1 ][line width=3.75]    (480.29,87.86) -- (510.45,87.84) ;
%Straight Lines [id:da12815983430074263] 
\draw [color={rgb, 255:red, 74; green, 74; blue, 74 }  ,draw opacity=1 ][line width=3.75]    (40.21,118.52) -- (510.83,117.83) ;
%Straight Lines [id:da6848828510614595] 
\draw [color={rgb, 255:red, 208; green, 2; blue, 27 }  ,draw opacity=1 ][line width=3.75]    (510.83,117.83) -- (540.99,117.82) ;
%Straight Lines [id:da15813074719216158] 
\draw [color={rgb, 255:red, 74; green, 144; blue, 226 }  ,draw opacity=0.3 ][line width=3.75]    (69.71,187.86) -- (420.29,187.86) ;
%Straight Lines [id:da35263834508188174] 
\draw [color={rgb, 255:red, 126; green, 211; blue, 33 }  ,draw opacity=0.4 ][line width=3.75]    (420.29,187.86) -- (450.45,187.84) ;
%Straight Lines [id:da8629827121359759] 
\draw [color={rgb, 255:red, 74; green, 144; blue, 226 }  ,draw opacity=0.3 ][line width=3.75]    (99.71,197.86) -- (450.29,197.86) ;
%Straight Lines [id:da21612604800365132] 
\draw [color={rgb, 255:red, 126; green, 211; blue, 33 }  ,draw opacity=0.4 ][line width=3.75]    (450.29,197.86) -- (480.45,197.84) ;
%Straight Lines [id:da1532761810904648] 
\draw [color={rgb, 255:red, 74; green, 144; blue, 226 }  ,draw opacity=0.3 ][line width=3.75]    (129.71,207.86) -- (480.29,207.86) ;
%Straight Lines [id:da49379949923739797] 
\draw [color={rgb, 255:red, 126; green, 211; blue, 33 }  ,draw opacity=0.4 ][line width=3.75]    (480.29,207.86) -- (510.45,207.84) ;
%Straight Lines [id:da7557552569446582] 
\draw [color={rgb, 255:red, 74; green, 74; blue, 74 }  ,draw opacity=1 ][line width=3.75]    (70.21,248.52) -- (540.83,247.83) ;
%Straight Lines [id:da6497367738155644] 
\draw [color={rgb, 255:red, 208; green, 2; blue, 27 }  ,draw opacity=1 ][line width=3.75]    (540.83,247.83) -- (570.99,247.82) ;
%Straight Lines [id:da16442298022326685] 
\draw [color={rgb, 255:red, 74; green, 144; blue, 226 }  ,draw opacity=1 ][line width=3.75]    (160.31,218.26) -- (510.89,218.26) ;
%Straight Lines [id:da13261300539291] 
\draw [color={rgb, 255:red, 126; green, 211; blue, 33 }  ,draw opacity=1 ][line width=3.75]    (510.89,218.26) -- (541.05,218.24) ;
%Straight Lines [id:da5275976205841877] 
\draw [color={rgb, 255:red, 155; green, 155; blue, 155 }  ,draw opacity=1 ] [dash pattern={on 4.5pt off 4.5pt}]  (69.71,65.86) -- (70,190.43) ;
%Straight Lines [id:da7064480348528277] 
\draw [color={rgb, 255:red, 155; green, 155; blue, 155 }  ,draw opacity=1 ] [dash pattern={on 4.5pt off 4.5pt}]  (420.29,65.57) -- (420.29,189.86) ;
%Straight Lines [id:da38626459618375697] 
\draw [color={rgb, 255:red, 155; green, 155; blue, 155 }  ,draw opacity=1 ] [dash pattern={on 4.5pt off 4.5pt}]  (450.29,65.57) -- (450.29,189.86) ;

% Text Node
\draw (484.5,19) node [anchor=north west][inner sep=0.75pt]  [font=\footnotesize] [align=left] {iteration $\displaystyle r$ };
% Text Node
\draw (484.5,158.5) node [anchor=north west][inner sep=0.75pt]  [font=\footnotesize] [align=left] {iteration $\displaystyle r+1$ };

\end{tikzpicture}
    \caption{Outer rolling window in forecast set-up and inner rolling windows procedure in the machine-learning based forecast reconciliation.}
    \label{fig:tikz_iterations}
\end{figure}
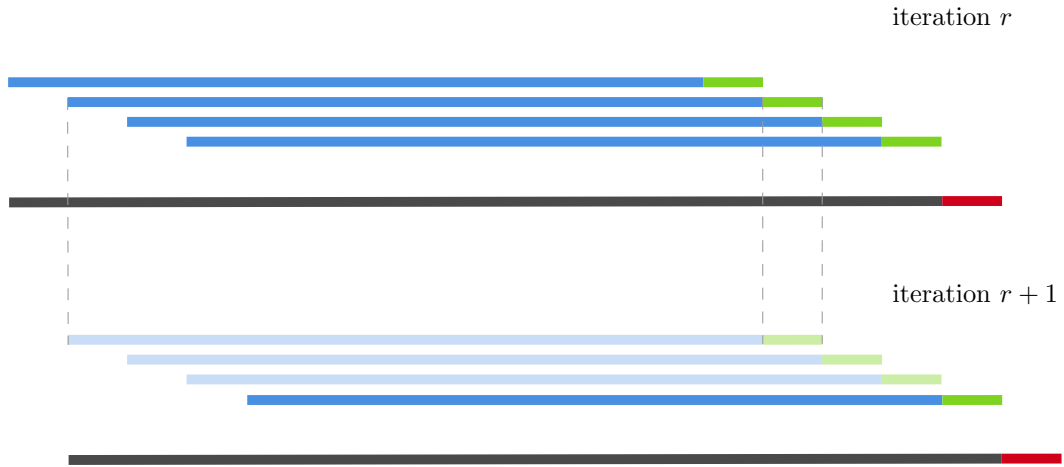

\subsection{Base Forecasts} \label{app:base-forecasts}
\noindent

We consider four base forecast models: (i) Naive, (ii) SARIMA, (iii) ETS and a (iv) forecast combination of the former three.  
\medskip

\noindent
{\bf Naive.}
The first base forecast model for temporal aggregation level $k$ of a specific cross-sectional time series is simply 
$x_j^{[k]} = x_{j-7m/k}^{[k]} + \varepsilon^{[k]}_j$,   which we call ``Naive"  since  it simply implies that the forecast of a specific time slot  (e.g., 30-min, hourly or daily) is the value observed in the same time slot and day of the  previous week.
This way of forecasting is popular in the industry since it is ultra fast, i.e.\ it does not require parameter estimation, and it works well in the case of strong seasonality patterns as observed in platform  data. 
\medskip\medskip

\noindent
{\bf SARIMA.}  Secondly, we consider ARIMA models, as implemented in the \texttt{forecast} package \citep{Rforecast-ref1, Rforecast-ref2} in \texttt{R}.  
For the daily time series, we search for each series  the optimal model  in the space of seasonal ARIMA models, thereby allowing for weekly seasonality (seasonal period equal to 7 
for daily data) and a maximal MA order $q=3$, and maximal AR order of $p=3$. For the seasonal components we set the maximal MA order to $Q=1$  and maximal AR order to $P = 1$. We restrict the maximum number of non-seasonal differences to $d = 1$.
All other arguments are kept at their default values. 

For the hourly and 30-min time series, we need to additionally account for the strong intra-day seasonal patterns in the platform data. Since the seasonal period is long in both cases ($m_1 = 34$ for 30-min data and $m_2 = 17$ for hourly data), we follow the common practice based on Fourier series to capture the seasonality. Specifically, for the 30-min time series $y_t$, we consider the regression model 
\begin{equation}
	y_t = \mu  + \sum_{s=1}^S \left[ \alpha_s \text{sin}(2\pi s t/m_1)+ \beta_s \text{cos}(2 \pi s t/m_1) \right] + \varepsilon_t, \label{fourier-eq}
\end{equation}
where $\varepsilon_t$ is subsequently modeled as an ARIMA process. Analogously for the hourly data thereby taking the seasonal period $m_2 = 17$. We select the optimal value of $S$ via the Bayesian Information Criterion. 

The above discussion applies to the last-mile delivery applications.
For the bicycle sharing application, we  
fit seasonal ARIMA models for the time series with 24-hour temporal frequency to capture the weekly seasonality, whereas for the higher-frequency time series (30-min, 1-hour, 1.5-hour, 2-hour, 3-hour, 4-hour, 6-hour, 8-hour, 12-hour) we apply a similar strategy based on Fourier series to capture the seasonality. More specifically, we capture intra-day seasonal patterns for frequencies 30-min, 1-hour, 1.5-hour, 2-hour, 3-hour, and weekly seasonal patterns for frequencies 4-hour, 6-hour, 8-hour, 12-hour. 

\medskip\medskip

\noindent
{\bf ETS.} Thirdly, we consider exponential smoothing (ETS; \citealp{hyndman2002state}) as another popular  model to produce the base forecasts. To this end, we use the \texttt{ets} function (with default arguments) as implemented in the \texttt{forecast} package. 
As above, for the 30-min and hourly series in the last-mile delivery application and the 30-min, 1-hour, 1.5-hour, 2-hour, 3-hour series in the bicycle sharing application, we first estimate the intra-day seasonality through the Fourier series approach (similarly, for the 4-hour, 6-hour, 8-hour, 12-hour series we capture the weekly seasonality) and subsequently model the error term in equation \eqref{fourier-eq} using ETS. 
\medskip\medskip

\noindent
{\bf Forecast Combination.} Our final base forecast model consists of a simple forecast combination of the previous three base forecast models. Forecast combination approaches have shown to perform well as base forecast methods in the hierarchical forecasting literature, thereby oftentimes providing an effective practice for improving forecast accuracy; see   for instance \cite{Hyndman_Rostami_2024} for an application on emergency medical services data. 
We opt for equal weights in the forecast combination, 
mainly because of its simplicity, ease in implementation and its established track-record of good performance in the forecast combination literature (e.g., \citealp{Elliott16}, Chapter 14).

\subsection{Machine Learning Methods} \label{app:ML-methods}

We consider three machine learning methods: (i) random forest, (ii) XGBoost and (iii) LightGBM.  
\medskip

\noindent
{\bf Random Forest.}
Random forests, proposed by  \cite{Breiman2001}, produce forecasts by combining regression trees. A regression tree is a nonparametric method  that partitions the feature space to compute local averages as forecasts, see \cite{Efron_Hastie_2016} for a  textbook treatment. The tuning parameters are the number of trees that are used in the forecast combination, the number of features to randomly select when constructing each regression tree split, and the minimum number of observations in each terminal node to compute the local forecasts.  
We use the standard implementation of the \texttt{randomForest} package \citep{randomForest} in \verb|R| with default settings for the hyperparamters (i.e.\ 
number of trees: \texttt{ntree} = 500, number of variables sampled at each split: $\texttt{mtry} = \text{\# of features}/3$, minimum size of terminal nodes: \texttt{nodesize} = 5) when reporting our main results.

%In Section %\ref{subsec:result:tuning}, 
We also investigate the performance of the random forest based forecast reconciliation method when the hyperparameters are tuned.
To this end, we follow \cite{spiliotis2021hierarchical} and use the package \texttt{mlr} in \texttt{R} (\citealp{mlr})
which implements a random grid search with cross-validation to find the optimal hyperparameters. In the cross-validation procedure, we leave each of the four subsequent weeks in the validation set once out as test sample and use the root mean squared error as cross-validation score to compute the optimal hyperparameters. 
Note that we tune the hyperparameters only once every four iterations (i.e. once every month) in the outer rolling window (so in 8 out of the 31 outer rolling windows) to keep the computational burden low; in between we use the optimal hyperparameters from the previous tuning round as fixed values. 
The lower and upper bounds for the hyperparameters are set to $(2, 50)$ for \texttt{mtry} and $(5,50)$ for \texttt{nodesize}. We set the bounds for \texttt{ntree} between 50 and 500 on interval steps of 10.

\medskip\medskip

\noindent
{\bf XGBoost.}
Gradient boosting, proposed by \cite{Friedman_AOS_2001}, constructs forecasts by sequentially fitting small regression trees, i.e.\ weak learners, to the residuals by the ensemble of the previous trees. This procedure results in a one final tree, constructed as a sum of trees, used for forecasting. 
Extreme gradient boosting (XGBoost), introduced by \cite{Chen_XGBoost_2016},  optimizes the implementation of the gradient boosting framework in terms of speed and flexibility. 
We use the  \texttt{xgboost} package \citep{chen2015xgboost} in \verb|R| with fixed choices of the tuning parameters when reporting our main results. 
We fix the hyperparameters to the default values as follows: 100 boosting iterations (\texttt{nrounds}), 6 as max tree depth (\texttt{max\_depth}),  0.3 as learning rate (\texttt{eta}), 
1 as subsample ratio (\texttt{subsample}), 
1 as subsample ratio of columns (\texttt{colsample\_bytree}), 
1 as minimum sum of instance weight (hessian) (\texttt{min\_child\_weight}) and
0 as minimum loss reduction (\texttt{gamma}).

Subsequently, to tune the hyperparameters, we use the same procedure as discussed for random forest but this time use Bayesian optimization to tune the hyperparameters since grid search is computationally very expensive in this case.
To this end, we use the \texttt{rBayesianOptimization} (\citealp{rBayesianOptimization-Rpackage}) 
in \texttt{R} and consider the following intervals with lower and upperbounds for each hyperparameter, in line with \cite{spiliotis2021hierarchical}: 
we set the values of \texttt{max\_depth} between (2, 10), the learning rate (\texttt{eta}) between (0.01, 0.05), \texttt{subsample} values between (0.3, 1), \texttt{colsample\_bytree} values between (0.3, 1), \texttt{min\_child\_weight} between (0, 10) and \texttt{gamma} between (0, 5). The values for the maximum number of boosting iterations (\texttt{nrounds}) are over the range of 50 and 200.

\medskip\medskip

\noindent
{\bf LightGBM.}
LightGBM, put forward by Microsoft in 2016, is a gradient boosting framework that uses tree-based learning algorithms like XGBoost but as the name suggests has computational advantages with respect to training speed, memory usage and parallelization. The implementation of gradient-based one-side sampling and exclusive feature bundling techniques allows  handling large training datasets. We use the  \texttt{lightgbm} package (\citealp{lightgbm-Rpackage}) in \verb|R| with fixed, default hyperparameters when reporting our main results: 100 boosting iterations (\texttt{nrounds}), 31 as maximum number of leaves (\texttt{num\_leaves}),  0.1 as learning rate (\texttt{eta}), 
1 as subsample ratio (\texttt{subsample}), 1 as subsample ratio of columns (\texttt{colsample\_bytree}), 0.001 as minimum sum of instance weight (hessian) (\texttt{min\_child\_weight})
and 0 as $\ell_1$-regularization (\texttt{lambda\_l1}) and no limit to the max depth (\texttt{max\_depth}) for the tree model. 

Finally, to tune the hyperparameters, we also use Bayesian optimization with the following similar prior values for the hyperparameters as with XGBoost. 
The maximum number of leaves (\texttt{num\_leaves}) fixed to the range of 5 and 31. The learning rate (\texttt{eta}) is set between (0.01, 0.05), \texttt{subsample} values between (0.3, 1), \texttt{colsample\_bytree} values are set between (0.3, 1), \texttt{min\_child\_weight} between (0, 10), \texttt{max\_depth} between (2, 10) and \texttt{lambda\_l1} between (0, 5). The values for the maximum number of boosting iterations (\texttt{nrounds}) are over the range of 50 and 200.

\renewcommand{\thetable}{B.\arabic{table}}
\setcounter{table}{0}

\renewcommand{\thefigure}{B.\arabic{figure}}
\setcounter{figure}{0}

\newpage

\section{Platform Applications: Forecast Performance in terms of MASE \label{app:platformapp}} 
\vspace{-0.25cm}

\begin{table}[h!]
	{
		\centering
		\caption{
			London forecast reconciliation results for the MASE accuracy index.}
		\resizebox{\textwidth}{!}{ \begin{tabular}{ll c c c|ccc|ccc}
				\hline
				\textbf{Base} & \textbf{Forecast} & \multicolumn{3}{c}{\textbf{Delivery Area}} & \multicolumn{3}{c}{\textbf{Zone}} & \multicolumn{3}{c}{\textbf{London}} \\
				\textbf{Forecasts} & \textbf{Method} \\
				\hline
				&& \textbf{30min} & \textbf{hour} & \textbf{day} & \textbf{30min} & \textbf{hour} & \textbf{day} &\textbf{30min} & \textbf{hour} & \textbf{day}  \\
				Naive & Base & 0.9824 & 0.9637 & 0.8748 & 0.9263 & 0.9027 & 0.8124 & 0.8456 & 0.8375 & 0.7903 \\ 
				& Random forest & \cellcolor{lightgray}0.7669 & \cellcolor{lightgray}0.7821 & \cellcolor{lightgray}0.8216 & \cellcolor{lightgray}0.8025 & \cellcolor{lightgray}0.8187 & \cellcolor{lightgray}0.8087 & \cellcolor{lightgray}0.8740 & \cellcolor{lightgray}0.8667 & \cellcolor{lightgray}0.8132 \\ 
				& XGBoost & 0.8362 & 0.8427 & 0.8522 & 0.8538 & 0.8573 & 0.8326 & 0.8976 & 0.8826 & 0.8310 \\ 
				& LightGBM & 0.8052 & 0.8147 & 0.8405 & 0.8338 & 0.8407 & 0.8267 & 0.8984 & 0.8827 & 0.8283 \\ 
				\\[-0.15cm]
				
				ETS & Base & 0.9092 & 0.9974 & 0.7213 & 1.2484 & 1.3634 & 0.6884 & 1.7459 & 1.7079 & 0.6707 \\ 
				& Bottom Up & 0.9092 & 1.0082 & 1.5758 & 1.1840 & 1.3201 & 1.7864 & 1.6391 & 1.6819 & 1.8977 \\ 
				& tcs & 0.8847 & 0.9704 & 1.4441 & 1.1357 & 1.2613 & 1.6694 & 1.5493 & 1.5908 & 1.7722 \\ 
				& cst & 0.8709 & 0.9518 & 1.3841 & 1.0890 & 1.2048 & 1.5389 & 1.4509 & 1.4903 & 1.6093 \\ 
				& ite & 0.8871 & 0.9766 & 1.4746 & 1.1451 & 1.2772 & 1.7080 & 1.5677 & 1.6159 & 1.8118 \\ 
				& oct & 0.8843 & 0.9675 & 1.4111 & 1.1337 & 1.2554 & 1.6346 & 1.5403 & 1.5770 & 1.7325 \\ 
				& Random forest & \cellcolor{lightgray}0.7849 & \cellcolor{lightgray}0.8130 & \cellcolor{lightgray}0.9005 & \cellcolor{lightgray}0.8624 & \cellcolor{lightgray}0.9061 & \cellcolor{lightgray}0.9303 & 1.0235 & 1.0362 & 0.9552 \\ 
				& XGBoost & 0.8827 & 0.9275 & 1.0175 & 0.9096 & 0.9464 & 0.9355 & \cellcolor{lightgray}0.9807 & \cellcolor{lightgray}0.9863 & \cellcolor{lightgray}0.8968 \\ 
				& LightGBM & 0.8291 & 0.8682 & 0.9886 & 0.8778 & 0.9175 & 0.9376 & 0.9868 & 0.9929 & 0.9117 \\ 
				\\[-0.15cm]
				
				SARIMA & Base & 0.8875 & 0.9662 & 0.7708 & 1.1520 & 1.2731 & 0.7226 & 1.6034 & 1.6229 & 0.7441 \\ 
				& Bottom Up & 0.8875 & 0.9744 & 1.4648 & 1.1624 & 1.2948 & 1.7398 & 1.6316 & 1.6738 & 1.8658 \\ 
				& tcs & 0.8670 & 0.9430 & 1.3555 & 1.1181 & 1.2395 & 1.6223 & 1.5476 & 1.5882 & 1.7350 \\ 
				& cst & 0.8536 & 0.9251 & 1.2891 & 1.0752 & 1.1882 & 1.5007 & 1.4538 & 1.4946 & 1.5827 \\ 
				& ite & 0.8677 & 0.9471 & 1.3779 & 1.1239 & 1.2518 & 1.6539 & 1.5589 & 1.6063 & 1.7651 \\ 
				& oct & 0.8641 & 0.9366 & 1.3144 & 1.1026 & 1.2171 & 1.5645 & 1.5163 & 1.5532 & 1.6797 \\ 
				& Random forest & \cellcolor{lightgray}0.7771 & \cellcolor{lightgray}0.7986 & \cellcolor{lightgray}0.8464 & \cellcolor{lightgray}0.8258 & \cellcolor{lightgray}0.8600 & 0.8384 & 0.9455 & 0.9590 & 0.8317 \\ 
				& XGBoost & 0.8759 & 0.9039 & 0.9472 & 0.8739 & 0.8951 & 0.8371 & 0.9026 & 0.9060 & 0.7779 \\ 
				& LightGBM & 0.8161 & 0.8440 & 0.9046 & 0.8331 & 0.8585 & \cellcolor{lightgray}0.8130 & \cellcolor{lightgray}0.8895 & \cellcolor{lightgray}0.8956 & \cellcolor{lightgray}0.7582 \\ 
				\\[-0.15cm]
				
				Forecast & Base & 0.8273 & 0.8622 & 0.7310 & 0.9582 & 1.0197 & 0.6855 & 1.2171 & 1.2186 & 0.6848 \\ 
				Combination & Bottom Up & 0.8273 & 0.8661 & 1.1351 & 0.9517 & 1.0196 & 1.2810 & 1.2049 & 1.2286 & 1.3533 \\ 
				& tcs & 0.8199 & 0.8557 & 1.0893 & 0.9370 & 1.0019 & 1.2392 & 1.1738 & 1.1989 & 1.3105 \\ 
				& cst & 0.8139 & 0.8468 & 1.0510 & 0.9151 & 0.9744 & 1.1657 & 1.1216 & 1.1464 & 1.2201 \\ 
				& ite & 0.8220 & 0.8599 & 1.1101 & 0.9456 & 1.0158 & 1.2689 & 1.1939 & 1.2242 & 1.3408 \\ 
				& oct & 0.8174 & 0.8505 & 1.0582 & 0.9262 & 0.9856 & 1.1948 & 1.1483 & 1.1688 & 1.2615 \\ 
				& Randomforest & \cellcolor{lightgray}0.7654 & \cellcolor{lightgray}0.7816 & 0.8671 & \cellcolor{lightgray}0.8129 & 0.8412 & 0.9145 & 0.9303 & 0.9380 & 0.9442 \\ 
				& XGBoost & 0.8346 & 0.8455 & 0.8710 & 0.8435 & 0.8546 & 0.8533 & 0.8917 & 0.8912 & 0.8487 \\ 
				& LightGBM & 0.8026 & 0.8146 & \cellcolor{lightgray}0.8533 & 0.8209 & \cellcolor{lightgray}0.8351 & \cellcolor{lightgray}0.8435 & \cellcolor{lightgray}0.8857 & \cellcolor{lightgray}0.8856 & \cellcolor{lightgray}0.8366 \\ 
				%\midrule
				\hline
		\end{tabular}}
		\label{tab:London_FR_results_mase}
		\raggedright
		\footnotesize
		$\;$\\[-0.1cm]
		Notes: This table shows forecast accuracy measured in mean absolute scaled error
		(MASE). 
		The forecast horizon is one week-ahead. 
		The best forecast reconciliation results for each base forecast (Naive, ETS, SARIMA, Forecast Combination) are highlighted in gray. 
		\\[-0.1cm]}
\end{table}

\begin{table}[htbp]
	
	\caption{
		Manchester forecast reconciliation results for the MASE accuracy index.}
	\centering
	\footnotesize
	%\resizebox{\textwidth}{!}{ 
		\begin{tabular}{ll c c c |ccc}
			\hline
			\textbf{Period} & \textbf{Forecast} & \multicolumn{3}{c}{\textbf{Delivery Area}} &  \multicolumn{3}{c}{\textbf{Market}} \\
			\textbf{} & \textbf{Method} \\
			\hline
			&& \textbf{30min} & \textbf{hour} & \textbf{day} &\textbf{30min} & \textbf{hour} & \textbf{day} \\
			\\
			
			Before shift & Base & 0.8717 & 0.9537 & 0.8450 & 1.4371 & 1.5384 & 0.9517 \\ 
			& Bottom Up & 0.8717 & 0.9618 & 1.4621 & 1.4855 & 1.6004 & 1.8757 \\ 
			& tcs & 0.8490 & 0.9265 & 1.3518 & 1.4082 & 1.5088 & 1.7642 \\ 
			& cst & 0.8417 & 0.9144 & 1.3061 & 1.3685 & 1.4599 & 1.6792 \\  
			& ite  & 0.8503 & 0.9275 & 1.3595 & 1.4143 & 1.5143 & 1.7824 \\ 
			& oct & 0.8512 & 0.9288 & 1.3497 & 1.4093 & 1.5117 & 1.7493 \\ 
			& Random forest & \cellcolor{lightgray}0.7792 & \cellcolor{lightgray}0.8063 & \cellcolor{lightgray}0.8768 & 0.9346 & 0.9677 & 0.8945 \\ 
			& XGBoost & 0.8809 & 0.9206 & 1.0030 & 0.9271 & 0.9518 & \cellcolor{lightgray}0.8713 \\ 
			& LightGBM & 0.8259 & 0.8642 & 0.9915 & \cellcolor{lightgray}0.9206 & \cellcolor{lightgray}0.9501 & 0.8921 \\ 
			\\
			\\
			During shift & Base & 0.9219 & 1.0624 & 0.8995 & 1.5797 & 1.7296 & 0.6676 \\ 
			& Bottom Up & 0.9219 & 1.0740 & 1.8529 & 1.6131 & 1.7767 & 2.0856 \\ 
			& tcs & 0.8940 & 1.0309 & 1.6943 & 1.4969 & 1.6401 & 1.8706 \\ 
			& cst & 0.8859 & 1.0170 & 1.6178 & 1.4309 & 1.5610 & 1.7293 \\ 
			& ite & 0.8988 & 1.0377 & 1.7151 & 1.5152 & 1.6587 & 1.9058 \\ 
			& oct & 0.8877 & 1.0214 & 1.6469 & 1.4652 & 1.6090 & 1.8169 \\ 
			& Randomforest & \cellcolor{lightgray}0.7977 & \cellcolor{lightgray}0.8658 & \cellcolor{lightgray}1.1074 & \cellcolor{lightgray}0.8420 & 0.8677 & 0.6978 \\ 
			& XGBoost & 0.8939 & 0.9771 & 1.2160 & 0.8479 & 0.8551 & \cellcolor{lightgray}0.6717 \\ 
			& LightGBM & 0.8242 & 0.8986 & 1.1424 & 0.8319 & \cellcolor{lightgray}0.8448 & 0.7108 \\ 
			\\
			\\
			After shift & Base & 0.7874 & 0.8700 & 0.5977 & 1.3320 & 1.4557 & 0.5892 \\ 
			& Bottom Up & 0.7874 & 0.8820 & 1.3808 & 1.3576 & 1.4816 & 1.7568 \\ 
			& tcs & 0.7652 & 0.8486 & 1.2104 & 1.2565 & 1.3596 & 1.5810 \\ 
			& cst & 0.7602 & 0.8387 & 1.1369 & 1.2075 & 1.3005 & 1.4573 \\ 
			& ite  & 0.7687 & 0.8536 & 1.2140 & 1.2724 & 1.3757 & 1.6030 \\ 
			& oct & 0.7598 & 0.8413 & 1.1807 & 1.2280 & 1.3296 & 1.5263 \\ 
			& Randomforest & \cellcolor{lightgray}0.6525 & \cellcolor{lightgray}0.6664 & \cellcolor{lightgray}0.6148 & 0.7296 & 0.7405 & \cellcolor{lightgray}0.5354 \\ 
			& XGBoost & 0.7340 & 0.7598 & 0.7115 & 0.7427 & 0.7363 & 0.5554 \\ 
			& LightGBM & 0.6854 & 0.7091 & 0.6905 & \cellcolor{lightgray}0.7211 & \cellcolor{lightgray}0.7208 & 0.5389 \\ 
			\hline
		\end{tabular}%}
	\label{tab:Manchester_reconiliation_MASE_results}
	$\;$\\[0.2cm]
	\raggedright
	\footnotesize
	Notes: This table shows forecast accuracy measured in weighted absolute percentage error (MASE) for SARIMA base forecasts. 
	The best forecast reconciliation results for the periods Before shift (prior to June 14, 2023), After shift (after July 14, 2023), and During shift (period in between) are highlighted in gray. 
	The linear benchmark methods are defined in Table \ref{tabel:linearbenchmarks}.  \\[0.1cm]
\end{table}

\begin{table}[htbp]
	\centering
	
	\caption{
		Citi Bike forecast reconciliation results for the H3 cells for the MASE accuracy index.}
	\resizebox{\textwidth}{!}{ \begin{tabular}{ll cccccccccc}
			\midrule
			\textbf{Base} & \textbf{Forecast} & \multicolumn{10}{c}{\textbf{Temporal Frequency}} \\
			\textbf{Forecasts} & \textbf{Method} \\
			\midrule
			&& \textbf{30min} & \textbf{1h} & \textbf{1.5h} & \textbf{2h} & \textbf{3h} & \textbf{4h} &\textbf{6h} & \textbf{8h} & \textbf{12h} & \textbf{24h} \\
			\\
			Naive & Base & 1.0521 & 1.0544 & 1.0573 & 1.0596 & 1.0612 & 1.0677 & 1.0693 & 1.0756 & 1.0838 & 1.0894 \\ 
			& Random forest & \cellcolor{lightgray}0.9883 & \cellcolor{lightgray}1.0030 & \cellcolor{lightgray}1.0099 & \cellcolor{lightgray}1.0112 & \cellcolor{lightgray}1.0135 & \cellcolor{lightgray}1.0151 & \cellcolor{lightgray}1.0032 & \cellcolor{lightgray}0.9886 & \cellcolor{lightgray}1.0017 & \cellcolor{lightgray}0.9797 \\ 
			& XGBoost & 1.0878 & 1.0996 & 1.1018 & 1.1025 & 1.0979 & 1.0969 & 1.0798 & 1.0559 & 1.0682 & 1.0334 \\ 
			& LightGBM & 1.0653 & 1.0816 & 1.0862 & 1.0895 & 1.0873 & 1.0913 & 1.0694 & 1.0577 & 1.0654 & 1.0302 \\ 
			\\
			\\
			ETS & Base & 1.4020 & 1.3554 & 1.3757 & 1.3809 & 1.3639 & 1.2197 & 1.1765 & 1.1796 & 1.1408 & 0.8585 \\ 
			& Bottom Up & 1.4020 & 1.4561 & 1.4847 & 1.5013 & 1.5163 & 1.5216 & 1.5216 & 1.5272 & 1.5378 & 1.4401 \\ 
			& tcs & 1.2976 & 1.3438 & 1.3680 & 1.3818 & 1.3887 & 1.3827 & 1.3802 & 1.3616 & 1.3551 & 1.2208 \\ 
			& cst & 1.3254 & 1.3784 & 1.4051 & 1.4212 & 1.4358 & 1.4345 & 1.4373 & 1.4278 & 1.4327 & 1.3147 \\  
			& ite & 1.3337 & 1.3854 & 1.4114 & 1.4269 & 1.4402 & 1.4394 & 1.4410 & 1.4346 & 1.4399 & 1.3182 \\ 
			& oct & 1.3118 & 1.3582 & 1.3826 & 1.3961 & 1.4033 & 1.3977 & 1.3950 & 1.3806 & 1.3764 & 1.2424 \\ 
			& Random forest & \cellcolor{lightgray}0.9687 & \cellcolor{lightgray}0.9795 & \cellcolor{lightgray}0.9838 & \cellcolor{lightgray}0.9832 & \cellcolor{lightgray}0.9793 & \cellcolor{lightgray}0.9702 & \cellcolor{lightgray}0.9458 & \cellcolor{lightgray}0.9496 & \cellcolor{lightgray}0.9462 & \cellcolor{lightgray}0.9296 \\ 
			& XGBoost & 1.0483 & 1.0571 & 1.0548 & 1.0538 & 1.0473 & 1.0408 & 1.0111 & 1.0111 & 1.0066 & 0.9757 \\ 
			& LightGBM & 1.0193 & 1.0326 & 1.0339 & 1.0350 & 1.0296 & 1.0274 & 0.9964 & 0.9918 & 0.9960 & 0.9642 \\ 
			\\
			\\
			SARIMA & Base & 1.2908 & 1.2452 & 1.2528 & 1.2680 & 1.1935 & 1.0547 & 0.9969 & 0.9857 & 0.8651 & 0.8587 \\ 
			& Bottom Up & 1.2908 & 1.3389 & 1.3626 & 1.3761 & 1.3778 & 1.3732 & 1.3529 & 1.3471 & 1.3231 & 1.2031 \\ 
			& tcs & 1.1860 & 1.2196 & 1.2351 & 1.2411 & 1.2279 & 1.2037 & 1.1789 & 1.1578 & 1.1169 & 0.9793 \\ 
			& cst & 1.1874 & 1.2230 & 1.2390 & 1.2441 & 1.2316 & 1.2062 & 1.1790 & 1.1643 & 1.1213 & 0.9836 \\ 
			& ite & 1.1863 & 1.2201 & 1.2349 & 1.2399 & 1.2256 & 1.1989 & 1.1706 & 1.1527 & 1.1101 & 0.9703 \\ 
			& oct & 1.1859 & 1.2192 & 1.2351 & 1.2416 & 1.2284 & 1.2045 & 1.1796 & 1.1584 & 1.1183 & 0.9809 \\ 
			& Random forest & \cellcolor{lightgray}0.9676 & \cellcolor{lightgray}0.9780 & \cellcolor{lightgray}0.9801 & \cellcolor{lightgray}0.9806 & \cellcolor{lightgray}0.9730 & \cellcolor{lightgray}0.9559 & \cellcolor{lightgray}0.9256 & \cellcolor{lightgray}0.9291 & \cellcolor{lightgray}0.9194 & \cellcolor{lightgray}0.8938 \\ 
			& XGBoost & 1.0392 & 1.0488 & 1.0443 & 1.0434 & 1.0313 & 1.0228 & 0.9883 & 0.9888 & 0.9882 & 0.9483 \\ 
			& LightGBM & 1.0082 & 1.0195 & 1.0200 & 1.0202 & 1.0103 & 1.0022 & 0.9677 & 0.9691 & 0.9643 & 0.9275 \\ 
			\\
			\\
			Forecast & Base & 1.0942 & 1.0615 & 1.0664 & 1.0744 & 1.0440 & 0.9599 & 0.9541 & 0.9418 & 0.8970 & 0.8459 \\ 
			Combination & Bottom Up & 1.0942 & 1.1268 & 1.1424 & 1.1536 & 1.1631 & 1.1600 & 1.1551 & 1.1558 & 1.1574 & 1.0833 \\ 
			& tcs & 1.0409 & 1.0664 & 1.0785 & 1.0862 & 1.0895 & 1.0785 & 1.0722 & 1.0652 & 1.0542 & 0.9696 \\ 
			& cst & 1.0589 & 1.0885 & 1.1024 & 1.1122 & 1.1183 & 1.1107 & 1.1049 & 1.1023 & 1.0946 & 1.0179 \\ 
			& ite & 1.0695 & 1.0986 & 1.1125 & 1.1223 & 1.1281 & 1.1215 & 1.1153 & 1.1141 & 1.1080 & 1.0314 \\ 
			& oct & 1.0385 & 1.0635 & 1.0755 & 1.0824 & 1.0850 & 1.0730 & 1.0665 & 1.0588 & 1.0468 & 0.9587 \\ 
			& Random forest & \cellcolor{lightgray}0.9316 & \cellcolor{lightgray}0.9414 & \cellcolor{lightgray}0.9452 & \cellcolor{lightgray}0.9483 & \cellcolor{lightgray}0.9489 & \cellcolor{lightgray}0.9460 & \cellcolor{lightgray}0.9285 & \cellcolor{lightgray}0.9363 & \cellcolor{lightgray}0.9360 & \cellcolor{lightgray}0.9178 \\ 
			& XGBoost & 1.0174 & 1.0261 & 1.0248 & 1.0251 & 1.0193 & 1.0207 & 0.9991 & 1.0049 & 1.0063 & 0.9821 \\ 
			& LightGBM & 0.9961 & 1.0099 & 1.0105 & 1.0136 & 1.0115 & 1.0136 & 0.9904 & 1.0009 & 1.0045 & 0.9774 \\ 
			
			\midrule
	\end{tabular}}
	\label{tab:Citibike_results_areas_MASE}
	\raggedright
	\footnotesize
	$\;$\\[-0.1cm]
	Notes: This table shows forecast accuracy measured in mean absolute scaled error (MASE). 
	The best forecast reconciliation results for each base forecast (Naive, ETS, SARIMA, Forecast Combination) are highlighted in gray.
	The linear benchmark methods are defined in Table \ref{tabel:linearbenchmarks}.  
	\\[0.1cm]
\end{table}

\begin{table}[htbp]
	\centering
	\caption{
		Citi Bike forecast reconciliation results for the market for the MASE accuracy index.}
	\resizebox{\textwidth}{!}{ \begin{tabular}{ll cccccccccc}
			\midrule
			\textbf{Base} & \textbf{Forecast} & \multicolumn{10}{c}{\textbf{Temporal Frequency}} \\
			\textbf{Forecasts} & \textbf{Method} \\
			\midrule
			&& \textbf{30min} & \textbf{1h} & \textbf{1.5h} & \textbf{2h} & \textbf{3h} & \textbf{4h} &\textbf{6h} & \textbf{8h} & \textbf{12h} & \textbf{24h} \\
			\\
			Naive & Base & 1.0962 & 1.0970 & 1.0989 & 1.1011 & 1.1026 & 1.1088 & 1.1124 & 1.1226 & 1.1300 & 1.1368 \\ 
			& Random forest & \cellcolor{lightgray}1.0760 & \cellcolor{lightgray}1.0720 & \cellcolor{lightgray}1.0706 & \cellcolor{lightgray}1.0674 & \cellcolor{lightgray}1.0645 & \cellcolor{lightgray}1.0613 & \cellcolor{lightgray}1.0489 & \cellcolor{lightgray}1.0318 & \cellcolor{lightgray}1.0370 & \cellcolor{lightgray}1.0186 \\ 
			& XGBoost & 1.1294 & 1.1222 & 1.1169 & 1.1119 & 1.1034 & 1.0988 & 1.0868 & 1.0664 & 1.0646 & 1.0399 \\ 
			& LightGBM & 1.1380 & 1.1321 & 1.1290 & 1.1270 & 1.1194 & 1.1211 & 1.1018 & 1.0822 & 1.0809 & 1.0497 \\ 
			\\
			\\
			ETS & Base & 1.8760 & 1.5007 & 1.4829 & 1.4706 & 1.4051 & 1.2403 & 1.1622 & 1.1718 & 1.1097 & 0.8897 \\ 
			& Bottom Up & 1.6934 & 1.7012 & 1.7100 & 1.7126 & 1.7005 & 1.6991 & 1.6914 & 1.6866 & 1.7156 & 1.6726 \\ 
			& tcs & 1.5184 & 1.5215 & 1.5262 & 1.5262 & 1.5066 & 1.4905 & 1.4813 & 1.4330 & 1.4347 & 1.3672 \\ 
			& cst & 1.5325 & 1.5366 & 1.5406 & 1.5402 & 1.5226 & 1.5053 & 1.4980 & 1.4514 & 1.4620 & 1.3818 \\ 
			& ite & 1.5684 & 1.5736 & 1.5784 & 1.5783 & 1.5620 & 1.5485 & 1.5402 & 1.5058 & 1.5238 & 1.4488 \\ 
			& oct & 1.5475 & 1.5518 & 1.5571 & 1.5571 & 1.5393 & 1.5257 & 1.5166 & 1.4754 & 1.4891 & 1.4147 \\ 
			& Random forest & \cellcolor{lightgray}1.0518 & \cellcolor{lightgray}1.0400 & \cellcolor{lightgray}1.0360 & \cellcolor{lightgray}1.0283 & \cellcolor{lightgray}1.0126 & \cellcolor{lightgray}0.9974 & \cellcolor{lightgray}0.9691 & \cellcolor{lightgray}0.9593 & \cellcolor{lightgray}0.9691 & \cellcolor{lightgray}0.9530 \\ 
			& XGBoost & 1.0877 & 1.0765 & 1.0701 & 1.0626 & 1.0459 & 1.0359 & 1.0060 & 0.9939 & 1.0071 & 0.9792 \\ 
			& LightGBM & 1.0722 & 1.0622 & 1.0561 & 1.0525 & 1.0341 & 1.0281 & 0.9942 & 0.9873 & 1.0004 & 0.9863 \\ 
			\\
			\\
			SARIMA & Base & 1.4241 & 1.3784 & 1.3821 & 1.3950 & 1.2470 & 1.0891 & 1.0604 & 1.0529 & 0.9037 & 0.9511 \\ 
			& Bottom Up & 1.3988 & 1.3967 & 1.3964 & 1.3937 & 1.3598 & 1.3250 & 1.3018 & 1.2380 & 1.2006 & 1.1089 \\ 
			& tcs & 1.3385 & 1.3347 & 1.3317 & 1.3260 & 1.2867 & 1.2465 & 1.2145 & 1.1630 & 1.1091 & 1.0042 \\ 
			& cst & 1.3236 & 1.3187 & 1.3152 & 1.3067 & 1.2641 & 1.2209 & 1.1852 & 1.1464 & 1.0817 & 0.9708 \\  
			& ite & 1.3393 & 1.3346 & 1.3314 & 1.3235 & 1.2818 & 1.2403 & 1.2039 & 1.1626 & 1.0997 & 0.9936 \\ 
			& oct & 1.3396 & 1.3358 & 1.3330 & 1.3271 & 1.2885 & 1.2481 & 1.2176 & 1.1653 & 1.1123 & 1.0076 \\ 
			& Random forest & \cellcolor{lightgray}1.0460 & \cellcolor{lightgray}1.0319 & \cellcolor{lightgray}1.0252 & \cellcolor{lightgray}1.0174 & \cellcolor{lightgray}0.9957 & \cellcolor{lightgray}0.9753 & \cellcolor{lightgray}0.9419 & \cellcolor{lightgray}0.9377 & \cellcolor{lightgray}0.9292 & \cellcolor{lightgray}0.9098 \\ 
			& XGBoost & 1.0529 & 1.0383 & 1.0281 & 1.0208 & 1.0034 & 0.9937 & 0.9601 & 0.9518 & 0.9568 & 0.9235 \\ 
			& LightGBM & 1.0414 & 1.0286 & 1.0190 & 1.0101 & 0.9918 & 0.9800 & 0.9496 & 0.9437 & 0.9444 & 0.9174 \\ 
			\\
			\\
			Forecast & Base & 1.2718 & 1.1534 & 1.1475 & 1.1505 & 1.0788 & 0.9933 & 0.9871 & 0.9771 & 0.9168 & 0.9045 \\ 
			Combination & Bottom Up & 1.2289 & 1.2324 & 1.2330 & 1.2357 & 1.2257 & 1.2148 & 1.2013 & 1.1833 & 1.1873 & 1.1493 \\ 
			& tcs & 1.1703 & 1.1716 & 1.1719 & 1.1726 & 1.1604 & 1.1407 & 1.1308 & 1.1039 & 1.0927 & 1.0417 \\ 
			& cst & 1.1876 & 1.1895 & 1.1897 & 1.1909 & 1.1801 & 1.1608 & 1.1502 & 1.1281 & 1.1186 & 1.0641 \\ 
			& ite & 1.2159 & 1.2188 & 1.2199 & 1.2222 & 1.2121 & 1.1949 & 1.1834 & 1.1661 & 1.1617 & 1.1101 \\ 
			& oct & 1.1654 & 1.1663 & 1.1669 & 1.1669 & 1.1542 & 1.1338 & 1.1250 & 1.0961 & 1.0825 & 1.0299 \\ 
			& Random forest & \cellcolor{lightgray}1.0026 & \cellcolor{lightgray}0.9950 & \cellcolor{lightgray}0.9920 & \cellcolor{lightgray}0.9904 & \cellcolor{lightgray}0.9829 & \cellcolor{lightgray}0.9722 & \cellcolor{lightgray}0.9531 & \cellcolor{lightgray}0.9649 & \cellcolor{lightgray}0.9559 & \cellcolor{lightgray}0.9567 \\ 
			& XGBoost & 1.0369 & 1.0276 & 1.0240 & 1.0207 & 1.0132 & 1.0012 & 0.9836 & 0.9913 & 0.9849 & 0.9813 \\ 
			& LightGBM & 1.0307 & 1.0227 & 1.0177 & 1.0197 & 1.0084 & 1.0027 & 0.9794 & 0.9897 & 0.9904 & 0.9795 \\ 
			
			\midrule
	\end{tabular}}
	\label{tab:Citibike_results_market_MASE}
	\raggedright
	\footnotesize
	$\;$\\[-0.1cm]
	Notes: This table shows forecast accuracy measured in mean absolute scaled error (MASE). 
	The best forecast reconciliation results for each base forecast (Naive, ETS, SARIMA, Forecast Combination) are highlighted in gray.
	The linear benchmark methods are defined in Table \ref{tabel:linearbenchmarks}.  
	\\[0.1cm]
\end{table}

\renewcommand{\thetable}{C.\arabic{table}}
\setcounter{table}{0}

\renewcommand{\thefigure}{C.\arabic{figure}}
\setcounter{figure}{0}

\clearpage
\newpage

\section{ Citi Bike Sensitivity Analyses \label{app:citibike}}

\begin{table}[h!]
	\centering
	\caption{
		Citi Bike forecast reconciliation results: sensitivity analysis features matrix.}
	\resizebox{\textwidth}{!}{ \begin{tabular}{ll cccccccccc}
			\midrule
			\textbf{Base} & \textbf{Forecast} & \multicolumn{10}{c}{\textbf{Temporal Frequency}} \\
			\textbf{Forecasts} & \textbf{Method} \\ 
			\midrule
			\multicolumn{2}{l}{\textbf{H3 cells results}}     & \textbf{30min} & \textbf{1h} & \textbf{1.5h} & \textbf{2h} & \textbf{3h} & \textbf{4h} &\textbf{6h} & \textbf{8h} & \textbf{12h} & \textbf{24h} \\
			\\
			Naive & Random forest & 0.9975 & 0.9985 & 0.9979 & 0.9996 & 0.9988 & 1.0019 & 0.9945 & 0.9960 & 0.9952 & 0.9936 \\ 
			& XGBoost & 0.9871 & 0.9895 & 0.9878 & 0.9866 & 0.9856 & 0.9887 & 0.9827 & 0.9914 & 0.9875 & 0.9846 \\ 
			& LightGBM & 0.9883 & 0.9899 & 0.9897 & 0.9886 & 0.9894 & 0.9932 & 0.9888 & 0.9909 & 0.9942 & 0.9954 \\ 
			\\
			%\\
			ETS & Random forest & 0.9752 & 0.9729 & 0.9725 & 0.9726 & 0.9719 & 0.9713 & 0.9705 & 0.9690 & 0.9747 & 0.9782 \\ 
			& XGBoost & 0.9595 & 0.9613 & 0.9620 & 0.9603 & 0.9605 & 0.9570 & 0.9520 & 0.9462 & 0.9569 & 0.9536 \\ 
			& LightGBM & 0.9599 & 0.9584 & 0.9585 & 0.9562 & 0.9568 & 0.9522 & 0.9497 & 0.9565 & 0.9564 & 0.9616 \\ 
			\\
			%\\
			SARIMA & Random forest & 0.9861 & 0.9827 & 0.9842 & 0.9850 & 0.9848 & 0.9896 & 0.9870 & 0.9948 & 0.9983 & 1.0038 \\ 
			& XGBoost & 0.9803 & 0.9829 & 0.9840 & 0.9819 & 0.9842 & 0.9789 & 0.9819 & 0.9816 & 0.9793 & 0.9903 \\ 
			& LightGBM & 0.9841 & 0.9867 & 0.9869 & 0.9871 & 0.9896 & 0.9909 & 0.9868 & 0.9953 & 0.9952 & 1.0061 \\ 
			\\
			%\\
			Forecast & Randomforest & 0.9882 & 0.9864 & 0.9852 & 0.9851 & 0.9837 & 0.9889 & 0.9865 & 0.9884 & 0.9888 & 0.9924 \\ 
			Combination & XGBoost & 0.9752 & 0.9754 & 0.9731 & 0.9736 & 0.9721 & 0.9705 & 0.9652 & 0.9705 & 0.9600 & 0.9640 \\ 
			& LightGBM & 0.9700 & 0.9696 & 0.9684 & 0.9676 & 0.9680 & 0.9670 & 0.9640 & 0.9662 & 0.9644 & 0.9738 \\ \\
			
			\midrule
			\multicolumn{2}{l}{\textbf{Market results}}     & \textbf{30min} & \textbf{1h} & \textbf{1.5h} & \textbf{2h} & \textbf{3h} & \textbf{4h} &\textbf{6h} & \textbf{8h} & \textbf{12h} & \textbf{24h} \\
			\\
			Naive & Randomforest & 1.0191 & 1.0194 & 1.0203 & 1.0210 & 1.0173 & 1.0212 & 1.0118 & 1.0155 & 1.0144 & 1.0102 \\ 
			& XGBoost & 1.0351 & 1.0351 & 1.0369 & 1.0367 & 1.0357 & 1.0396 & 1.0281 & 1.0372 & 1.0460 & 1.0511 \\ 
			& LightGBM & 1.0245 & 1.0265 & 1.0267 & 1.0269 & 1.0265 & 1.0274 & 1.0184 & 1.0285 & 1.0357 & 1.0411 \\ 
			\\
			%\\
			ETS & Random forest & 1.0041 & 1.0021 & 1.0016 & 1.0015 & 1.0039 & 0.9997 & 1.0001 & 1.0055 & 0.9951 & 1.0033 \\ 
			& XGBoost & 0.9889 & 0.9903 & 0.9913 & 0.9923 & 0.9948 & 0.9868 & 0.9860 & 0.9849 & 0.9775 & 0.9780 \\ 
			& LightGBM & 0.9987 & 0.9985 & 0.9991 & 0.9960 & 1.0025 & 0.9888 & 0.9952 & 0.9974 & 0.9905 & 0.9812 \\ 
			\\
			%\\
			SARIMA & Random forest & 1.0262 & 1.0239 & 1.0248 & 1.0275 & 1.0291 & 1.0310 & 1.0311 & 1.0325 & 1.0336 & 1.0357 \\ 
			& XGBoost & 1.0381 & 1.0431 & 1.0458 & 1.0461 & 1.0450 & 1.0343 & 1.0402 & 1.0443 & 1.0273 & 1.0387 \\ 
			& LightGBM & 1.0468 & 1.0500 & 1.0523 & 1.0558 & 1.0562 & 1.0524 & 1.0509 & 1.0512 & 1.0461 & 1.0542 \\ 
			\\
			%\\
			Forecast & Random forest & 1.0137 & 1.0131 & 1.0134 & 1.0149 & 1.0104 & 1.0180 & 1.0149 & 1.0115 & 1.0146 & 1.0082 \\ 
			Combination & XGBoost & 1.0172 & 1.0177 & 1.0147 & 1.0192 & 1.0106 & 1.0227 & 1.0109 & 1.0151 & 1.0171 & 0.9953 \\ 
			& LightGBM & 1.0209 & 1.0211 & 1.0222 & 1.0208 & 1.0182 & 1.0202 & 1.0191 & 1.0152 & 1.0198 & 1.0175 \\ 
			\midrule
	\end{tabular}}
	\label{tab:sensitivity_bigX_citibike_results_areas}
	\raggedright
	\footnotesize
	$\;$\\[-0.1cm]
	Notes: This table shows relative forecast accuracy measured in weighted absolute percentage error (WAPE) when using the complete features matrix instead of the compact features matrix. A value below one indicates better performance when using the former compared to the latter.   
	\\[0.1cm]
\end{table}

\begin{table}[htbp]
	\caption{
		Citi Bike forecast reconciliation results: sensitivity analysis temporal frequencies.}
	\centering
	\footnotesize
	\begin{tabular}{ll c c c |ccc}
		\hline
		\textbf{Period} & \textbf{Forecast} & \multicolumn{3}{c}{\textbf{H3 cells}} &  \multicolumn{3}{c}{\textbf{Market}} \\
		\textbf{} & \textbf{Method} \\
		\hline
		&& \textbf{30min} & \textbf{1h} & \textbf{24h} &\textbf{30min} & \textbf{1h} & \textbf{24h} \\
		\\
		Naive & Base & 1.0000 & 1.0000 & 1.0000 & 1.0000 & 1.0000 & 1.0000 \\ 
		& Random forest & 0.9708 & 0.9750 & 0.9874 & 0.9674 & 0.9739 & 0.9924 \\ 
		& XGBoost & 0.9426 & 0.9503 & 0.9217 & 0.9453 & 0.9551 & 0.9538 \\ 
		& LightGBM & 0.9409 & 0.9459 & 0.9294 & 0.9529 & 0.9614 & 0.9615 \\ 
		\\
		\\
		ETS &Base & 1.0000 & 1.0000 & 1.0000 & 1.0000 & 1.0000 & 1.0000 \\ 
		&Bottom Up & 1.0000 & 1.0000 & 1.0000 & 1.0000 & 1.0000 & 1.0000 \\ 
		&tcs & 0.9464 & 0.9419 & 0.8682 & 0.9256 & 0.9239 & 0.8648 \\ 
		&cst & 0.9318 & 0.9270 & 0.8581 & 0.9164 & 0.9148 & 0.8487 \\ 
		&ite & 0.9306 & 0.9254 & 0.8540 & 0.9235 & 0.9221 & 0.8671 \\ 
		&oct & 0.9550 & 0.9515 & 0.8910 & 0.9330 & 0.9316 & 0.8777 \\ 
		&Random forest & 0.9218 & 0.9163 & 0.9544 & 0.9003 & 0.8967 & 0.9739 \\ 
		&XGBoost & 0.8984 & 0.8987 & 0.8865 & 0.8911 & 0.8947 & 0.9371 \\ 
		&LightGBM & 0.8930 & 0.8893 & 0.8809 & 0.8840 & 0.8865 & 0.9468 \\ 
		\\
		\\
		SARIMA& Base & 1.0000 & 1.0000 & 1.0000 & 1.0000 & 1.0000 & 1.0000 \\ 
		&Bottom Up & 1.0000 & 1.0000 & 1.0000 & 1.0000 & 1.0000 & 1.0000 \\ 
		& tcs & 0.9558 & 0.9516 & 0.9140 & 0.9749 & 0.9750 & 0.9528 \\ 
		& cst & 0.9534 & 0.9497 & 0.9123 & 0.9685 & 0.9683 & 0.9447 \\ 
		& ite & 0.9537 & 0.9492 & 0.9094 & 0.9759 & 0.9760 & 0.9579 \\ 
		& oct & 0.9556 & 0.9516 & 0.9070 & 0.9751 & 0.9750 & 0.9516 \\ 
		& Random forest & 0.8958 & 0.8881 & 0.9228 & 0.8553 & 0.8494 & 0.9141 \\ 
		& XGBoost & 0.8720 & 0.8719 & 0.8829 & 0.8339 & 0.8351 & 0.9014 \\ 
		& LightGBM & 0.8687 & 0.8627 & 0.8637 & 0.8271 & 0.8281 & 0.8941 \\ 
		\\
		\\
		Forecast & Base & 1.0000 & 1.0000 & 1.0000 & 1.0000 & 1.0000 & 1.0000 \\ 
		Combination & Bottom Up & 1.0000 & 1.0000 & 1.0000 & 1.0000 & 1.0000 & 1.0000 \\ 
		& tcs & 0.9588 & 0.9536 & 0.9047 & 0.9510 & 0.9492 & 0.9067 \\ 
		& cst & 0.9470 & 0.9416 & 0.8932 & 0.9394 & 0.9376 & 0.8958 \\ 
		& ite & 0.9476 & 0.9418 & 0.8944 & 0.9463 & 0.9449 & 0.9104 \\ 
		& oct & 0.9700 & 0.9665 & 0.9292 & 0.9669 & 0.9654 & 0.9306 \\ 
		& Random forest & 0.9588 & 0.9586 & 0.9720 & 0.9337 & 0.9345 & 0.9548 \\ 
		& XGBoost & 0.9276 & 0.9296 & 0.9092 & 0.9038 & 0.9063 & 0.9117 \\ 
		& LightGBM & 0.9302 & 0.9316 & 0.9117 & 0.9080 & 0.9118 & 0.9041 \\ 
		\midrule
	\end{tabular}%}
\label{tab:sensitvity_3freq_citibike_results}
% \footnotesize
$\;$\\[0.2cm]
\raggedright
Notes: This table shows relative forecast accuracy measured in weighted absolute percentage error (WAPE)  when  using all ten temporal frequencies for reconciliation compared to using only the temporal frequencies of interest. A value below one indicates better performance when using the former compared to the latter.  \\[0.1cm]
\end{table}

\newpage
\clearpage
\setcounter{table}{0}
\setcounter{figure}{0}

%%%%%
\end{appendices}
\end{document}